%% file: Biophys1.tex
\documentstyle[11pt]{article}
\newcommand{\writingdate}{May 26th, 2000}
\input biophyscommands.tex

\input epsf.tex
\begin{document}
\title{\Large\bf Statistical Mechanics
of Recurrent Neural Networks\\
I. Statics}
\author{\large\bf A.C.C. Coolen\\[3mm]
Department of Mathematics, King's College London\\
Strand, London WC2R 2LS, UK}
\date{\writingdate}
\maketitle

\tableofcontents


\clearpage\section{Introduction}

Statistical mechanics deals with large systems of stochastically
interacting microscopic elements (particles, atomic magnets, polymers, etc.).
The strategy of statistical mechanics is to abandon any ambition to solve models of such systems at the
microscopic level of individual elements,
but to use the microscopic laws to
calculate equations describing the behaviour of a suitably chosen set of {\em macroscopic} observables.
The toolbox of statistical mechanics consists of methods
to perform this reduction from the microscopic to a macroscopic level, which are all based
on efficient ways to do the bookkeeping of probabilities. The experience and intuition
that has been built up over the last century tells us what to expect,
and serves as a guide in finding  the macroscopic observables and in seeing the
difference between relevant mathematical subtleties and irrelevant ones.
As in any statistical theory, clean and transparent mathematical laws
can be expected to emerge only for large (preferably infinitely large) systems.
In this limit one often encounters phase transitions, i.e. drastic
changes in the system's macroscopic behaviour at specific values
of global control parameters.

Recurrent neural networks, i.e. neural networks with synaptic feedback loops, appear to meet the
criteria for statistical mechanics to apply, provided we indeed restrict ourselves to
large systems. Here the microscopic stochastic dynamical variables are the
firing states of the neurons or their membrane potentials, and one is mostly interested in
quantities such as average state correlations and global information processing quality,
which are indeed measured by macroscopic observables.
In contrast to layered networks, one cannot simply write down
the values of successive neuron states for models of recurrent neural
networks; here they must be solved from (mostly stochastic) coupled dynamic equations.
Under special conditions (`detailed balance'), which usually translate
into the requirement of synaptic symmetry, the stochastic process of evolving neuron states leads towards an
equilibrium situation where the microscopic state probabilities are
known, and where the techniques of {\em equilibrium statistical mechanics} can be
applied in one form or another. The equilibrium distribution found, however, will
not always be of the conventional Boltzmann form. For non-symmetric networks, where
the asymptotic (stationary) statistics are not known,
dynamical techniques from {\em non-equilibrium statistical mechanics} are the only tools available
for analysis. The `natural' set of macroscopic quantities
(or `order parameters') to be calculated can be defined in practice as the smallest
set which will obey closed deterministic
equations in the limit of an infinitely large network.

\begin{figure}[t]
\vspace*{-3mm}
\begin{center}
\epsfxsize=40mm\epsfbox{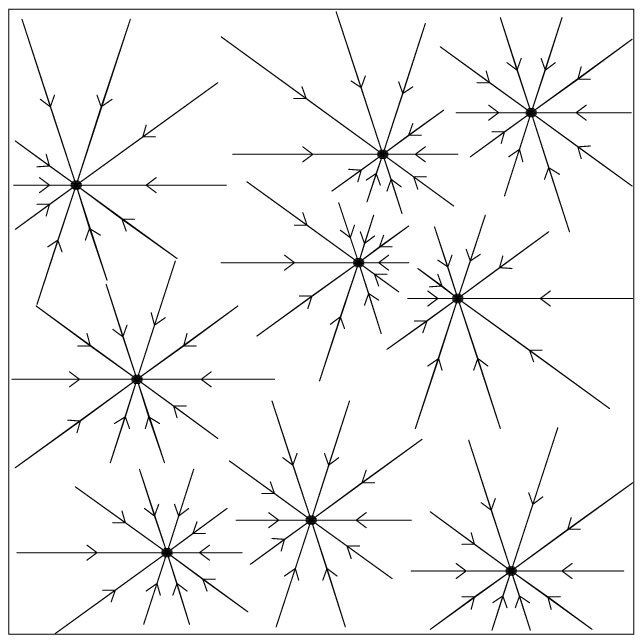}
\hspace*{20mm}
\epsfxsize=40mm\epsfbox{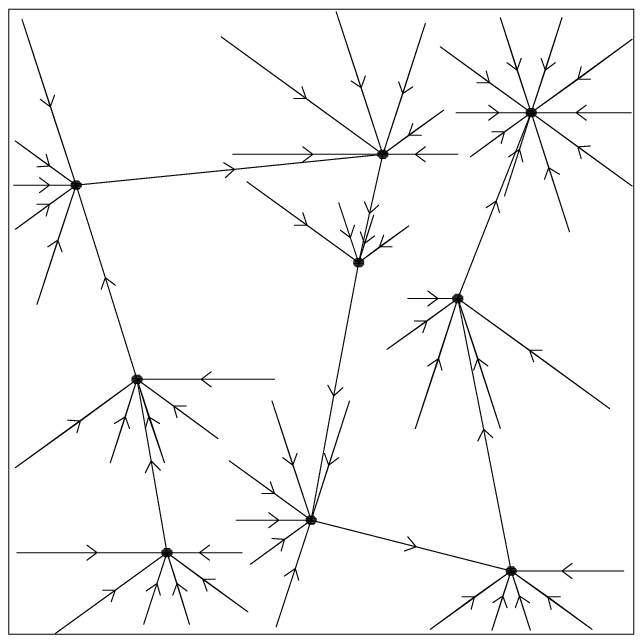}
\end{center}
\vspace*{-3mm}
\caption{Information processing by recurrent neural
networks through the creation and manipulation of attractors in state space.
Patterns stored: the microscopic states $\bullet$.
If the synapses are symmetric we will generally find that the attractors will
have to be fixed-points (left picture). With non-symmetric synapses, the attractors can also be
sequences of microscopic states (right picture).}
\label{fig:attractors}
\end{figure}

Being high-dimensional non-linear systems with extensive
feedback, the dynamics of recurrent neural networks are
generally dominated by a wealth of attractors (fixed-point attractors,
limit-cycles, or even more exotic types), and the practical use of
recurrent neural networks (in both biology and engineering) lies in the potential for
creation and manipulation of these attractors through
adaptation of the network parameters (synapses and thresholds).
Input fed into a recurrent neural network usually serves to
induce a specific initial configuration (or firing pattern) of the
neurons, which serves as a cue, and the `output' is given by the (static or dynamic)
attractor which has been triggered by this cue.
The most familiar types of recurrent neural network models, where the idea
of creating and manipulating attractors has been worked out and applied explicitly,
are the so-called attractor
neural networks for associative memory, designed to store and retrieve information in the form
of neuronal firing patterns and/or sequences
of neuronal firing patterns.
Each pattern to be stored is represented as a microscopic state vector.
One then constructs synapses and  thresholds such that
the dominant attractors of the network are precisely the pattern vectors
(in the case of static recall), or where, alternatively, they are trajectories in which the patterns
 are successively generated microscopic system states.
From an initial configuration (the `cue', or input pattern to be recognised) the system
is allowed to evolve in time autonomously, and
the final state (or trajectory) reached can be interpreted as the
pattern (or pattern sequence) recognized by network from the input
(see figure \ref{fig:attractors}).
For such programmes to work one clearly needs recurrent neural networks with extensive
`ergodicity breaking': the state vector will during the course of the dynamics
(at least on finite time-scales) have to be
confined to a restricted region of state space (an `ergodic component'), the location of which is
to depend strongly on the initial conditions.
Hence our interest will mainly be in systems with many attractors.
This, in turn, has implications at a theoretical/mathematical
level: solving models of recurrent neural networks with extensively many attractors
requires advanced tools from disordered systems theory, such as
replica theory (statics) and generating functional analysis
(dynamics).
It will turn out that a crucial issue is whether or not the
synapses are symmetric. Firstly, synaptic asymmetry is found to rule out microscopic
equilibrium, which has
implications for the mathematical techniques which are available:
studying models of recurrent networks with non-symmetric synapses
requires solving the dynamics, even if one is only interested in the stationary state.
Secondly, the degree of synaptic asymmetry turns out to be a deciding factor in
determining to what extent the dynamics will be glassy, i.e. extremely
slow and non-trivial, close to saturation (where one has an extensive
number of attractors).

In this paper (on statics) and its sequel (on dynamics) I will discuss only the statistical
mechanical analysis of neuronal firing processes in recurrent networks with
static synapses, i.e. network operation as opposed to network learning. I will
also restrict myself to networks with either full or randomly
diluted connectivity, the area in which the main progress has
been made during the last few decades. Apart from these
restrictions, the text aims to be reasonably comprehensive and
self-contained. Even within the confined area of the operation of
recurrent neural networks a truly impressive amount has been achieved, and many
of the constraints on mathematical models which were once thought to be essential for retaining
solvability but which were regrettable from a biological point
of view (such as  synaptic symmetry, binary neuron states,
instantaneous neuronal communication, a small number of attractors, etc.)
have by now been removed with success.
At the beginning of the new millennium we know much more
about the dynamics and statics of recurrent neural networks than ever before.
I aim to cover in a more or less unified manner the most important models and
techniques which have been launched over the years, ranging from simple symmetric and non-symmetric
networks with only a finite number of attractors, to the more complicated
ones with an extensive number, and I will explain in detail the
techniques which have been designed and used to solve them.

In the present paper I will first discuss and
solve various members of the simplest class of models: those where all
synapses are the same. Then I turn to
the Hopfield model, which is the archetypical model to describe the
functioning of symmetric neural networks as associative memories
(away from saturation, where the number of attractors is finite), and to a coupled oscillator model
storing phase patterns (again away from saturation).
Next I will discuss a model with Gaussian synapses, where the number of attractors diverges,
in order to introduce
the so-called replica method, followed by a section on the solution of the
Hopfield model near saturation.
I close this paper with a guide to further references and an assessment of the
past and future deliverables of the equilibrium statistical mechanical analysis
of recurrent neural networks.


\section{Definitions \& Properties of Microscopic Laws}

In this section I define the most common microscopic models
for recurrent neural networks, I show how one can derive the corresponding
descriptions of the stochastic evolution in terms of evolving state probabilities,
and I discuss some fundamental statistical mechanical properties.

\subsection{Stochastic Dynamics of Neuronal Firing States}

{\em Microscopic Definitions for Binary Neurons.}
The simplest non-trivial definition of a recurrent neural network is that where
$N$ binary neurons $\sigma_i\in\{-1,1\}$ (in which the states `1' and `-1' represent firing and rest, respectively)
 respond iteratively and synchronously to post-synaptic potentials (or local fields) $h_i(\bsigma)$,
with $\bsigma=(\sigma_1,\ldots,\sigma_N)$.
The fields are
assumed to
depend linearly on the instantaneous neuron states:
\be
{\sl Parallel:}~~~~~~~~~~
\sigma_i(\ell\plus 1) =\sgn\left[h_i(\bsigma(\ell))+T\eta_i(\ell)\right]
~~~~~~~~~~
h_i(\bsigma)=\sum_{j} J_{ij}\sigma_j+\theta_i
\label{eq:Ising_parallel}
\ee
The stochasticity is in the independent random numbers
$\eta_i(\ell)\in\Re$ (representing threshold noise), which are all drawn according
to some distribution $w(\eta)$.
The parameter $T$ is introduced to control the amount of noise. For $T=0$ the process
(\ref{eq:Ising_parallel}) is deterministic: $\sigma_i(\ell+1)=\sgn[h_i(\bsigma(\ell))]$.
The opposite extreme is choosing $T=\infty$, here the system evolution is fully random. The external
fields $\theta_i$ represent neural thresholds and/or external
stimuli, $J_{ij}$ represents the synaptic efficacy at the junction
$j\to i$ ($J_{ij}>0$ implies excitation, $J_{ij}<0$
inhibition).
Alternatively we could decide that at each iteration step $\ell$
only a single randomly drawn neuron
$\sigma_{i_\ell}$
is to undergo an update of
the type (\ref{eq:Ising_parallel}):
\be
{\sl Sequential:}~~~~~~~~~~
\begin{array}{ll}
i\neq i_\ell: & \sigma_i(\ell\plus 1)=\sigma_i(\ell) \\[-1mm]
i=i_\ell: &
\sigma_i(\ell\plus 1) =\sgn\left[h_i(\bsigma(\ell))+T\eta_i(\ell)\right]
\room
\end{array}
\label{eq:Ising_sequential}
\ee
with the local fields as in (\ref{eq:Ising_parallel}).
The stochasticity is now both in the independent random numbers
$\eta_i(\ell)$ (the threshold noise) and in the site $i_\ell$ to be updated,
drawn randomly from the set $\{1,\ldots,N\}$.
For simplicity we assume $w(-\eta)=w(\eta)$, and define
\bd
g[z]=2\int_0^z\!d\eta~w(\eta):~~~~~~~~~~g[-z]=-g[z],~~~~~\lim_{z\to\pm\infty}g[z]=\pm
1,~~~~~\frac{d}{dz}g[z]\geq 0
\ed
Popular choices for the threshold noise
distributions are
\bd
w(\eta)=(2\pi)^{-\frac{1}{2}} e^{-\frac{1}{2}\eta^2}:~~~g[z]={\rm
Erf}[z/\sqrt{2}],~~~~~~~~~~~~~~~~~~~
w(\eta)=\frac{1}{2}[1\minus \tanh^2(\eta)]:~~~g[z]=\tanh(z)
\ed
\vsp

\noindent{\em From Stochastic Equations to Evolving Probabilities}.
From the microscopic equations (\ref{eq:Ising_parallel},\ref{eq:Ising_sequential}), which are
suitable for numerical simulations,  we can
derive an equivalent but mathematically more convenient
description in terms of
microscopic state probabilities $p_\ell(\bsigma)$.
Equations (\ref{eq:Ising_parallel},\ref{eq:Ising_sequential}) state
that, if the system  state $\bsigma(\ell)$  is given,
a neuron $i$ to be updated will obey
\be
{\rm Prob}\left[\sigma_{i}(\ell\plus 1)\right]=
\frac{1}{2}\left[1+\sigma_{i}(\ell\plus1) ~g[\beta h_{i}(\bsigma(\ell))]\right]
\label{eq:single_update}
\ee
with $\beta=T^{-1}$.
In the case (\ref{eq:Ising_parallel}) this rule applies to all neurons, and thus
we simply get $p_{\ell+1}(\bsigma)
=\prod_{i=1}^{N}\frac{1}{2}\left[1\plus\sigma_i ~g[\beta
h_i(\bsigma(\ell))]\right]$.
If, on the other hand, instead of $\bsigma(\ell)$ only the probability distribution
$p_\ell(\bsigma)$ is given, this expression for $p_{\ell+1}(\bsigma)$ is to be averaged over the possible states at time $\ell$:
\be
{\sl Parallel:}~~~~~~~~~
p_{\ell+1}(\bsigma)=\sum_{\bsigma^{\prime}}
W\left[\bsigma;\bsigma^{\prime}\right]
p_\ell(\bsigma^{\prime})
~~~~~~~~~
W\left[\bsigma;\bsigma^{\prime}\right]=
\prod_{i=1}^{N}\frac{1}{2}\left[1+\sigma_i~ g[\beta
h_i(\bsigma^\prime)]\right]
\label{eq:Ising_parallel_Markov}
\ee
This is the standard representation of a Markov chain.
Also the sequential process (\ref{eq:Ising_sequential}) can
be formulated in terms of probabilities, but
here expression (\ref{eq:single_update}) applies only to the randomly drawn candidate
$i_\ell$.
After averaging over all possible realisations of the sites $i_\ell$ we obtain:
\bd
p_{\ell+1}(\bsigma)
=\frac{1}{N}\sum_{i}\left\{
[\prod_{j\neq i}\delta_{\sigma_j,\sigma_j(\ell)}]~
\frac{1}{2}\left[1+\sigma_i~ g[\beta
h_i(\bsigma(\ell))]\right]\right\}
\ed
(with the Kronecker symbol: $\delta_{ij}=1$ if $i=j$, $\delta_{ij}=0$ otherwise).
If, instead of $\bsigma(\ell)$, the probabilities
$p_\ell(\bsigma)$ are given, this expression is to be averaged over
the possible states at time $\ell$, with the result:
\bd
p_{\ell+1}(\bsigma)=
\frac{1}{N}\sum_{i}
\frac{1}{2}\left[1+\sigma_i~g[\beta
h_i(\bsigma)]\right]p_\ell(\bsigma)
+
\frac{1}{N}\sum_{i}
\frac{1}{2}\left[1+\sigma_i ~g[\beta
h_i(F_i\bsigma)]\right] p_\ell(F_i\bsigma)
\ed
with the state-flip operators
$F_i\Phi(\bsigma)=\Phi(\sigma_1,\ldots,\sigma_{i-1},\minus \sigma_i,\sigma_{i+1},\ldots,\sigma_N)$.
This equation can again be written in the standard form $p_{\ell+1}(\bsigma)=\sum_{\bsigma^{\prime}}
W\left[\bsigma;\bsigma^{\prime}\right]p_\ell(\bsigma^\prime)$, but now with
the transition matrix
\be
{\sl Sequential:}~~~~~~~~~~
W\left[\bsigma;\bsigma^{\prime}\right]=\delta_{\bsigma,\bsigma^{\prime}}
+\frac{1}{N}\sum_{i}\left\{
w_i(F_i\bsigma)\delta_{\bsigma,F_i\bsigma^{\prime}}-
w_i(\bsigma)\delta_{\bsigma,\bsigma^{\prime}}\right\}
\label{eq:Ising_sequential_Markov}
\ee
where $\delta_{\bsigma,\bsigma^\prime}=\prod_{i} \delta_{\sigma_i,\sigma_i^\prime}$
and
\be
w_i(\bsigma)=\frac{1}{2}\left[1\minus\sigma_i\tanh\left[\beta
h_i(\bsigma)\right]\right]
\label{eq:transitionrates}
\ee
Note that, as soon as $T>0$, the two transition matrices $W[\bsigma;\bsigma^\prime]$
in (\ref{eq:Ising_parallel_Markov},\ref{eq:Ising_sequential_Markov})
both describe {\em
ergodic} systems: from any initial state $\bsigma^\prime$ one can reach
any final state $\bsigma$ with nonzero probability in a
finite number of steps (being one in the parallel case, and $N$ in the sequential case).
It now follows from the standard theory of stochastic processes
(see e.g. \cite{VanKampen,Gardiner}) that in both cases the system evolves
towards a unique stationary distribution $p_\infty(\bsigma)$,
where all probabilities $p_{\infty}(\bsigma)$ are non-zero.
\vsp

\noindent{\em From Discrete to Continuous Times}.
The above processes have the (mathematically and biologically)
less appealing property that time is measured in discrete units.
For the sequential case we will now assume that the
{\em duration} of each of the iteration steps is
a continuous random number (for parallel dynamics this would make little sense,
since all updates would still be made in full synchrony). The statistics of the
durations are described by a function $\pi_\ell(t)$, defined as the
probability that at time $t$ precisely $\ell$ updates have been
made. Upon denoting the previous discrete-time probabilities as $\hat{p}_\ell(\bsigma)$,
our new process (which now includes the randomness in step duration) will be described by
\bd
p_t(\bsigma)=\sum_{\ell\geq 0}\pi_\ell(t)\hat{p}_\ell(\bsigma)
=
\sum_{\ell\geq 0}\pi_\ell(t)
\sum_{\bsigma^{\prime}}W^\ell\left[\bsigma;\bsigma^{\prime}\right]p_0(\bsigma^{\prime})
\ed
and time has become a continuous variable. For $\pi_\ell(t)$ we make the
Poisson
choice $\pi_\ell(t)=\frac{1}{\ell!}(\frac{t}{\Delta})^\ell
e^{-t/\Delta}$.
From $\bra \ell\ket_\pi=t/\Delta$ and $\bra \ell^2\ket_\pi=t/\Delta+t^2/\Delta^2$
it follows that $\Delta$ is the average duration of an iteration step, and that
the relative deviation in $\ell$ at a given $t$ vanishes for $\Delta\to 0$ as
 $\sqrt{\bra\ell^2\ket_\pi-\bra\ell\ket_\pi^2}/\bra
 \ell\ket_\pi=\sqrt{\Delta/t}$.
The nice properties of the Poisson distribution under temporal derivation allow us to derive:
\bd
\Delta\frac{d}{dt}p_t(\bsigma)
=\sum_{\bsigma^{\prime}}W\left[\bsigma;\bsigma^{\prime}\right]p_t(\bsigma^{\prime})- p_t(\bsigma)
\ed
For sequential dynamics we choose $\Delta=\frac{1}{N}$ so that, as in the parallel case,
in one time unit each
neuron will on average be updated once. The master
equation corresponding to (\ref{eq:Ising_sequential_Markov}) acquires the
form
\be
\frac{d}{dt}p_t(\bsigma)=\sum_{i}\left\{w_i(F_i\bsigma)p_t(F_i\bsigma)-
w_i(\bsigma)p_t(\bsigma)\right\}
\label{eq:sequentialmaster}
\ee
The $w_i(\bsigma)$ (\ref{eq:transitionrates}) now play the role of {\em
transition rates}.
The choice $\Delta=\frac{1}{N}$ implies $\sqrt{\bra\ell^2\ket_\pi-\bra\ell\ket_\pi^2}/\bra
 \ell\ket_\pi=\sqrt{1/Nt}$, so we will still for $N\to\infty$ no longer
have uncertainty in where we are on the $t$ axis.
\vsp

\noindent{\em Microscopic Definitions for Continuous Neurons.}
Alternatively,
we could start with continuous neuronal
variables $\sigma_i$ (representing e.g. firing frequencies or oscillator
phases), where $i=1,\ldots,N$, and with stochastic equations of the form
\be
\sigma_i(t\plus \Delta)= \sigma_i(t)+\Delta f_i(\bsigma(t))+\sqrt{2T\Delta}\xi_i(t)
\label{eq:discrete}
\ee
Here we have introduced (as yet unspecified) deterministic state-dependent forces
$f_i(\bsigma)$, and uncorrelated Gaussian distributed random forces
$\xi_i(t)$ (the noise), with
$\bra \xi_i(t)\ket = 0$ and
$\bra \xi_i(t)\xi_j(t^\prime)\ket =
\delta_{ij}\delta_{t,t^\prime}$.
As before, the parameter $T$ controls the amount of noise in the system,
ranging from $T=0$ (deterministic dynamics) to $T=\infty$
(completely random dynamics).
If we take the limit $\Delta\to 0$ in (\ref{eq:discrete}) we find
a Langevin equation (with a continuous time variable):
\be
\frac{d}{dt}\sigma_i(t)
=f_i(\bsigma(t))+\eta_i(t)
\label{eq:langevin}
\ee
This equation acquires its meaning only as the limit $\Delta\to 0$ of
(\ref{eq:discrete}). The moments of the new noise variables $\eta_i(t)=\xi_i(t)\sqrt{2T/\Delta}$
in (\ref{eq:langevin}) are given
by $\bra \eta_i(t)\ket = 0$ and
$\bra \eta_i(t)\eta_j(t^\prime)\ket=2T\delta_{ij}\delta(t\minus
t^\prime)$. This can be derived from the moments of the
$\xi_i(t)$. For instance:
\bd
\bra \eta_i(t)\eta_j(t^\prime)\ket=
\lim_{\Delta\to 0}
\frac{2T}{\Delta}\bra \xi_i(t)\xi_j(t^\prime)\ket
=2T\delta_{ij}\lim_{\Delta\to 0}\frac{1}{\Delta}\delta_{t,t^\prime}
=2TC\delta_{ij}\delta(t\minus t^\prime)
\ed
The constant $C$ is found by summing
over $t^\prime$, before taking the limit $\Delta\to 0$, in the above equation:
\bd
\int\!dt^\prime~\bra \eta_i(t)\eta_j(t^\prime)\ket=
\lim_{\Delta\to 0}
2T\sum_{t^\prime=-\infty}^\infty \bra \xi_i(t)\xi_j(t^\prime)\ket
=2T\delta_{ij}
\lim_{\Delta\to 0} \sum_{t^\prime=-\infty}^{\infty}\delta_{t,t^\prime}
=2T\delta_{ij}
\ed
Thus $C=1$, which indeed implies
$\bra \eta_i(t)\eta_j(t^\prime)\ket=2T\delta_{ij}\delta(t\minus
t^\prime)$.
More directly, one can also calculate the moment generating function
\be
\bra e^{i\int\!dt\sum_i \psi_i(t)\eta_i(t)}\ket=\lim_{\Delta\to 0}
\prod_{i,t}
\int\!\!\frac{dz}{\sqrt{2\pi}}~e^{-\frac{1}{2}z^2+iz\psi_i(t)\sqrt{2T
\Delta}}
=\lim_{\Delta\to 0}
\prod_{i,t} e^{-T\Delta\psi^2_i(t)}=e^{-T \int\!dt\sum_i\psi^2_i(t)}
\label{eq:generator}
\ee
\vsp

\noindent{\em From Stochastic Equations to Evolving Probabilities}.
A mathematically more convenient description of the process
(\ref{eq:langevin}) is provided by the Fokker-Planck equation for
the microscopic
state probability density $p_t(\bsigma)
=\bra \delta[\bsigma\minus\bsigma(t)]\ket$, which we will now derive.
For the discrete-time process (\ref{eq:discrete}) we expand
the $\delta$-distribution in the definition of
$p_{t+\Delta}(\bsigma)$ (in a distributional sense):
\bd
p_{t+\Delta}(\bsigma)-p_{t}(\bsigma)
= \bra
\delta\left[\bsigma\minus
\bsigma(t)\minus
\Delta\bbf(\bsigma(t))\minus \sqrt{2T\Delta}\bxi(t)
\right]\ket-\bra \delta[\bsigma\minus\bsigma(t)]\ket
\ed
\bd
=
-
\sum_i\frac{\partial}{\partial\sigma_i}
\bra \delta[\bsigma\minus\bsigma(t)]
\left[\Delta f_i(\bsigma(t))\plus \sqrt{2T\Delta}\xi_i(t)
\right]\ket
+T\Delta\sum_{ij}\frac{\partial^2}{\partial\sigma_i\partial\sigma_j}
\bra \delta[\bsigma\minus
\bsigma(t)]\xi_i(t)\xi_j(t)\ket
+\order(\Delta^{\frac{3}{2}})
\ed
The variables $\bsigma(t)$ depend only on noise variables
$\xi_j(t^\prime)$ with $t^\prime<t$, so that for any function
$A$:
$\bra A[\bsigma(t)]\xi_i(t)\ket= \bra A[\bsigma(t)]\ket\bra
\xi_i(t)\ket=0$, and
$\bra A[\bsigma(t)]\xi_i(t)\xi_j(t)\ket=\delta_{ij}\bra A[\bsigma(t)]\ket$.
 As a consequence:
\bd
\frac{1}{\Delta}\left[p_{t+\Delta}(\bsigma)-p_{t}(\bsigma)\right]
=
-\sum_i\frac{\partial}{\partial\sigma_i}\bra \delta[\bsigma\minus
\bsigma(t)]f_i(\bsigma(t))\ket
+T\sum_{i}\frac{\partial^2}{\partial\sigma^2_i}
\bra\delta[\bsigma\minus\bsigma(t)]\ket
+\order(\Delta^{\frac{1}{2}})
\ed
\bd
=
-\sum_i\frac{\partial}{\partial\sigma_i}\left[p_t(\bsigma)
f_i(\bsigma)\right]
+T\sum_i\frac{\partial^2}{\partial\sigma^2_i}p_t(\bsigma)
+\order(\Delta^{\frac{1}{2}})
\ed
By taking the limit $\Delta\to 0$ we then arrive at the Fokker-Planck
equation:
\be
\frac{d}{dt}p_t(\bsigma)=-\sum_i
\frac{\partial}{\partial\sigma_i}\left[p_t(\bsigma)
f_i(\bsigma)\right]+T\sum_i\frac{\partial^2}{\partial\sigma_i^2}p_t(\bsigma)
\label{eq:fokkerplanck}
\ee
\vsp

\noindent{\em Examples: Graded Response Neurons and Coupled Oscillators.}
In the case of graded response neurons the continuous variable $\sigma_i$
represents the membrane potential of
neuron $i$, and (in their simplest form) the deterministic forces
are given by $f_i(\bsigma)=\sum_{j} J_{ij}\tanh[\gamma
\sigma_j]-\sigma_i+\theta_i$, with $\gamma>0$ and with the $\theta_i$
representing injected currents.
Conventional notation is restored by putting $\sigma_i\to
u_i$. Thus equation
(\ref{eq:langevin}) specialises to
\be
\frac{d}{dt}u_i(t)
=\sum_{j} J_{ij}\tanh[\gamma u_j(t)]-u_i(t)+\theta_i +\eta_i(t)
\label{eq:graded_response}
\ee
One often chooses $T=0$ (i.e. $\eta_i(t)=0$), the rationale being that  threshold
noise is already assumed to have been incorporated via the non-linearity in (\ref{eq:graded_response}).

In our second example the variables $\sigma_i$ represent the phases of
 coupled neural oscillators, with forces of the form
 $f_i(\bsigma)=\sum_{j}J_{ij}
\sin(\sigma_j\minus \sigma_i)+\omega_i$. Individual synapses $J_{ij}$ now
try to enforce either pair-wise synchronisation ($J_{ij}>0$) or
pair-wise anti-synchronisation ($J_{ij}<0$), and the $\omega_i$
represent the natural frequencies of the individual oscillators.
Conventional notation dictates $\sigma_i\to\phi_i$, giving
\be
\frac{d}{dt}\phi_{i}(t)=\omega_{i}+
\sum_{j}J_{ij}
\sin[\phi_j(t)\minus \phi_i(t)]
+\eta_i(t)
\label{eq:oscillators}
\ee

\subsection{Synaptic Symmetry \& Lyapunov Functions}

\noindent{\em Noise-free Symmetric Networks of Binary Neurons.}
In the deterministic limit $T\rightarrow 0$ the
rules (\ref{eq:Ising_parallel}) for networks of synchronously evolving
binary neurons reduce to the deterministic map
\be
\sigma_i(\ell+1)=\sgn\left[h_i(\bsigma(\ell))\right]
\label{eq:paralleldet}
\ee
It turns out that for systems with symmetric interactions, $J_{ij}=J_{ji}$ for all
$(ij)$, one can construct a Lyapunov function, i.e. a function of
$\bsigma$ which during the dynamics  decreases monotonically and is bounded from below
(see e.g. \cite{Khalil}):
\be
{\sl Binary~\&~Parallel:}~~~~~~~~~~
L[\bsigma]=-\sum_{i}|h_i(\bsigma)|-\sum_{i}\sigma_i\theta_i
\label{eq:Lyapunov_par}
\ee
Clearly  $L\geq \minus \sum_{i}[\sum_j|J_{ij}|\plus |\theta_i|]\minus \sum_i|\theta_i|$.
During
iteration of (\ref{eq:paralleldet}) we find:
\bd
L[\bsigma(\ell+1)]-L[\bsigma(\ell)]=-\sum_{i}|h_i(\bsigma(\ell\plus1))|
+\sum_{i}\sigma_i(\ell\plus1)[\sum_{j}J_{ij}\sigma_j(\ell)\plus\theta_i]
-\sum_{i}\theta_i\left[\sigma_i(\ell\plus1)\minus \sigma_i(\ell)\right]
\vspace*{-3mm}
\ed
\bd
=-\sum_{i}|h_i(\bsigma(\ell\plus1))|
+\sum_{i}\sigma_i(\ell)h_i(\bsigma(\ell\plus1))
=-\sum_{i}|h_i(\bsigma(\ell\plus1))|\left[1\minus \sigma_i(\ell\plus2)\sigma_i(\ell)\right]~\leq~0
\ed
(where we used (\ref{eq:paralleldet}) and $J_{ij}=J_{ji}$).
So $L$ decreases monotonically until a stage is reached where
$\sigma_i(\ell+2)=\sigma_i(\ell)$ for all $i$. Thus, with symmetric
interactions this system will in the deterministic limit always end up in
a limit cycle with period $\leq2$.
A similar result is found for networks with binary neurons and
sequential dynamics.
In the limit $T\to 0$ the
rules (\ref{eq:Ising_sequential}) reduce to the map
\be
\sigma_i(\ell+1)=\delta_{i,i_\ell}\sgn\left[h_i(\bsigma(\ell))\right]+[1-\delta_{i,i_\ell}]\sigma_i(\ell)
\label{eq:sequentialdet}
\ee
(in which we still have randomness in the choice of site to be
updated).
For systems with symmetric interactions and
without self-interactions, i.e. $J_{ii}=0$ for all $i$,  we again find a Lyapunov function:
\be
{\sl Binary~\&~Sequential:}~~~~~~~~~~
L[\bsigma]=
-\frac{1}{2}\sum_{ij}\sigma_i J_{ij}\sigma_j-\sum_{i}\sigma_i\theta_i
\label{eq:Lyapunov_seq}
\ee
This quantity is bounded from below: $L\geq -\frac{1}{2}\sum_{ij}|J_{ij}|-\sum_i|\theta_i|$.
Upon  calling the site $i_\ell$ selected for update at step $\ell$
simply $i$, the change in $L$ during
iteration of (\ref{eq:sequentialdet}) can be written as:
\bd
L[\bsigma(\ell+1)]-L[\bsigma(\ell)]=-\theta_i[\sigma_i(\ell\plus 1)\minus \sigma_i(\ell)]
-\frac{1}{2}\sum_{k}J_{ik}[\sigma_i(\ell\plus 1)\sigma_k(\ell\plus 1)-\sigma_i(\ell)\sigma_k(\ell)]
\vspace*{-3mm}
\ed
\bd
-\frac{1}{2}\sum_{j}J_{ji}[\sigma_j(\ell\plus 1)\sigma_i(\ell\plus 1)-\sigma_j(\ell)\sigma_i(\ell)]
\ed
\bd
=[\sigma_i(\ell)-\sigma_i(\ell\plus1)][
\sum_{j}J_{ij}\sigma_j(\ell)
+\theta_i]
=-|h_i(\bsigma(\ell))|\left[1-\sigma_i(\ell)\sigma_i(\ell\plus1)\right]~\leq~0
\ed
Here we used (\ref{eq:sequentialdet}), $J_{ij}=J_{ji}$, and absence of self-interactions.
Thus $L$ decreases monotonically until
$\sigma_i(t\plus 1)=\sigma_i(t)$ for all $i$. With symmetric
synapses, but without diagonal terms, the sequentially evolving
binary neurons system
will in the deterministic limit always end up in
a stationary state.
\vsp

\noindent{\em Noise-free Symmetric Networks of Continuous Neurons.}
One can derive similar results for models with continuous variables.
Firstly, in the deterministic limit the graded response equations
(\ref{eq:graded_response}) simplify to
\be
\frac{d}{dt}u_i(t)
=\sum_{j} J_{ij}\tanh[\gamma u_j(t)]-u_i(t)+\theta_i
\label{eq:gradeddet}
\ee
Symmetric networks again admit a Lyapunov function (there is no need to eliminate
self-interactions):
\bd
{\sl Graded~Response:}~~~~~~~~
L[\bu]=-\frac{1}{2}\sum_{ij}J_{ij}\tanh[\gamma u_i]\tanh[\gamma
u_j]+\sum_i\left[\gamma \!\int_0^{u_i}\!\!dv~v[1\minus \tanh^2[\gamma v]]
\minus \theta_i \tanh[\gamma u_i]\right]
\ed
Clearly $L\geq -\frac{1}{2}\sum_{ij}|J_{ij}|\minus\sum_i |\theta_i|$ (the term in $L[\bu]$ with the integral is non-negative).
During the
noise-free dynamics (\ref{eq:gradeddet}) one can use the identity
$\partial L/\partial u_i=\minus\gamma [1\minus\tanh^2[\gamma u_i]] (d u_i/dt)$, valid only
when $J_{ij}=J_{ji}$, to derive
\bd
\frac{d}{dt}L=\sum_i\frac{\partial L}{\partial u_i} \frac{d u_i}{dt}
=-\gamma\sum_i [1\minus\tanh^2[\gamma u_i]]~
[\frac{d}{dt}u_i]^2\leq 0
\ed
Again $L$ is found to decrease monotonically, until $du_i/dt=0$ for all $i$, i.e. until we are at a fixed-point.

Finally, the  coupled oscillator equations (\ref{eq:oscillators}) reduce in the
noise-free limit to
\be
\frac{d}{dt}\phi_{i}(t)=\omega_{i}+
\sum_{j}J_{ij}
\sin[\phi_j(t)\minus \phi_i(t)]
\label{eq:oscillatordet}
\ee
Note that self-interactions $J_{ii}$ always drop out automatically.
For symmetric oscillator networks, a construction
of the type followed for the graded response equations
would lead us to propose
\be
{\sl Coupled~Oscillators:}~~~~~~~~~~
L[\bphi]=-\frac{1}{2}\sum_{ij}J_{ij}\cos[\phi_i\minus\phi_j]-\sum_i
\omega_i\phi_i
\label{eq:lyapunov_osc}
\ee
This function indeed decreases monotonically, due to
$\partial L/\partial \phi_i= -d \phi_i/dt$ :
\bd
\frac{d}{dt}L=\sum_i\frac{\partial L}{\partial \phi_i} \frac{d \phi_i}{dt}
=-\sum_i
[\frac{d}{dt}\phi_i]^2\leq 0
\ed
In fact
(\ref{eq:oscillatordet}) describes gradient descent on the surface
$L[\bphi]$.
However, due to the term with the natural frequencies $\omega_i$
the function $L[\bphi]$ is not bounded, so it cannot
be a Lyapunov function. This could have been
expected; when $J_{ij}=0$ for all $(i,j)$, for instance, one finds continually increasing
phases $\phi_i(t)=\phi_i(0)\plus \omega_i
t$. Removing the $\omega_i$, in contrast, gives the bound $L\geq
\minus\sum_j|J_{ij}|$. Now the system must go to a fixed-point.
In the special case $\omega_i=\omega$ ($N$ identical natural
frequencies) we can transform away the $\omega_i$
by putting $\phi(t)=\tilde{\phi}_i(t)\plus\omega t$, and
find the relative phases $\tilde{\phi}_i$ to go to a
fixed-point.

\subsection{Detailed Balance \& Equilibrium Statistical Mechanics}

\noindent{\em Detailed Balance for Binary Networks.}
The results obtained above indicate that networks with symmetric
synapses are a special class. We now show how synaptic symmetry
is closely related to the detailed
balance property, and derive a number of consequences. An ergodic Markov
chain of the form
(\ref{eq:Ising_parallel_Markov},\ref{eq:Ising_sequential_Markov}),
i.e.
\be
p_{\ell+1}(\bsigma)=\sum_{\bsigma^{\prime}}
W\left[\bsigma;\bsigma^{\prime}\right]
p_\ell(\bsigma^{\prime})
\label{eq:discrete_markov}
\ee
is said to obey detailed balance if
its (unique) stationary solution $p_{\infty}(\bsigma)$ has the
property
\be
W\left[\bsigma;\bsigma^{\prime}\right]
p_{\infty}(\bsigma^{\prime})
=W\left[\bsigma^{\prime};\bsigma\right]
p_{\infty}(\bsigma)~~~~~{\rm for~all}~\bsigma,\bsigma^{\prime}
\label{eq:discrete_detailedbalance}
\ee
All $p_{\infty}(\bsigma)$ which satisfy
(\ref{eq:discrete_detailedbalance}) are stationary solutions of
(\ref{eq:discrete_markov}), this is easily verified by substitution.
The converse is not true.
Detailed balance states that, in addition to $p_\infty(\bsigma)$ being stationary,
one has {\em equilibrium}: there is no
net probability current between any two microscopic system states.

It is not a trivial matter to investigate systematically for which
choices of the threshold noise distribution $w(\eta)$ and the synaptic matrix
$\{J_{ij}\}$ detailed balance holds. It can be shown that, apart
from trivial cases (e.g. systems with self-interactions only) a
Gaussian distribution $w(\eta)$ will not support detailed balance.
Here we will work out details only for the choice $w(\eta)=\frac{1}{2}[1\minus
\tanh^2(\eta)]$, and for $T>0$ (where both discrete systems are
ergodic).
For parallel dynamics the transition matrix is given in
(\ref{eq:Ising_parallel_Markov}), now with $g[z]=\tanh[z]$, and the detailed balance condition
(\ref{eq:discrete_detailedbalance}) becomes
\be
\frac{e^{\beta\sum_{i}\sigma_i h_i(\bsigma^{\prime})}p_{\infty}(\bsigma^{\prime})}{\prod_{i}\cosh[\beta
h_i(\bsigma^{\prime})]}
=
\frac{e^{\beta\sum_{i}\sigma^{\prime}_i h_i(\bsigma)}p_{\infty}(\bsigma)}{\prod_{i}\cosh[\beta
h_i(\bsigma)]}
~~~~~{\rm for~all}~\bsigma,\bsigma^{\prime}
\label{eq:parequil}
\ee
All  $p_{\infty}(\bsigma)$ are non-zero (ergodicity), so we may safely put
 $p_{\infty}(\bsigma)= e^{\beta[\sum_{i}\theta_i\sigma_i+K(\bsigma)]}\prod_{i}\cosh[\beta
h_i(\bsigma)]$,
which, in combination with definition (\ref{eq:Ising_parallel})
simplifies the detailed balance condition to:
\be
K(\bsigma)-K(\bsigma^{\prime})=\sum_{ij}\sigma_i\left[ J_{ij}-J_{ji}\right]\sigma_j^{\prime}
~~~~~{\rm for~all}~\bsigma,\bsigma^{\prime}
\label{eq:parallel_reduced}
\ee
Averaging (\ref{eq:parallel_reduced}) over all possible $\bsigma^{\prime}$ gives
$K(\bsigma)=\bra K(\bsigma^{\prime})\ket_{\bsigma^{\prime}}$ for all
$\bsigma$, i.e. $K$ is a constant, whose value follows from normalising $p_\infty(\bsigma)$.
So, if detailed balance holds
the equilibrium distribution must be:
\be
p_{\rm eq}(\bsigma)~\sim~e^{\beta\sum_{i}\theta_i\sigma_i}\prod_{i}\cosh[\beta
h_i(\bsigma)]
\label{eq:peretto}
\ee
For symmetric systems detailed balance indeed holds:
(\ref{eq:peretto})  solves
(\ref{eq:parequil}), since $K(\bsigma)=K$ solves the
reduced problem (\ref{eq:parallel_reduced}).
For non-symmetric systems, however,
there can be no equilibrium. For $K(\bsigma)=K$
the condition (\ref{eq:parallel_reduced}) becomes
$\sum_{ij}\sigma_i\left[ J_{ij}\minus J_{ji}\right]\sigma_j^{\prime}=0$ for all
$\bsigma,\bsigma^{\prime}\in\{\minus 1,1\}^N$.
For $N\geq 2$ the vector pairs $(\bsigma,\bsigma^{\prime})$ span the
space of all $N\times N$ matrices, so $J_{ij}\minus J_{ji}$ must be zero.
For $N=1$ there simply exists no non-symmetric synaptic
matrix. In conclusion: for binary networks with parallel dynamics, interaction symmetry implies detailed
balance, and vice versa.

For sequential dynamics, with $w(\eta)=\frac{1}{2}[1\minus \tanh^2(\eta)]$, the transition matrix is given by
(\ref{eq:Ising_sequential_Markov}) and the detailed balance condition
(\ref{eq:discrete_detailedbalance})
simplifies to
\bd
\frac{e^{\beta\sigma_i h_i(F_i\bsigma)}p_{\infty}(F_i\bsigma)}{\cosh\left[\beta h_i(F_i\bsigma)\right]}=
\frac{e^{-\beta\sigma_i h_i(\bsigma)}p_{\infty}(\bsigma)}{\cosh\left[\beta h_i(\bsigma)\right]}
~~~~~{\rm for~all}~\bsigma~{\rm and~all}~i
\ed
Self-interactions $J_{ii}$, inducing $h_i(F_i\bsigma)\neq h_i(\bsigma)$,
complicate matters. Therefore we first consider
systems where all $J_{ii}=0$.
All stationary
probabilities $p_{\infty}(\bsigma)$ being non-zero (ergodicity), we may write:
\be
p_{\infty}(\bsigma)=
e^{\beta[\sum_{i}\theta_i\sigma_i+\frac{1}{2}\sum_{i\neq
j}\sigma_i J_{ij}\sigma_j+K(\bsigma)]}
\label{eq:seqequil}
\ee
Using relations like
$\sum_{k\neq l}J_{kl}F_i(\sigma_k\sigma_l)=
\sum_{k\neq l}J_{kl}\sigma_k\sigma_l\minus 2\sigma_i\sum_{k\neq
i}\left[J_{ik}\plus J_{ki}\right]\sigma_k$
we can simplify the detailed balance condition to
$K(F_i\bsigma)\minus K(\bsigma)=\sigma_i\sum_{k\neq
i}\left[J_{ik}\minus J_{ki}\right]\sigma_k$ for all $\bsigma$  and all
$i$.
If to this expression  we apply the general identity
$\left[1\minus F_i\right]f(\bsigma)=2\sigma_i\bra
\sigma_i f(\bsigma)\ket_{\sigma_i}$
we find for $i\neq j$:
\bd
[F_j\minus 1][F_i\minus 1]K(\bsigma)=-2\sigma_i\sigma_j\left[J_{ij}\minus J_{ji}\right]
~~~~~{\rm for~all}~\bsigma~{\rm and~all}~i\neq j
\ed
The left-hand side is symmetric under permutation of the pair $(i,j)$,
which implies that the interaction matrix must also be
symmetric: $J_{ij}=J_{ji}$ for all $(i,j)$.
We now find the trivial
solution $K(\bsigma)=K$ (constant),
detailed balance holds and the  corresponding equilibrium distribution is
\be
p_{\rm eq}(\bsigma)~\sim~e^{-\beta H(\bsigma)}~~~~~~~~~~~
H(\bsigma)= -\frac{1}{2}\sum_{i\neq j}\sigma_i
J_{ij}\sigma_j\minus \sum_{i}\theta_i\sigma_i
\label{eq:gibbs}
\ee
In conclusion: for binary networks with sequential dynamics, but without
self-interactions,
interaction symmetry implies detailed balance, and vice versa.
In the case of self-interactions the situation is more
complicated. However, here one can still show that non-symmetric models with
detailed balance  must be  pathological,
since the requirements can be met only for very specific choices for the $\{J_{ij}\}$.
\vsp

\noindent{\em Detailed Balance for Networks with Continuous Neurons.}
Let us finally turn to the question of when we find microscopic equilibrium
(stationarity without probability currents) in continuous models described by a
Fokker-Planck equation (\ref{eq:fokkerplanck}).
Note that
(\ref{eq:fokkerplanck}) can be seen as a continuity equation
for the density of a conserved quantity: $\frac{d}{dt}p_t(\bsigma)+\sum_i\frac{\partial}{\partial
\sigma_i}J_i(\bsigma,t)=0$. The components $J_i(\bsigma,t)$ of the
current density are given by
\bd
J_i(\bsigma,t)=
[f_i(\bsigma)-T\frac{\partial}{\partial\sigma_i}]p_t(\bsigma)
\ed
Stationary distributions $p_\infty(\bsigma)$ are
those which give $\sum_i\frac{\partial}{\partial\sigma_i}
J_i(\bsigma,\infty)=0$ (divergence-free currents). Detailed balance
implies the stronger statement
$J_i(\bsigma,\infty)=0$ for all $i$ (zero currents),
so $f_i(\bsigma)=T\partial\log
p_\infty(\bsigma)/\partial\sigma_i$, or
\be
f_i(\bsigma)=-\partial H(\bsigma)/\partial\sigma_i,
~~~~~~~~~~~~~~
p_\infty(\bsigma)~\sim~e^{-\beta H(\bsigma)}
\label{eq:langevin_conservative}
\ee
for some $H(\bsigma)$, i.e. the forces $f_i(\bsigma)$ must be
conservative. However, one can have conservative
forces without a normalisable equilibrium distribution.
Just take $H(\bsigma)=0$, i.e. $f_i(\bsigma,t)=0$:
here  we have $p_{\rm eq}(\bsigma)=C$, which is not normalisable for
$\bsigma\in\Re^N$. For this particular case
equation (\ref{eq:fokkerplanck}) is solved easily:
$p_t(\bsigma)=[4\pi Tt]^{-N/2}\int\!d\bsigma^\prime~p_0(\bsigma^\prime)
e^{-[\bsigma-\bsigma^\prime]^2/4Tt}$, so the limit
$\lim_{t\to\infty}p_t(\bsigma)$ indeed does not exist.
One can prove the following (see e.g. \cite{Zinn-Justin}).
If the forces are conservative
and if  $p_\infty(\bsigma)\sim e^{-\beta H(\bsigma)}$ is normalisable,
then it is
the unique stationary solution of the Fokker-Planck equation,
to which the system converges for all initial distributions
$p_0\in L^1[\Re^N]$ which obey
$\int_{\Re^N}\!d\bsigma ~e^{\beta H(\bsigma)}p^2_0(\bsigma)<\infty$.

Assessing when our two particular model examples of graded response neurons
or coupled oscillators obey detailed balance has thus been reduced mainly to
checking whether the associated deterministic forces $f_i(\bsigma)$ are
conservative. Note that conservative forces must obey
\be
{\rm for~all}~\bsigma,~{\rm for~all}~i\neq j:~~~~~~~~~~\partial f_i(\bsigma)/\partial \sigma_j-
\partial f_j(\bsigma)/\partial\sigma_i=0
\label{eq:db_test}
\ee
In the graded response equations (\ref{eq:gradeddet}) the
deterministic forces are $f_i(\bu)=\sum_{j} J_{ij}\tanh[\gamma
u_j]\minus u_i\plus \theta_i$.
Here $\partial f_i(\bu)/\partial u_j\minus \partial f_j(\bu)/\partial
u_i=\gamma\{ J_{ij}[1\minus\tanh^2[\gamma u_j]\minus J_{ji}[1\minus\tanh^2[\gamma
u_i]\}$. At $\bu=\bnul$ this reduces to $J_{ij}\minus
J_{ji}$, i.e. the interaction matrix must be symmetric. For symmetric matrices
we find away from $\bu=\bnul$:
$\partial f_i(\bu)/\partial u_j\minus \partial f_j(\bu)/\partial
u_i=\gamma J_{ij}\{\tanh^2[\gamma u_i]\minus\tanh^2[\gamma
u_j]\}$. The only way for this to be zero for any $\bu$ is
by having $J_{ij}=0$ for all $i\neq j$, i.e. all neurons are disconnected
(in this trivial case the system (\ref{eq:gradeddet}) does indeed obey detailed
balance). Network models of interacting graded-response neurons of the
type (\ref{eq:gradeddet}) apparently never reach equilibrium, they will
always violate detailed balance and exhibit microscopic probability currents.
In the case of coupled oscillators (\ref{eq:oscillators}), where
the deterministic forces are
$f_i(\bphi)=\sum_{j}J_{ij}
\sin[\phi_j\minus \phi_i]\plus\omega_i$ one finds the
left-hand side of condition (\ref{eq:db_test}) to give
$\partial f_i(\bphi)/\partial\phi_j\minus \partial f_j(\bphi)/\partial \phi_i=
[J_{ij}\minus J_{ji}]\cos[\phi_j\minus\phi_i]$. Requiring this to
be zero for any $\bphi$ gives the condition $J_{ij}=J_{ji}$ for
any $i\neq j$. We have already seen that symmetric oscillator
networks indeed have conservative forces: $f_i(\bphi)=-\partial
H(\bphi)/\partial\phi_i$, with
$H(\bphi)=\minus \frac{1}{2}\sum_{ij}J_{ij}\cos[\phi_i\minus\phi_j]\minus\sum_i
\omega_i\phi_i$.
If in addition we choose all $\omega_i=0$ the function $H(\bsigma)$
will also be bounded from below, and, although $p_\infty(\bphi)\sim e^{-\beta H(\bphi)}$
is still not normalisable on $\bphi\in\Re^N$, the full $2\pi$-periodicity of the function $H(\bsigma)$
now allows us to identify $\phi_i\plus 2\pi\equiv\phi_i$ for all $i$, so
that now $\bphi\in[\minus\pi,\pi]^N$ and $\int\!d\bphi~e^{-\beta H(\bphi)}$
does exist. Thus symmetric coupled oscillator networks with zero natural frequencies obey
detailed balance. In the case of non-zero natural frequencies, in contrast,
detailed balance does not hold.
\vsp

\noindent{\em Equilibrium Statistical Mechanics}.
The above results establish the link with equilibrium
statistical mechanics (see e.g. \cite{StatMech1,StatMech2}).
For binary systems with symmetric synapses (in the sequential
case: without self-interactions) and with threshold
noise distributions of the form $w(\eta)=\frac{1}{2}[1\minus\tanh^2(\eta)]$,
detailed balance holds and we know
the equilibrium distributions.
For sequential dynamics it has the Boltzmann form (\ref{eq:gibbs}) and we can apply standard
equilibrium statistical mechanics.
The parameter $\beta$ can formally be
identified with the inverse `temperature' in equilibrium,
$\beta=T^{-1}$, and the function $H(\bsigma)$ is the usual Ising spin
Hamiltonian.
In particular we can define the
partition function $Z$ and the free energy $F$:
\be
p_{\rm eq}(\bsigma)=\frac{1}{Z}e^{-\beta H(\bsigma)}~~~~~~~~~~
H(\bsigma)= -\frac{1}{2}\sum_{i\neq j}\sigma_i J_{ij}\sigma_j-\sum_{i}\theta_i\sigma_i
\nsp
\label{eq:seq_thermodynamics1}
\ee
\be
Z=\sum_{\bsigma}e^{-\beta H(\bsigma)}~~~~~~~~~~
F=-\beta^{-1}\log Z
\label{eq:seq_thermodynamics2}
\ee
The free energy can be used as the generating function for equilibrium
averages. Taking derivatives with
respect to external fields $\theta_i$ and interactions $J_{ij}$, for instance, produces
$\bra\sigma_i\ket=\minus \partial F/\partial\theta_i$ and
$\bra\sigma_i\sigma_j\ket=\minus \partial F/\partial
J_{ij}$,
whereas equilibrium averages of arbitrary state variable
$f(\bsigma)$ can be obtained by adding suitable generating terms to
the Hamiltonian:
$H(\bsigma)\rightarrow H(\bsigma)\plus\lambda f(\bsigma)$,
$~\bra f\ket=\lim_{\lambda\rightarrow 0}\partial F/\partial
\lambda$.

In the
parallel case (\ref{eq:peretto})  we can again formally write the equilibrium probability
distribution in the Boltzmann form \cite{perettopaper} and define a corresponding partition function
$\tilde{Z}$ and a free energy $\tilde{F}$:
\be
p_{\rm eq}(\bsigma)=\frac{1}{Z}e^{-\beta\tilde{H}(\bsigma)}~~~~~~~~~~
\tilde{H}(\bsigma)=
-\sum_{i}\theta_i\sigma_i-\frac{1}{\beta}\sum_{i}\log 2 \cosh[\beta
h_i(\bsigma)]
\nsp
\label{eq:par_thermodynamics1}
\ee
\be
\tilde{Z}=\sum_{\bsigma}e^{-\beta \tilde{H}(\bsigma)}~~~~~~~~~~
\tilde{F}=-\beta^{-1}\log \tilde{Z}
\label{eq:par_thermodynamics2}
\ee
which again serve to generate averages:
$\tilde{H}(\bsigma)\rightarrow \tilde{H}(\bsigma)\plus\lambda
f(\bsigma)$,
$~\bra f\ket=\lim_{\lambda\rightarrow 0}\partial \tilde{F}/\partial
\lambda$.
However, standard thermodynamic relations involving
derivation with respect to $\beta$ need no longer be valid, and
derivation with respect to fields or interactions generates different
types of averages, such as
\bd
-\partial\tilde{F}/\partial\theta_i=\bra\sigma_i\ket+\bra\tanh[\beta
h_i(\bsigma)]\ket
~~~~~~~~~~
-\partial\tilde{F}/\partial J_{ii}=
\bra\sigma_i\tanh[\beta
h_i(\bsigma)]\ket
\ed
\bd
i\neq j:~~~~-\partial\tilde{F}/\partial J_{ij}=
\bra\sigma_i\tanh[\beta h_j(\bsigma)]\ket+
\bra\sigma_j\tanh[\beta h_i(\bsigma)]\ket
\ed
One can use $\bra \sigma_i\ket=\bra\tanh[\beta
h_i(\bsigma)]\ket$, which can be derived directly from the equilibrium equation
$p_{\rm eq}(\bsigma)=\sum_{\bsigma^\prime}W[\bsigma;\bsigma^\prime]p_{\rm eq}(\bsigma)$,
to simplify the first of these
identities.

A connected network of graded-response neurons
can never be in an equilibrium state, so our only model example
with continuous neuronal variables for which we can set up the
equilibrium statistical mechanics formalism is the system of
coupled oscillators (\ref{eq:oscillators}) with symmetric synapses and absent (or
uniform) natural frequencies $\omega_i$. If we define the phases
as $\phi_i\in[\minus \pi,\pi]$ we have again an equilibrium distribution of the
Boltzmann form, and we can define the standard thermodynamic quantities:
\be
p_{\rm eq}(\bphi)=\frac{1}{Z}e^{-\beta H(\bphi)}~~~~~~~~~~
H(\bphi)=- \frac{1}{2}\sum_{ij}J_{ij}\cos[\phi_i\minus\phi_j]
\label{eq:oscill_thermodynamics1}
\ee
\nsp
\be
Z=\int_{-\pi}^{\pi}\!\cdots\!\int_{-\pi}^{\pi}\!\!d\bphi ~
e^{-\beta H(\bphi)}~~~~~~~~~~
F=-\beta^{-1}\log Z
\label{eq:oscill_thermodynamics2}
\ee
These generate equilibrium
averages in the usual manner. For instance
$\bra\cos[\phi_i\minus\phi_j]\ket=
\minus \partial F/\partial J_{ij}$,
whereas averages of arbitrary state variables
$f(\bphi)$  follow, as before, upon introducing suitable generating terms:
$H(\bphi)\rightarrow H(\bphi)\plus\lambda f(\bphi)$,
$~\bra f\ket=\lim_{\lambda\rightarrow 0}\partial F/\partial
\lambda$.

In this chapter we restrict ourselves to symmetric networks which
obey detailed balance, so that we know the equilibrium probability
distribution and equilibrium statistical mechanics applies.
In the case of sequential dynamics we will accordingly not
allow for the presence of self-interactions.


\section{Simple Recurrent Networks with Binary Neurons}

\subsection{Networks with Uniform Synapses}

We now turn to a simple toy model to show how equilibrium statistical mechanics
is used for solving neural network models, and to illustrate similarities and
differences between the different dynamics types.
We choose uniform infinite-range synapses
and zero external fields, and calculate the free energy for the
binary systems (\ref{eq:Ising_parallel},\ref{eq:Ising_sequential}),
parallel and sequential, and with threshold noise distribution
$w(\eta)=\frac{1}{2}[1\minus \tanh^2(\eta)]$:
\bd
J_{ij}=J_{ji}=J/N~~~(i\neq j),~~~~~~~~~~~J_{ii}=\theta_i=0~~~{\rm
for~all}~i
\ed
The free energy is an extensive object, $\lim_{N\to\infty}F/N$
is finite.
For the models (\ref{eq:Ising_parallel},\ref{eq:Ising_sequential}) we now obtain:
\bd
{\sl Binary ~\&~ Sequential:}
~~~~~~~~~~~~
\lim_{N\rightarrow\infty}F/N=-\lim_{N\rightarrow\infty}(\beta
N)^{-1}\log \sum_{\bsigma}
e^{\beta N \left[\frac{1}{2}Jm^2(\bsigma)\right]}
~~~~~
\ed
\bd
{\sl Binary~\&~Parallel:}
~~~~~~~~~~~~
\lim_{N\rightarrow\infty}\tilde{F}/N=
-\lim_{N\rightarrow\infty}(\beta N)^{-1}\log
\sum_{\bsigma}
e^{N\left[\log 2 \cosh[\beta
Jm(\bsigma)]\right]}
\ed
with the average activity
$m(\bsigma)=\frac{1}{N}\sum_{k}\sigma_k$. We
have to count
the number of states $\bsigma$ with a prescribed average activity $m=2n/N-1$
($n$ is the number of neurons $i$ with $\sigma_i=1$), in expressions of the form
\bd
\frac{1}{N}\log\sum_{\bsigma}e^{N U\left[m(\bsigma)\right]}
=
\frac{1}{N}\log\sum_{n=0}^{N}\left(\!\begin{array}{c}N\\n\end{array}\!\right)
e^{NU[2n/N-1]}
=
\frac{1}{N}\log\int_{-1}^{1}dm~e^{N\left[\log2-c^*(m)+U[m]\right]}
\ed
\bd
\lim_{N\to\infty}
\frac{1}{N}\log\sum_{\bsigma}e^{N U\left[m(\bsigma)\right]}
=\log 2 + \max_{m\in[-1,1]} \left\{ U[m]-c^*(m)\right\}
\ed
with the entropic function
$c^*(m)=\frac{1}{2}(1\plus m)\log(1\plus m)\plus \frac{1}{2}(1\minus m)\log(1\minus
m)$. In order to get there we used Stirling's formula to obtain the leading term of
the factorials (only terms which are exponential in $N$ survive the limit
$N\rightarrow\infty$), we converted (for $N\to\infty$) the summation over $n$
into an integration over $m=2n/N\minus 1\in[\minus 1,1]$, and we carried
out the integral over $m$ via saddle-point integration (see e.g. \cite{Peretto}).
This leads to a saddle-point problem whose solution gives the free energies:
\begin{figure}[t]
\begin{center}\vspace*{-10mm}
\epsfxsize=55mm\epsfbox{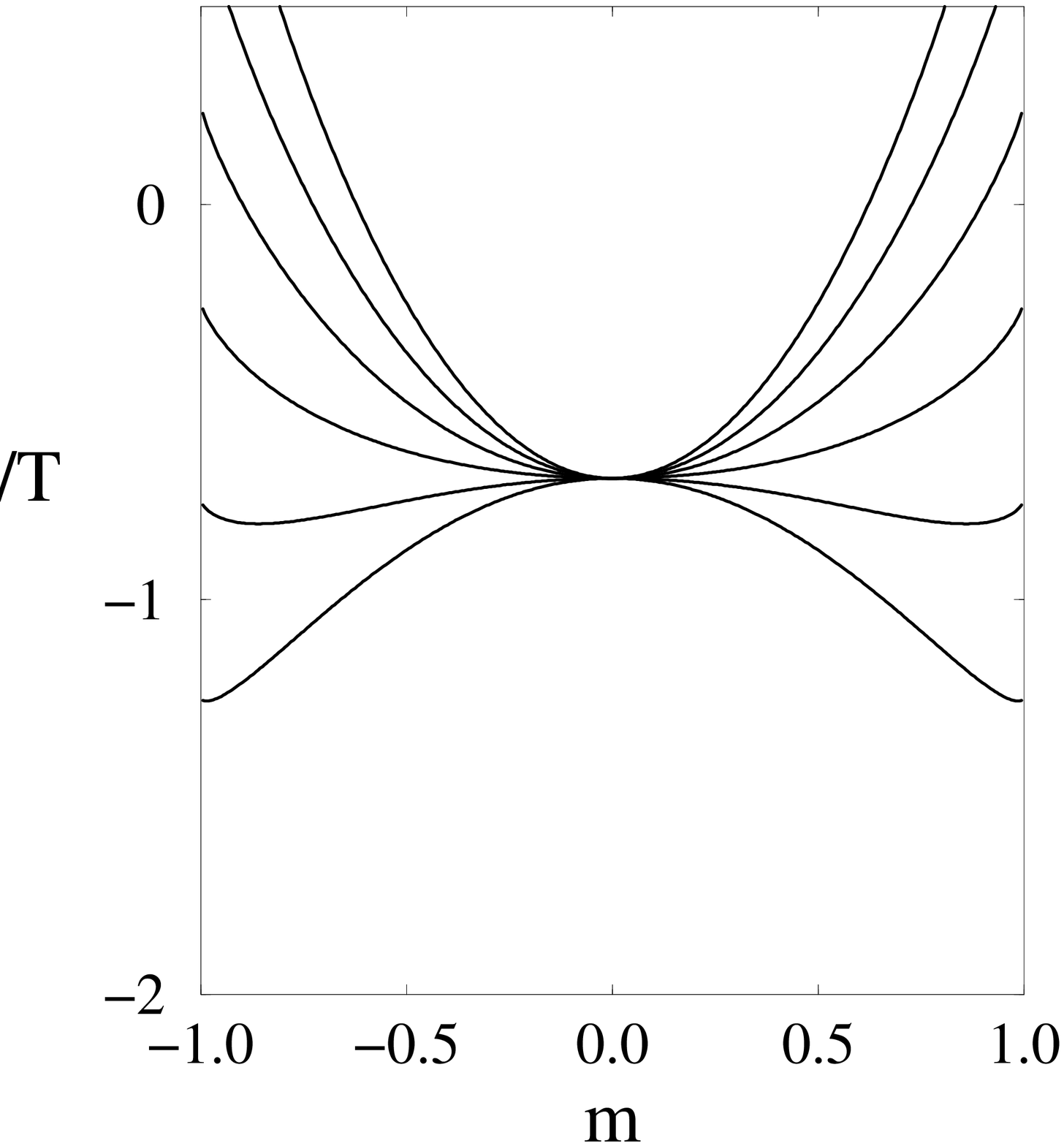}
\epsfxsize=55mm\epsfbox{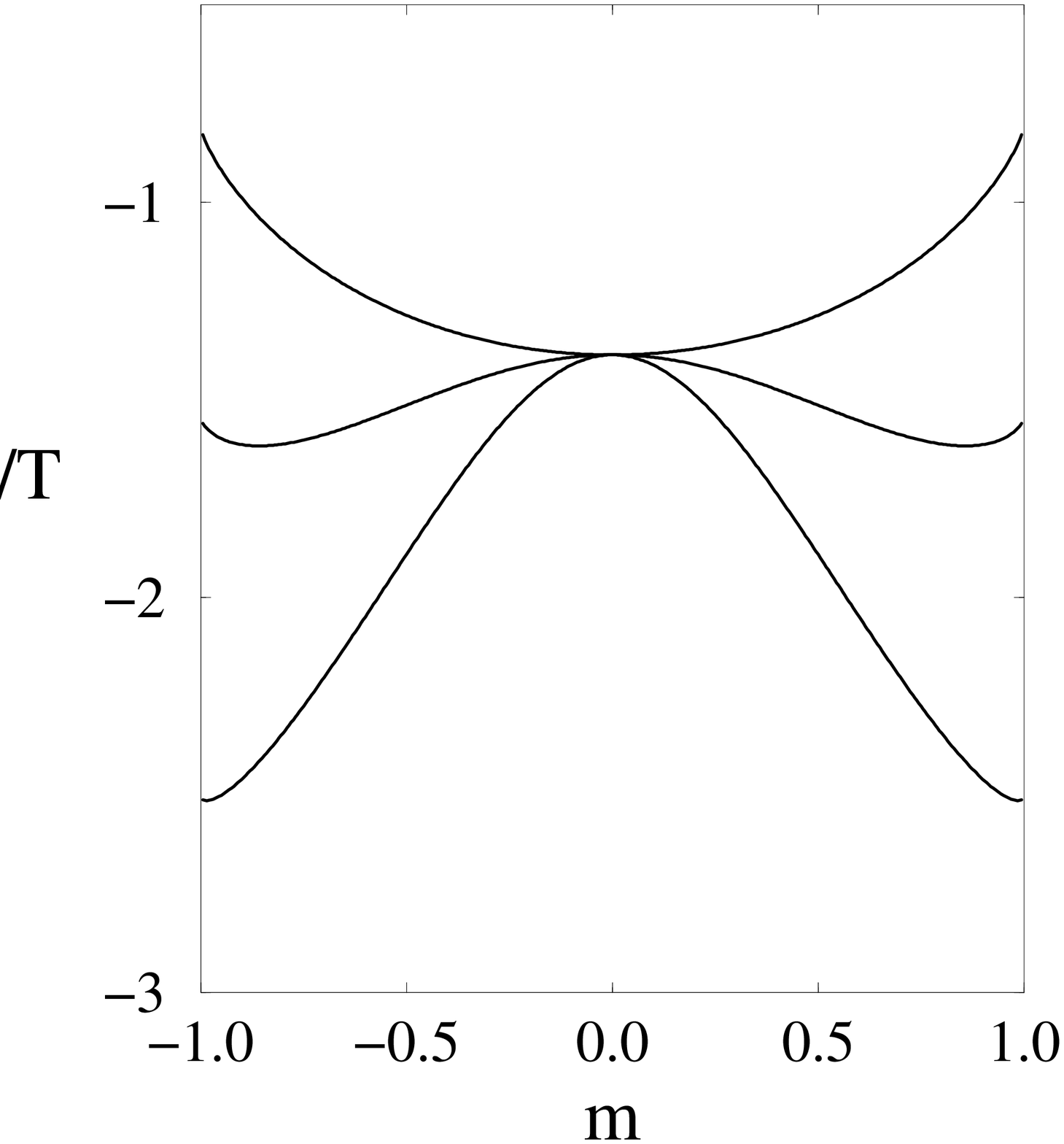}
\epsfxsize=55mm\epsfbox{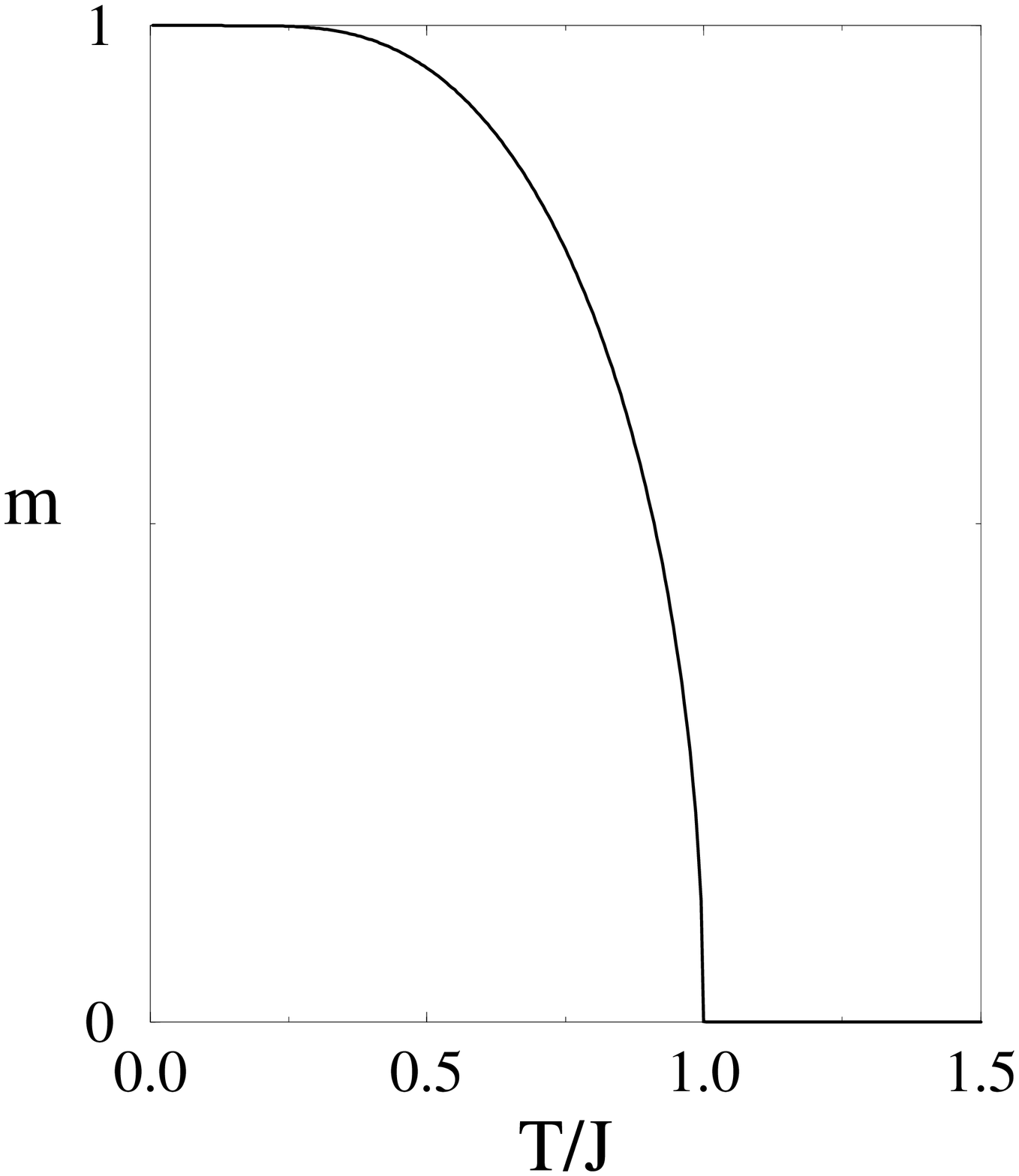}
\end{center}
\vspace*{-5mm}
\caption{The functions $f_{\rm seq}(m)/T$ (left) and $f_{\rm par}(m)/T$ (middle)
for networks of binary neurons and uniform synapses, and for different choices of
the re-scaled interaction strength $J/T$ ($T=\beta^{-1}$).
Left picture (sequential dynamics):
$J/T=\minus\frac{5}{2},\minus\frac{3}{2},\minus\frac{1}{2},\frac{1}{2},\frac{3}{2},\frac{5}{2}$ (from top to bottom).
Middle picture (parallel dynamics): $J/T=\pm \frac{5}{2},\pm\frac{3}{2},\pm\frac{1}{2}$
(from top to bottom, here the free energy is independent of the
sign of $J$). The right picture gives, for $J>0$, the location of
the non-negative minimum of $f_{\rm seq}(m)$ and $f_{\rm par}(m)$ (which is identical to the average activity
in thermal equilibrium) as a function of
$T/J$. A phase transition to states with non-zero average activity occurs at $T/J=1$. }
\label{fig:energygraphs}
\end{figure}
\be
\lim_{N\rightarrow\infty}F/N=\min_{m\in[-1,1]} f_{\rm
seq}(m)~~~~~~~~~~~~~~~
\beta f_{\rm seq}(m)=
c^*(m)-\log 2 - \frac{1}{2}\beta Jm^2
~~~~~~~~~
\label{eq:sequentialsaddle}
\ee
\be
\lim_{N\rightarrow\infty}\tilde{F}/N=\min_{m\in[-1,1]}
f_{\rm par}(m)~~~~~~~~~~~~~~
\beta f_{\rm par}(m)=c^*(m)-2\log 2 -\log \cosh[\beta Jm]
\label{eq:parallelsaddle}
\ee
The functions to be minimised
are shown in figure \ref{fig:energygraphs}.
The equations from which to solve the minima are easily obtained by
differentiation, using $\frac{d}{dm}c^*(m)=\tanh^{-1}(m)$. For
sequential dynamics we find
\be
{\sl Binary~\&~Sequential:}~~~~~~~~~~
m=\tanh[\beta Jm]
\label{eq:curieweisssequential}
\ee
(the so-called Curie-Weiss law). For parallel dynamics we find
\bd
m=\tanh\left[\beta J\tanh[\beta Jm]\right]
\ed
One finds that the solutions of the latter equation again
obey a Curie-Weiss law. The definition $\hat{m}=\tanh[\beta|J|m]$
transforms it into the coupled
equations
$m=\tanh[\beta|J|\hat{m}]$ and
$\hat{m}\!=\!\tanh [\beta|J|m]$, from which we derive
$0\leq[m\minus \hat{m}]^2=[m\minus \hat{m}]\left[\tanh[\beta|J|\hat{m}]\minus \tanh[\beta
|J|m]\right]
\leq0$. Since $\tanh[\beta |J|m]$ is a monotonically increasing
function of $m$,
this implies $\hat{m}=m$, so
\be
{\sl Binary~\&~Parallel:}~~~~~~~~~~
m=\tanh[\beta|J|m]
\label{eq:curieweissparallel}
\ee
Our study of the toy models has thus been reduced to analysing the
non-linear equations (\ref{eq:curieweisssequential}) and
(\ref{eq:curieweissparallel}).
If $J\geq0$ (excitation) the two types of dynamics lead to the same
behaviour. At high noise levels, $T>J$, both minimisation problems are
solved by
$m=0$ (see figure \ref{fig:energygraphs}), describing a disorganised (paramagnetic) state.
This can be seen upon writing the right-hand side of
(\ref{eq:curieweisssequential}) in integral form:
\bd
m^2=m\tanh[\beta J m]=\beta J m^2\int_0^{1}\!dz~[1\minus\tanh^2[\beta J m
z]]\leq \beta J m^2
\ed
So $m^2[1\minus \beta J]\leq 0$, which gives $m=0$ as soon as $\beta
J<1$. A phase transition occurs at $T=J$ (a bifurcation of non-trivial solutions of (\ref{eq:curieweisssequential})),
and
for $T<J$ the equations for $m$ are solved
by the two non-zero solutions of (\ref{eq:curieweisssequential}),
describing a state where either all neurons tend to be firing
($m>0$) or where they tend to be quiet ($m<0$).
This becomes clear when we expand
(\ref{eq:curieweisssequential}) for small $m$:
$m=\beta J m+\order(m^3)$, so precisely at $\beta J=1$ one finds a
de-stabilisation of the trivial solution $m=0$, together with the
creation of (two) stable non-trivial ones (see also figure
\ref{fig:energygraphs}).
Furthermore, using the identity $c^*(\tanh x)=x\tanh x\minus \log \cosh x$,
we obtain from (\ref{eq:sequentialsaddle},\ref{eq:parallelsaddle}) the
relation
$\lim_{N\rightarrow\infty}\tilde{F}/N=2\lim_{N\rightarrow\infty}F/N$.
For $J<0$ (inhibition), however, the two types of dynamics give quite
different results. For sequential dynamics the relevant minimum is located at $m=0$ (the
paramagnetic state). For parallel dynamics, the minimisation problem is
invariant under $J\rightarrow\minus J$, so the behaviour is again of the
Curie-Weiss type (see figure
\ref{fig:energygraphs} and equation (\ref{eq:curieweissparallel})),
with a paramagnetic state for $T>|J|$,
a phase transition at $T=|J|$, and order for $T<|J|$.
This difference between the two types of dynamics
for $J<0$ is explained by studying dynamics.
As we will see in a subsequent
chapter, for the present (toy) model in the limit
$N\rightarrow\infty$  the average activity evolves in time
according to the deterministic laws
\bd
\frac{d}{dt}m=\tanh[\beta Jm]-m~~~~~~~~~~~~~~~~
m(t\plus 1)=\tanh[\beta Jm(t)]
\ed
for sequential and parallel dynamics, respectively. For $J<0$ the sequential system
always decays towards
the trivial state $m=0$,
whereas for sufficiently large $\beta$ the parallel system enters the stable limit-cycle
$m(t)=M_{\beta} (-1)^t$ (where $M_{\beta}$ is the non-zero solution of
(\ref{eq:curieweissparallel})).
The concepts of `distance' and `local
minima' are quite different for the two dynamics types; in contrast
to the sequential case, parallel dynamics allows the system to make the transition
$m\rightarrow-m$ in equilibrium.

\subsection{Phenomenology of Hopfield Models}

\noindent{\em The Ideas Behind the Hopfield Model.}
The Hopfield model \cite{Hopfield} is a
network of binary neurons of the type
(\ref{eq:Ising_parallel},\ref{eq:Ising_sequential}), with threshold noise
$w(\eta)=\frac{1}{2}[1\minus\tanh^2(\eta)]$, and  with a
specific recipe for the synapses $J_{ij}$ aimed at storing patterns,
motivated by suggestions made
in the late nineteen-forties \cite{Hebb}. The original model was
in fact defined more narrowly, as the zero noise limit of the
system (\ref{eq:Ising_sequential}), but the term has since then
been accepted to cover a larger network class.
Let us first consider the simplest case and try  to store a
single pattern $\bxi\in\{-1,1\}^N$ in noise-less
infinite-range binary networks. Appealing candidates for interactions and
thresholds would be
$J_{ij}=\xi_i\xi_j$ and $\theta_i=0$
(for sequential dynamics we put $J_{ii}=0$ for all $i$).
With this choice the Lyapunov function
(\ref{eq:Lyapunov_seq}) becomes:
\bd
L_{\rm seq}[\bsigma]=\frac{1}{2}N-\frac{1}{2}[\sum_{i}\xi_i\sigma_i]^2
\ed
It will have to decrease monotonically
during the dynamics,
 from which we
immediately deduce
\bd
\sum_{i}\xi_i\sigma_i(0)>0:~~~~\bsigma(\infty)=\bxi,
~~~~~~~~~~~~~~~~~~~~
\sum_{i}\xi_i\sigma_i(0)<0:~~~~\bsigma(\infty)=-\bxi
\ed
This system indeed reconstructs dynamically the original pattern $\bxi$
from an input vector $\bsigma(0)$, at least for sequential dynamics.
However, {\em en
passant} we have created an additional attractor: the state $-\bxi$.
This property is shared by all binary models in which the external fields
are zero, where the Hamiltonians $H(\bsigma)$
(\ref{eq:seq_thermodynamics1}) and $\tilde{H}(\bsigma)$ (\ref{eq:par_thermodynamics1})
are invariant under an overall sign change
$\bsigma\rightarrow-\bsigma$. A second feature common to several
(but not all) attractor neural networks is that {\em each} initial
state will lead to pattern reconstruction, even nonsensical (random)
ones.

\begin{figure}[t]
\begin{center}\vspace*{-4mm}
\epsfxsize=70mm\epsfbox{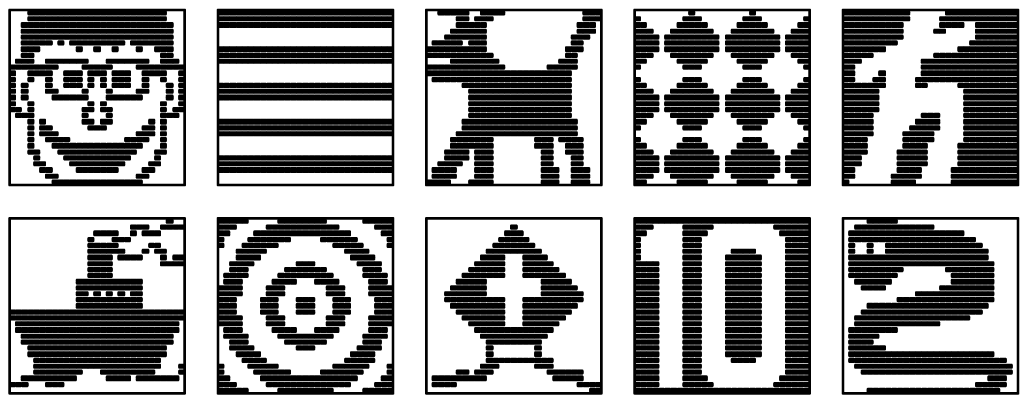}
\end{center}
\vspace*{-2mm}
\caption{Information represented as specific microscopic neuronal firing patterns $\bxi$ in an $N=841$
Hopfield network and drawn as images in the plane (black pixels: $\xi_i=1$, white pixels: $\xi_i=\minus 1$).}
\label{fig:patterns}
\end{figure}

\begin{figure}[t]
\begin{center}\vspace*{-3mm}
\epsfxsize=80mm\epsfbox{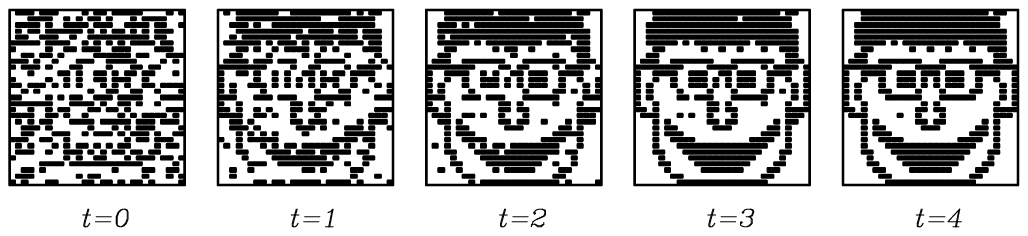}\hsp\hsp\hsp
\epsfxsize=80mm\epsfbox{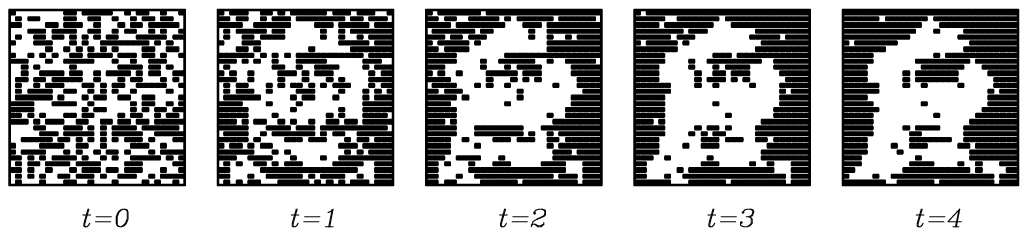}\\[5mm]
\epsfxsize=80mm\epsfbox{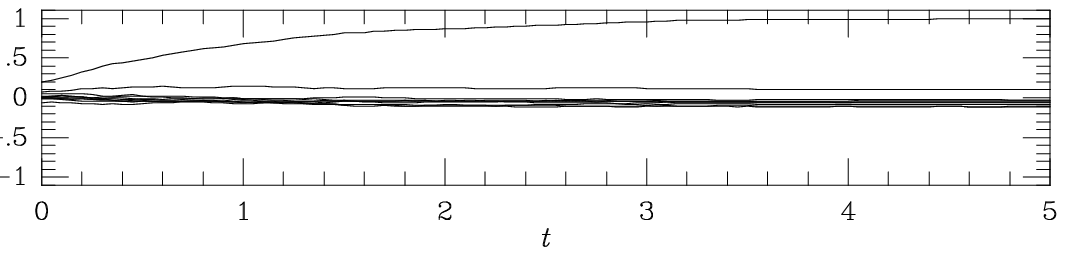}\hsp\hsp\hsp
\epsfxsize=80mm\epsfbox{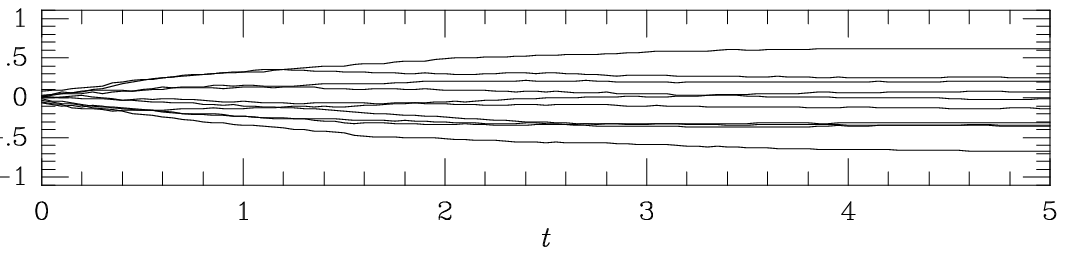}
\end{center}
\vspace*{-5mm}
\caption{Information processing in a sequential dynamics
Hopfield model with $N=841$, $p=10$ and $T=0.1$, and with the
$p=10$ stored patterns shown in figure
\ref{fig:patterns}. Left pictures: dynamic reconstruction
of a stored pattern from an initial
state which is a corrupted version thereof. Top left: snapshots
of the system state at times $t=0,1,2,3,4$
iterations/neuron. Bottom left: values of the overlap order
parameters as functions of time. Right pictures: evolution towards a spurious state
from a randomly drawn initial
state. Top right: snapshots
of the microscopic system state at times $t=0,1,2,3,4$
iterations/neuron. Bottom right: values of the overlap order
parameters as functions of time.}
\label{fig:recall_random}
\end{figure}

The Hopfield  model is obtained by generalising the previous
simple one-pattern
recipe to the case of an arbitrary number $p$
of binary patterns $\bxi^\mu\!=\!(\xi_1^\mu,\ldots,\xi^\mu_N)\!\in\!\{-1,1\}^N$:
\be
J_{ij}=\frac{1}{N}\sum_{\mu=1}^p
\xi_i^{\mu}\xi_j^{\mu}, ~~~~~~~\theta_i=0~~{\rm for~all}~i
~~~~~~~~~~({\rm sequential~dynamics:~} J_{ii}\to 0~~{\rm
for~all}~i)
\label{eq:hebbinteractions}
\ee
The prefactor $N^{-1}$ has been inserted to ensure that the limit $N\to\infty$
will exist in future expressions. The process of interest is that where, triggered by
correlation between the initial state and a
stored pattern $\bxi^\lambda$, the state vector $\bsigma$ evolves
towards $\bxi^\lambda$. If this
happens, pattern $\bxi^\lambda$ is said to be recalled.
The similarity between a state vector and the
stored patterns is measured by so-called overlaps
\be
m_{\mu}(\bsigma)=\frac{1}{N}\sum_{i}\xi_i^{\mu}\sigma_i
\label{eq:overlaps}
\ee
Numerical simulations illustrate the
functioning of the Hopfield model as an associative memory, and the
description of the recall process in terms of overlaps.
Our simulated system is an
$N=841$ Hopfield model, in which $p=10$ patterns have been stored (see figure
\ref{fig:patterns}) according to prescription
(\ref{eq:hebbinteractions}). The two-dimensional arrangement of the
neurons in this example is just a guide to the eye; since the network is fully connected
the physical location of the neurons is irrelevant.
The dynamics is as given by (\ref{eq:Ising_sequential}), with $T=0.1$.
In figure \ref{fig:recall_random} we first show (left column)
the result of letting the system evolve in time from an initial state,
which is a noisy version of one of the stored patterns (here $40 \%$
of the neuronal states $\sigma_i$ where corrupted, according to $\sigma_i\to\minus\sigma_i$). The top left
row of graphs shows snapshots of
the microscopic state as the system evolves in time.
The bottom left row shows the values of the $p=10$
overlaps $m_{\mu}$, as defined in (\ref{eq:overlaps}), as functions of
time; the one which evolves towards the value 1 corresponds to the pattern
being reconstructed.
The right column of figure \ref{fig:recall_random} shows a similar experiment,
but here the initial
state is drawn at random.
The system subsequently evolves towards a mixture of the stored patterns, which
is found to be very stable, due to the fact that the patterns involved (see figure \ref{fig:patterns})
are significantly correlated.
It will be clear that, although the idea of information storage via the creation of attractors
does work, the choice (\ref{eq:hebbinteractions})
for the synapses is still too simple to be optimal; in addition to the
desired states $\bxi^{\mu}$ and their mirror images
$-\bxi^{\mu}$, even more unwanted spurious attractors are created.
Yet this
model will already push the analysis to the limits, as soon as we
allow for the storage of an extensive number of patterns $\bxi^{\mu}$.
\vsp

\noindent{\em Issues Related to Saturation: Storage Capacity \&
Non-Trivial Dynamics.} In our previous simulation example the
loading of the network was modest; a total of $\frac{1}{2}N(N\minus 1)=353,\!220$ synapses were used to store
just $pN=8,\!410$ bits of information.
Let us now investigate the
behaviour of the network when the number of patterns scales with
the system size as $p=\alpha N$ ($\alpha>0$); now for large $N$ the
number of bits stored per synapse will be $pN/\frac{1}{2}N(N\minus 1)\approx
2\alpha$. This is called the saturation regime.
Again numerical simulations, but now with finite $\alpha$,
illustrate the main features
and complications of recall dynamics in the saturation regime.
In our example the dynamics are given by (\ref{eq:Ising_parallel}) (parallel updates),
with $T=0.1$ and threshold noise distribution $w(\eta)=\frac{1}{2}[1\minus
\tanh^2(\eta)]$; the patterns are chosen randomly.
\begin{figure}[t]
\begin{center}\vspace*{-7mm}
\epsfxsize=60mm\epsfbox{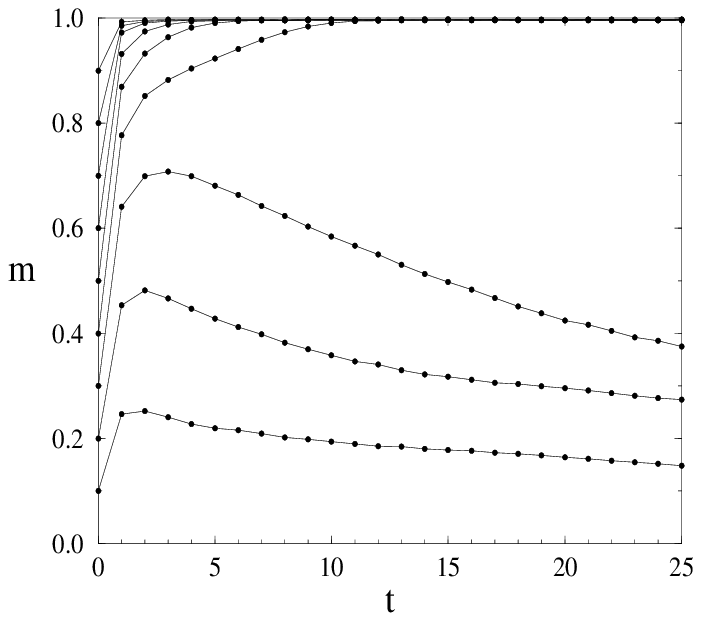}\hsp\hsp\hsp
\epsfxsize=60mm\epsfbox{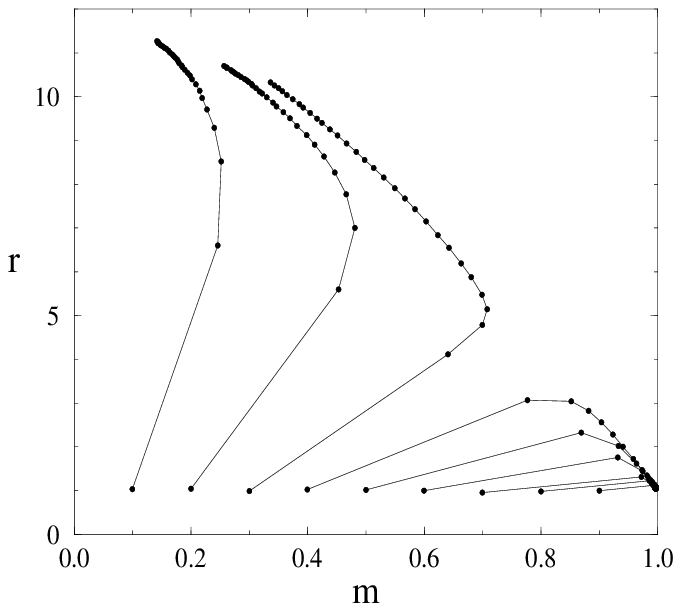}
\end{center}
\vspace*{-5mm}
\caption{Simulations of a parallel dynamics Hopfield model with $N\!=\!30,\!000$
and $\alpha\!=\!T\!=\!0.1$, and with random patterns.
Left: overlaps $m=m_1(\bsigma)$ with pattern one
as functions of time, following initial states correlated
with pattern one only, with $m_1(\bsigma(0))\in\{0.1,\ldots,0.9\}$.
Right: corresponding flow in the $(m,r)$ plane,
with  $r=\alpha^{-1}\sum_{\mu>1}m_\mu^2(\bsigma)$ measuring the
overlaps with non-nominated patterns.}
\label{fig:flows}
\end{figure}
Figure \ref{fig:flows} shows the result of measuring in such simulations the
two quantities
\be
m=m_1(\bsigma) ~~~~~~~~~~~~
r=\alpha^{-1}\sum_{\mu>1}m_\mu^2(\bsigma)
\label{eq:m_and_r}
\ee
following initial states which are correlated with pattern $\bxi^1$
only.
For large $N$ we can distinguish structural overlaps,
where $m_\mu(\bsigma)=\order(1)$, from accidental ones, where $m_\mu(\bsigma)=\order(N^{-\frac{1}{2}})$
(as for a randomly drawn $\bsigma$).
Overlaps with non-nominated patterns are seen to remain
$\order(N^{-\frac{1}{2}})$, i.e. $r(t)=\order(1)$.
We observe competition between pattern recall ($m\to 1$)
and interference of non-nominated patterns
($m\to 0$, with $r$ increasing),
and a profound slowing down of the process for non-recall
trajectories.
The initial overlap (the `cue') needed to
trigger recall is found to increase with increasing $\alpha$ (the loading)
and increasing $T$ (the noise). Further numerical experimentation,
with random patterns,
reveals that at any noise level $T$ there is a critical storage
level $\alpha_c(T)$ above which recall is impossible, with an absolute upper limit
of $\alpha_c=\max_{T}\alpha_c(T)=\alpha_c(0)\approx 0.139$.
The competing forces at work are easily recognised when working
out the local fields (\ref{eq:Ising_parallel}), using
(\ref{eq:hebbinteractions}):
\be
h_i(\bsigma)=\xi^1_i m_1(\bsigma) + \frac{1}{N}\sum_{\mu>1}\xi_i^\mu \sum_{j\neq i} \xi_j^\mu \sigma_j +\order(N^{-1})
\label{eq:fieldsinoverlaps}
\ee
The first term in (\ref{eq:fieldsinoverlaps}) drives $\bsigma$ towards pattern $\bxi^1$
as soon as $m_1(\bsigma)>0$. The second terms represent
interference, caused by correlations between $\bsigma$ and
non-nominated patterns.
One easily shows (to be demonstrated later) that for $N\to\infty$ the fluctuations in the
values of the recall overlap $m$ will vanish, and that
for the present types of initial states and threshold noise
the overlap $m$ will obey
\be
m(t\plus 1)=\int\!dz~P_t(z) \tanh[\beta(m(t)\plus z)]~~~~~~~~~~
P_t(z)=\lim_{N\to \infty} \frac{1}{N}\sum_i \bra \delta[z-
\frac{1}{N}\sum_{\mu>1}\xi_i^1\xi_i^\mu \sum_{j\neq i} \xi_j^\mu
\sigma_j(t)]\ket
~~
\label{eq:equation_m}
\ee
If all $\sigma_i(0)$ are drawn independently, ${\rm Prob}[\sigma_i(0)\!=\!\pm \xi_i^1]\!=\!\frac{1}{2}[1\pm
m(0)]$, the central limit theorem states that $P_0(z)$ is Gaussian.
One easily derives $\bra z\ket_0=0$ and
$\bra z^2\ket_0=\alpha$, so at $t=0$ equation (\ref{eq:equation_m})
gives
\be
m(1)=\int\!\frac{dz}{\sqrt{2\pi}}~e^{-\frac{1}{2}z^2} \tanh[\beta(m(0)+z\sqrt{\alpha})]
\label{eq:first_step}
\ee
The above ideas, and equation (\ref{eq:first_step}) in
particular, go back to \cite{Amari77}.
For times $t>0$, however, the independence of the states $\sigma_i$ need no longer hold.
\begin{figure}[t]
\begin{center}\vspace*{-7mm}
\epsfxsize=68mm\epsfbox{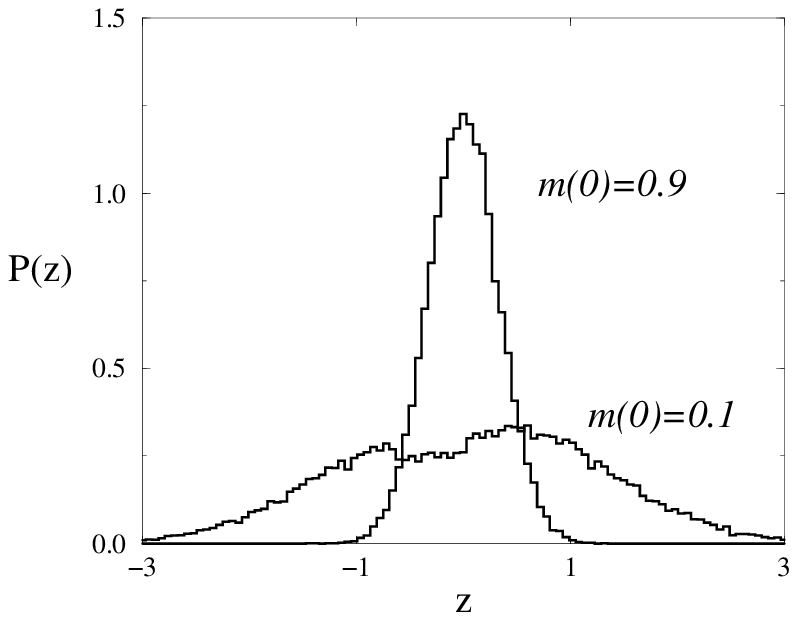}
\end{center}
\vspace*{-7mm}
\caption{Distributions of interference noise variables
$z_i=\frac{1}{N}\sum_{\mu>1}\xi_i^1\xi_i^\mu \sum_{j\neq i} \xi_j^\mu
\sigma_j$, as measured in the simulations of figure \ref{fig:flows}, at $t=10$.
Uni-modal histogram: noise distribution following
$m(0)=0.9$ (leading to recall). Bi-model histogram: noise
distribution following $m(0)=0.1$ (not leading to recall).}
\label{fig:noisedistribution}
\end{figure}
As a simple approximation one could just assume that the $\sigma_i$
remain uncorrelated at all times, i.e. ${\rm Prob}[\sigma_i(t)\!=\!\pm
\xi_i^1]\!=\!\frac{1}{2}[1\pm m(t)]$ for all $t\geq 0$,
such that the argument given for $t=0$
would hold generally, and where (for randomly drawn patterns)
the mapping (\ref{eq:first_step}) would describe
the overlap evolution at {\rm all} times:
\be
m(t+1)=\int\!\frac{dz}{\sqrt{2\pi}}~e^{-\frac{1}{2}z^2} \tanh[\beta(m(t)+z\sqrt{\alpha})]
\label{eq:extremely_diluted}
\ee
This equation, however, must be generally incorrect. Firstly, figure
\ref{fig:flows} already shows that knowledge of $m(t)$ {\em only} does
not yet permit prediction of $m(t\plus 1)$.
Secondly, upon working out its bifurcation properties one finds that
equation (\ref{eq:extremely_diluted}) predicts a storage capacity of $\alpha_c=2/\pi\approx
0.637$, which is no way near to what is actually being observed.
We will see in the paper on dynamics that only for certain types of extremely diluted networks
(where most of the synapses are cut) equation
(\ref{eq:extremely_diluted}) is indeed correct on finite
times;
in these networks the time it takes for correlations between neuron states
to build up diverges with $N$, so that correlations are simply not yet noticable on finite
times.

For fully connected Hopfield networks storing random patterns near saturation, i.e. with
$\alpha>0$, the complicated correlations building up between the microscopic variables
in the course of the dynamics generate an  interference noise distribution which is intrinsically
non-Gaussian, see e.g. figure \ref{fig:noisedistribution}. This
leads to a highly non-trivial dynamics which is fundamentally different from that in
the $\lim_{N\to\infty}p/N=0$ regime.
Solving models of recurrent neural networks in the saturation regime boils down to calculating
this non-Gaussian noise distribution, which
requires advanced mathematical techniques (in statics
and dynamics), and constitutes
the main challenge to the theorist.
The simplest way to evade this challenge is to study
situations where the interference noise
is either trivial (as with asymmetric extremely diluted models)
or where it vanishes, which happens in fully connected networks when
$\alpha\!=\!\lim_{N\to\infty}p/N\!=\!0$ (as with finite $p$). The
latter $\alpha=0$ regime is the one we will explore first.

\subsection{Analysis of Hopfield Models Away From Saturation}

\noindent{\em Equilibrium Order Parameter Equations}.
A binary Hopfield network with parameters given by
(\ref{eq:hebbinteractions}) obeys detailed balance, and
the Hamiltonian $H(\bsigma)$ (\ref{eq:seq_thermodynamics1}) (corresponding to sequential
dynamics) and the pseudo-Hamiltonian $\tilde{H}(\bsigma)$
(\ref{eq:par_thermodynamics1}) (corresponding to parallel
dynamics) become
\be
H(\bsigma)=
-\frac{1}{2}N\sum_{\mu=1}^{p}m_{\mu}^2(\bsigma)+\frac{1}{2}p
~~~~~~~~~~~~~~~~~~
\tilde{H}(\bsigma)=
-\frac{1}{\beta}\sum_{i}\log 2\cosh[\beta\sum_{\mu=1}^{p}\xi_i^{\mu}m_{\mu}(\bsigma)]
\label{eq:HHtilde}
\ee
with the overlaps (\ref{eq:overlaps}). Solving the statics
 implies calculating the free energies $F$ and $\tilde{F}$:
\bd
F=-\frac{1}{\beta}\log\sum_{\bsigma}e^{-\beta H(\bsigma)}~~~~~~~~~~~~
\tilde{F}=
-\frac{1}{\beta}\log\sum_{\bsigma}
e^{-\beta\tilde{H}(\bsigma)}
\ed
Upon introducing the short-hand notation
$\bm=(m_1,\ldots,m_p)$ and $\bxi_i=(\xi_i^{1},\ldots,\xi_i^{p})$,
both free energies can be expressed in terms of the density of states ${\cal
D}(\bm)=
2^{-N}\sum_{\bsigma}\delta[\bm\minus \bm(\bsigma)]$:
\be
F/N=-\frac{1}{\beta}\log2
-\frac{1}{\beta N}\log\int\!d\bm~ {\cal D}(\bm)~ e^{\frac{1}{2}\beta
N\bm^2}+\frac{p}{2N}
\label{eq:hopfreeenergy}
\ee
\be
\tilde{F}/N=-\frac{1}{\beta}\log2
-\frac{1}{\beta N}\log \int\!d\bm~ {\cal D}(\bm)~e^{\sum_{i=1}^{N}\log
2\cosh\left[\beta\bxi_i\cdot\bm\right]}
\label{eq:parhopfreeenergy}
\ee
(note: $\int\!d\bm~\delta[\bm\minus\bm(\bsigma)]=1$).
In order to proceed we need to specify how the number of patterns
$p$ scales with the system size $N$. In this section we will follow \cite{AGS} (equilibrium analysis following
sequential dynamics)
and \cite{Fontanari} (equilibrium analysis following parallel dynamics), and
assume $p$ to
be finite. One can now easily calculate the leading
contribution to the density of states, using the integral
representation of the $\delta$-function and keeping in mind that
according to (\ref{eq:hopfreeenergy},\ref{eq:parhopfreeenergy}) only
terms exponential in $N$ will retain statistical relevance
for $N\to\infty$:
\bd
\lim_{N\rightarrow\infty}\frac{1}{N}\log {\cal D}(\bm) =
\lim_{N\rightarrow\infty} \frac{1}{N}\log \int\! d\bx~e^{iN\bx\cdot\bm}\bra
e^{-i\sum_{i=1}^{N}\sigma_i\bxi_i\cdot\bx}\ket_{\bsigma}
\ed
\bd
=
\lim_{N\rightarrow\infty} \frac{1}{N}\log
\int\!d\bx~
e^{N[i\bx\cdot\bm+\bra\log\cos[\bxi\cdot\bx]\ket_{\bxi}]}
\ed
with the abbreviation
$\bra
\Phi(\bxi)\ket_{\bxi}=\lim_{N\rightarrow\infty}\frac{1}{N}\sum_{i=1}^N
\Phi(\bxi_i)$.
The leading contribution to both free energies can be expressed as a
finite-dimensional integral, for large $N$ dominated by that
saddle-point (extremum) for which the extensive exponent is real and maximal:
\bd
\lim_{N\rightarrow\infty}F/N=
-\frac{1}{\beta N}\log\int\!d\bm d\bx~
e^{- N\beta f(\bm,\bx)}
= {\rm extr}_{\bx,\bm}~ f(\bm,\bx)
\nsp
\ed
\bd
\lim_{N\rightarrow\infty}\tilde{F}/N=
-\frac{1}{\beta N}\log\int\!d\bm d\bx~
e^{- N\beta \tilde{f}(\bm,\bx)}= {\rm extr}_{\bx,\bm}~ \tilde{f}(\bm,\bx)
\ed
with
\bd
\begin{array}{ll}
f(\bm,\bx)\! & =
-\frac{1}{2}\bm^2-i\bx\cdot\bm-\beta^{-1}\bra\log2\cos\left[\beta\bxi\cdot\bx\right]\ket_{\bxi}\\[3mm]
\tilde{f}(\bm,\bx)\!& =
-\beta^{-1}\bra\log
2\cosh\left[\beta\bxi\cdot\bm\right]\ket_{\bxi}
-i\bx\cdot\bm-\beta^{-1}\bra\log2\cos\left[\beta \bxi\cdot\bx\right]\ket_{\bxi}
\end{array}
\ed
The saddle-point equations for $f$ and $\tilde{f}$ are given by:
\bd
\begin{array}{lllll}
f: & &  \bx=i\bm, & &
   i\bm=\bra\bxi\tan\left[\beta\bxi\cdot\bx\right]\ket_{\bxi}
   \\[2mm]
\tilde{f}: & &
\bx=i\bra\bxi\tanh\left[\beta\bxi\cdot\bm\right]\ket_{\bxi}, & &
i\bm=\bra\bxi\tan\left[\beta\bxi\cdot\bx\right]\ket_{\bxi}
\end{array}
\ed
In saddle-points $\bx$ turns out to be purely imaginary.
However, after a shift of the integration contours, putting $\bx=i\bx^\star(\bm)+\by$ (where $i\bx^\star(\bm)$ is the
imaginary saddle-point, and where $\by\in\Re^p$) we can eliminate $\bx$
in favor of $\by\in\Re^p$ which does have
a real saddle-point, by construction.\footnote{Our functions to be integrated have no poles,
but strictly speaking we still have to verify that the integration segments linking the original integration regime to the shifted one will
not contribute to the integrals. This is generally a tedious and distracting
task, which is often skipped. For simple models, however (e.g. networks with uniform synapses),
the verification can be carried out properly, and all is found to be safe.}
We then obtain\footnote{Here we used the equation $\partial f(\bm,\bx)/\partial\bm=\bnul$
to express $\bx$ in terms of $\bm$, because this
is simpler. Strictly speaking we should have used $\partial f(\bm,\bx)/\partial\bx=\bnul$
for this purpose; our short-cut could in principle generate additional solutions.
In the present model, however, we can check explicitly that this is not the case.
Also, in view of the imaginary saddle-point
$\bx$, we can not be certain that, upon elimination of $\bx$, the relevant saddle-point
of the remaining function $f(\bm)$ must be a minimum. This will have to be checked, for instance by inspection
of the $T\to\infty$ limit.}
\bd
\begin{array}{lll}
{\sl Sequential~Dynamics}: & & \bm=\bra\bxi\tanh[\beta\bxi\cdot\bm]\ket_{\bxi} \\[3mm]
{\sl Parallel~Dynamics}:
& & \bm=\bra\bxi\tanh[\beta\bxi\cdot[\bra\bxi^{\prime}\tanh[\beta\bxi^{\prime}\cdot\bm]\ket_{\bxi^{\prime}}]]\ket_{\bxi}
\end{array}
\ed
(compare to e.g.
(\ref{eq:curieweisssequential},\ref{eq:curieweissparallel})).
The solutions of the above two equations will in general be identical.
To see this, let us denote
$\hat{\bm}=\bra\bxi\tanh\left[\beta\bxi\cdot\bm\right]\ket_{\bxi}$, with which the saddle
point equation for $\tilde{f}$ decouples into:
\bd
\bm=\bra\bxi\tanh\left[\beta\bxi\cdot\hat{\bm}\right]\ket_{\bxi}~~~~~~~~~~~~~~~~
\hat{\bm}=\bra\bxi\tanh\left[\beta\bxi\cdot\bm\right]\ket_{\bxi}
\ed
so
\bd
\left[\bm\minus\hat{\bm}\right]^2=
\bra\left[(\bxi\cdot\bm)\minus(\bxi\cdot\hat{\bm})\right]
\left[\tanh(\beta\bxi\cdot\hat{\bm})\minus\tanh(\beta\bxi\cdot\bm)\right]\ket_{\bxi}
\ed
Since $tanh$ is a monotonicaly increasing function, we must have
$\left[\bm\minus\hat{\bm}\right]\cdot\bxi=0$ for each $\bxi$ that
contributes to the averages $\bra\ldots\ket_{\bxi}$. For all choices
of patterns where the covariance matrix
$C_{\mu\nu}=\bra\xi_{\mu}\xi_{\nu}\ket_{\bxi}$ is positive definite,
we thus obtain $\bm=\hat{\bm}$.
The final result is: for both types of dynamics (sequential and
parallel) the overlap order parameters in equilibrium are given by the
solution $\bm^{*}$ of
\be
\bm=\bra\bxi\tanh\left[\beta\bxi\cdot\bm\right]\ket_{\bxi}
\nsp
\label{eq:overlapeqns}
\ee
which minimises\footnote{We here indeed know the relevant
saddle-point to be a minimum: the only solution of the
saddle-point equations at high temperatures, $\bm=\bnul$, is seen to
minimise $f(\bm)$, since $f(\bm)\plus \beta^{-1}\log 2=\frac{1}{2}\bm^2(1\minus
\beta)\plus\order(\beta^3)$.}
\be
f(\bm)=
\frac{1}{2}\bm^2-\frac{1}{\beta}\bra\log2\cosh\left[\beta\bxi\cdot\bm\right]\ket_{\bxi}
\label{eq:freeenergysurface}
\ee
The free energies of the ergodic compoments are
$\lim_{N\rightarrow\infty}F/N=f(\bm^{*})$ and
$\lim_{N\rightarrow\infty}\tilde{F}/N=2f(\bm^{*})$.
Adding generating terms of the form $H\rightarrow H\plus \lambda
g[\bm(\bsigma)]$ to the Hamiltonians allows us identify
$\bra g[\bm(\bsigma)]\ket_{\rm eq}=\lim_{\lambda\rightarrow0}\partial F/\partial
\lambda=g[\bm^{*}]$.
Thus, in equilibrium the fluctuations in the overlap order
parameters $\bm(\bsigma)$ (\ref{eq:overlaps}) vanish for
$N\rightarrow\infty$. Their deterministic values are simply given by $\bm^{*}$.
Note that in the case of sequential dynamics we could also have
used linearisation with Gaussian integrals (as used previously for coupled oscillators
with uniform synapses) to arrive at this solution, with $p$ auxiliary integrations,
but that for parallel dynamics this would not have been possible.
\vsp

\noindent{\em Analysis of Order Parameter Equations: Pure States \& Mixture States}.
We will restrict our further discussion to the case of randomly drawn
patterns, so
\bd
\bra\Phi(\bxi)\ket_{\bxi}=2^{-p}\!\!\sum_{\bxi\in\{-1,1\}^p}\Phi(\bxi),~~~~~~~~~~~~~~
\bra \xi_{\mu}\ket_{\bxi}=0,~~~
\bra \xi_{\mu}\xi_{\nu}\ket_{\bxi}=\delta_{\mu\nu}
\ed
(generalisation to correlated patterns is in principle
straightforward).
We first establish an upper bound for the temperature for where
non-trivial solutions $\bm^{*}$ could exist, by writing
(\ref{eq:overlapeqns}) in integral form:
\bd
m_{\mu}=
\beta\bra\xi_{\mu}(\bxi\cdot\bm)\int_0^1\!d\lambda[1\minus \tanh^2[\beta\lambda\bxi\cdot\bm]]\ket_{\bxi}
\nsp
\ed
from which we deduce
\bd
0 =\bm^2\minus \beta\bra(\bxi\cdot\bm)^2\int_0^1\!d\lambda[1\minus \tanh^2
[\beta\lambda\bxi\cdot\bm]]\ket_{\bxi}
 \geq \bm^2\minus \beta\bra(\bxi\cdot\bm)^2\ket_{\bxi}=\bm^2(1\minus
 \beta)
\ed
For $T>1$ the only
solution of (\ref{eq:overlapeqns}) is the paramagnetic state
$\bm=0$, which gives for the
free energy per neuron $-T\log2$ and
$-2T\log2$ (for sequential and parallel dynamics,
respectively).
At $T=1$ a phase transition occurs, which follows from expanding
(\ref{eq:overlapeqns}) for small $|\bm|$ in powers of $\tau=\beta\minus 1$:
\bd
m_{\mu} =(1\plus\tau)m_{\mu}
-\frac{1}{3}\sum_{\nu\rho\lambda} m_{\nu}m_{\rho}m_{\lambda}\bra\xi_{\mu}\xi_{\nu}\xi_{\rho}\xi_{\lambda}\ket_{\bxi}
+\order(\bm^5\!,\tau\bm^3)
=m_{\mu}[1\plus \tau \minus \bm^2\plus \frac{2}{3}m_{\mu}^2]
+\order(\bm^5\!,\tau\bm^3)
\ed
The new saddle-point scales as
$m_{\mu}\!=\tilde{m}_{\mu}\tau^{1/2}\!+ \order(\tau^{3/2})$, with for each $\mu$:
$\tilde{m}_{\mu}\!=0$ or $0=1\minus
\tilde{\bm}^2\!+\frac{2}{3}\tilde{m}_{\mu}^2$.\newline
The solutions are of the form
$\tilde{m}_{\mu}\in\{-\tilde{m},0,\tilde{m}\}$.
If we denote with $n$ the number of non-zero components in the vector
$\tilde{\bm}$, we derive from the above identities:
$\tilde{m}_{\mu}=0$ or $\tilde{m}_{\mu}=\pm\sqrt{3}/\sqrt{3n\minus 2}$.
These saddle-points are called {\em mixture states}, since they
correspond to microscopic configurations correlated equally with a
finite number $n$ of the stored patterns (or their negatives).
Without loss of generality we can always perform gauge transformations
on the set of stored patterns (permutations and reflections), such that the mixture states
acquire the form
\be
\bm=m_n(\overbrace{1,\ldots,1}^{n~{\rm
times}},\overbrace{0,\ldots,0}^{p-n~{\rm times}})~~~~~~~~~~~~~~
m_n=[\frac{3}{3n\minus 2}]^{\frac{1}{2}}(\beta-1)^{1/2}+\ldots
\label{eq:mixturestates}
\ee
These states are in fact saddle-points of the surface $f(\bm)$
(\ref{eq:freeenergysurface}) for any finite temperature, as
can be verified by substituting (\ref{eq:mixturestates}) as an {\em
ansatz} into (\ref{eq:overlapeqns}):
\bd
\mu\leq n:~~~ m_n=\bra \xi_{\mu}\tanh[\beta m_n \sum_{\nu\leq
n}\xi_{\nu}]\ket_{\bxi}
~~~~~~~~~~~~~~~~
\mu> n:~~~ 0 =\bra \xi_{\mu}\tanh[\beta m_n \sum_{\nu\leq
n}\xi_{\nu}]\ket_{\bxi}
\ed
The second equation is automatically satisfied since the average
factorises. The first equation leads to a condition determining the
amplitude $m_n$ of the mixture states:
\be
 m_n=\bra [\frac{1}{n}\sum_{\mu\leq n}\xi_{\mu}]\tanh[\beta m_n \sum_{\nu\leq
n}\xi_{\nu}]\ket_{\bxi}
\label{eq:mixtureamplitude}
\ee
The corresponding values of $f(\bm)$, to be denoted by $f_n$, are
\be
f_n=\frac{1}{2}n m_n^2-\frac{1}{\beta}\bra\log 2\cosh[\beta
m_n\sum_{\nu\leq n}\xi_{\nu}]\ket_{\bxi}
\label{eq:mixtureenergy}
\ee
The relevant question at this stage is whether or not these saddle-points
correspond to local minima of the surface $f(\bm)$
(\ref{eq:freeenergysurface}). The second derivative of $f(\bm)$ is
given by
\be
\frac{\partial^2 f(\bm)}{\partial m_{\mu}\partial m_{\nu}}
=\delta_{\mu\nu}-
\beta\bra\xi_{\mu}\xi_{\nu}\left[1-\tanh^2\left[\beta\bxi\cdot\bm\right]\right]\ket_{\bxi}
\label{eq:curvature}
\ee
(a local minimum corresponds to a positive definite second derivative).
In the trivial saddle-point $\bm=0$ this gives simply
$\delta_{\mu\nu}(1\minus \beta)$, so at $T=1$ this state destabilises.
In a mixture state of the type (\ref{eq:mixturestates}) the second
derivative becomes:
\bd
D^{(n)}_{\mu\nu}=
\delta_{\mu\nu}-
\beta\bra\xi_{\mu}\xi_{\nu}[1\minus \tanh^2[\beta m_n
\sum_{\rho\leq n}\xi_{\rho}]]\ket_{\bxi}
\ed
\begin{figure}[t]
\begin{center}\vspace*{-4mm}
\epsfxsize=75mm\epsfbox{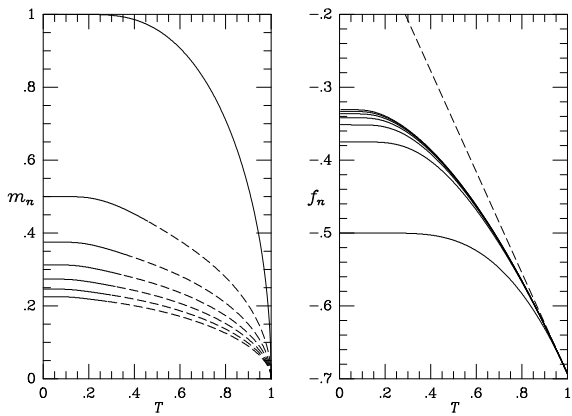}
\end{center}
\vspace*{-5mm}
\caption{Left picture: Amplitudes $m_n$ of the mixture states as
functions of temperature. From top to bottom: $n=1,3,5,7,9,11,13$.
Solid: region where they are stable (local minima of $f$). Dashed:
region where they are unstable. Right picture: corresponding `free
energies' $f_n$. From bottom to top: $n=1,3,5,7,9,11,13$. Dashed line:
`free energy' of the paramagnetic state ${\protect\boldmath m}=\bnul$ (for comparison).}
\label{fig:mixturestates}
\end{figure}
Due to the symmetries in the problem the spectrum of the matrix
$D^{(n)}$ can
be calculated. One finds the following eigenspaces, with
$Q=\bra\tanh^2[\beta m_n
\sum_{\rho\leq n}\xi_{\rho}]\ket_{\bxi}$ and $R=\bra\xi_1\xi_2\tanh^2[\beta m_n
\sum_{\rho\leq n}\xi_{\rho}]\ket_{\bxi}$:
\bd
\begin{array}{lll}
& {\rm Eigenspace:} & {\rm Eigenvalue:} \\[1mm]
I: & \bx=(0,\ldots,0,x_{n+1},\ldots,x_{p}) & 1\minus\beta[1\minus Q] \\
I\!I:& \bx=(1,\ldots,1,0,\ldots,0) & 1\minus\beta[1\minus Q\plus (1\minus n)R] \\
I\!I\!I: & \bx=(x_1,\ldots,x_n,0,\ldots,0),~
\sum_{\mu}x_{\mu}\!=\!0 & 1\minus\beta[1\minus Q\plus R]
\end{array}
\ed
Eigenspace $I\!I\!I$ and the quantity $R$ only come into play for
$n>1$. To find the smallest eigenvalue we need to know the sign of
$R$. With the abbreviation $M_{\bxi}=\sum_{\rho\leq n}\xi_{\rho}$ we find:
\bd
\begin{array}{ll}
n(n\minus 1)R\! &
= \bra M_{\bxi}^2\tanh^2[\beta m_n M_{\bxi}]\ket_{\bxi}- n\bra\tanh^2[\beta m_n M_{\bxi}]\ket_{\bxi}
\\[3mm] &
=\bra[ M^2_{\bxi}\minus \bra M^2_{\bxi^\prime}\ket_{\bxi^\prime}]
\tanh^2[\beta m_n |M_{\bxi}|]\ket_{\bxi}
\\[2mm] &
=\bra[ M^2_{\bxi}\minus \bra M^2_{\bxi^\prime}\ket_{\bxi^\prime}]
\left\{
\tanh^2[\beta m_n\sqrt{M^2_{\bxi}}]-
\tanh^2[\beta m_n\sqrt{\bra M^2_{\bxi^\prime}\ket_{\bxi^\prime}}]\right\}\ket_{\bxi}\geq 0
\end{array}
\ed
We may now identify the conditions for an $n$-mixture state to
be a local minimum of $f(\bm)$. For $n=1$ the relevant
eigenvalue is $I$, now the quantity $Q$
simplifies considerably. For $n>1$ the relevant eigenvalue is
$I\!I\!I$, here we can combine $Q$ and $R$ into one
single average:
\bd
\begin{array}{ll}
n=1: & 1\minus\beta[1\minus\tanh^2[\beta m_1]] >0  \\[1mm]
n=2: & 1\minus\beta >0  \\[1mm]
n\geq3: & 1\minus\beta[1\minus\bra\tanh^2[\beta m_n
\sum_{\rho=3}^{n}\xi_{\rho}]\ket_{\bxi}]>0
\end{array}
\ed
The $n=1$ states, correlated with one pattern only, are the
desired solutions. They are stable for all $T<1$, since partial
differentiation with respect to $\beta$ of the $n=1$
amplitude equation (\ref{eq:mixtureamplitude}) gives
\bd
m_1=\tanh[\beta m_1] ~~\rightarrow~~
1\minus \beta[1\minus\tanh^2[\beta
m_1]]=
m_1[1\minus \tanh^2[\beta m_1]](\partial m_1/\partial \beta)^{-1}
\ed
(clearly $\sgn[m_1]=\sgn[\partial m_1/\partial\beta]$). The $n=2$
mixtures are always unstable. For $n\geq3$ we have to solve the
amplitude equations (\ref{eq:mixtureamplitude}) numerically to
evaluate their stability. The result is shown in figure
\ref{fig:mixturestates}, together with the corresponding `free
energies' $f_n$ (\ref{eq:mixtureenergy}).
It turns out that only for odd $n$ will there
be a critical temperature below which the $n$-mixture states are local
minima of $f(\bm)$. From figure \ref{fig:mixturestates} we
can also conclude that, in terms of the network functioning as an associative
memory, noise is actually beneficial in the sense that it can be used
to eliminate the unwanted $n>1$ ergodic components (while retaining the
relevant ones: the pure $n=1$ states).
In fact the overlap equations (\ref{eq:overlapeqns}) do also allow for
stable solutions
different from the $n$-mixture states discussed here. They are in turn
found
to be continuously bifurcating mixtures of the mixture states.
However, for random (or uncorrelated) patterns they come into existence only near $T=0$
and play a marginal role; phase space is dominated
by the odd $n$-mixture states.

We have now solved the model in equilibrium for finite $p$ and $N\to\infty$.
Most of the relevant
information on when and to what extent stored random patterns will be recalled
is summarised in figure \ref{fig:mixturestates}.
For non-random patterns one simply has to study the bifurcation properties of
equation (\ref{eq:overlapeqns}) for
the new pattern statistics at hand; this is only qualitatively
different from the random pattern analysis explained above.
The occurrence of multiple saddle-points corresponding to local minima
of the free energy signals ergodicity breaking.
Although among these only the {\em global}  minimum will correspond to
the thermodynamic equilibrium state, the non-global minima correspond to
 true ergodic components, i.e. on finite time-scales they will be just as
relevant as the global minimum.

\section{Simple Recurrent Networks of Coupled Oscillators}

\subsection{Coupled Oscillators with Uniform Synapses}

Models with continuous variables involve integration over states, rather than
summation. For a coupled oscillator network (\ref{eq:oscillators})
with uniform synapses $J_{ij}=J/N$ and zero frequencies $\omega_i=0$
(which is a simple version of the model in \cite{Kuramoto})
we obtain for the free energy per oscillator:
\bd
\lim_{N\rightarrow\infty}F/N
=-\lim_{N\to\infty}\frac{1}{\beta N}\log \int_{-\pi}^{\pi}\!\cdots\!\int_{-\pi}^{\pi}\!\!d\bphi ~
e^{(\beta J/2N)
\left[[\sum_i\cos(\phi_i)]^2+[\sum_i \sin(\phi_i)]^2\right]}
\ed
We would now have to `count'
microscopic states with prescribed average cosines and sines.
A faster route exploits
auxiliary Gaussian integrals, via the identity
\be
e^{\frac{1}{2}y^2}=\int\!Dz~e^{yz}
\label{eq:gaussians}
\ee
with the short-hand
$Dx=(2\pi)^{-\frac{1}{2}}e^{-\frac{1}{2}x^2}dx$
(this alternative would also have been open to us in the
binary case; my aim in this section is to explain both methods):
\bd
\lim_{N\rightarrow\infty}F/N
=-\lim_{N\to\infty}\frac{1}{\beta N}\log\int_{-\pi}^{\pi}\!\cdots\!\int_{-\pi}^{\pi}\!\!d\bphi
\int\!DxDy~e^{ \sqrt{\beta J/N}
\left[x\sum_i\cos(\phi_i)+y\sum_i \sin(\phi_i)\right]}
\ed
\bd
=-\lim_{N\to\infty}\frac{1}{\beta N}\log\int\!DxDy \left[
\int_{-\pi}^{\pi}\!\!d\phi ~
e^{ \cos(\phi)\sqrt{\beta J(x^2+y^2)/N}}
\right]^N
\ed
\bd
=-\lim_{N\to\infty}\frac{1}{\beta N}\log\int_0^\infty\!dq~q e^{-\frac{1}{2}N\beta |J|q^2} \left[
\int_{-\pi}^{\pi}\!\!d\phi ~
e^{\beta |J|q \cos(\phi)\sqrt{{\rm sgn}(J)}}
\right]^N
\ed
where we have transformed to polar coordinates, $(x,y)=q\sqrt{\beta
|J|N}(\cos\theta,\sin\theta)$, and where we have already
eliminated (constant) terms which will not survive the limit
$N\to\infty$.
Thus, saddle-point integration gives us, quite similar to the previous
cases (\ref{eq:sequentialsaddle},\ref{eq:parallelsaddle}):
\begin{figure}[t]
\begin{center}\vspace*{-10mm}
\epsfxsize=55mm\epsfbox{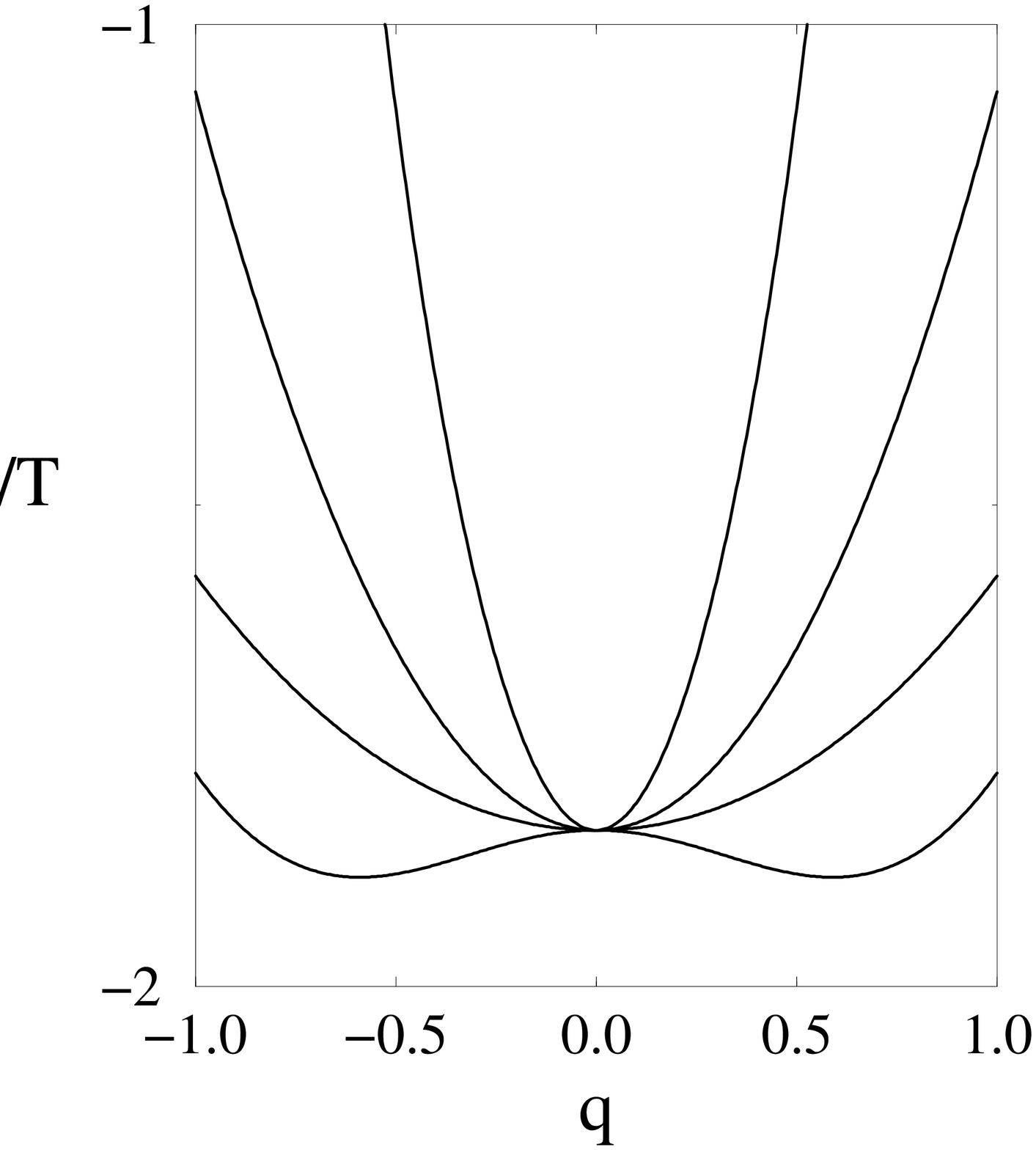}
\epsfxsize=55mm\epsfbox{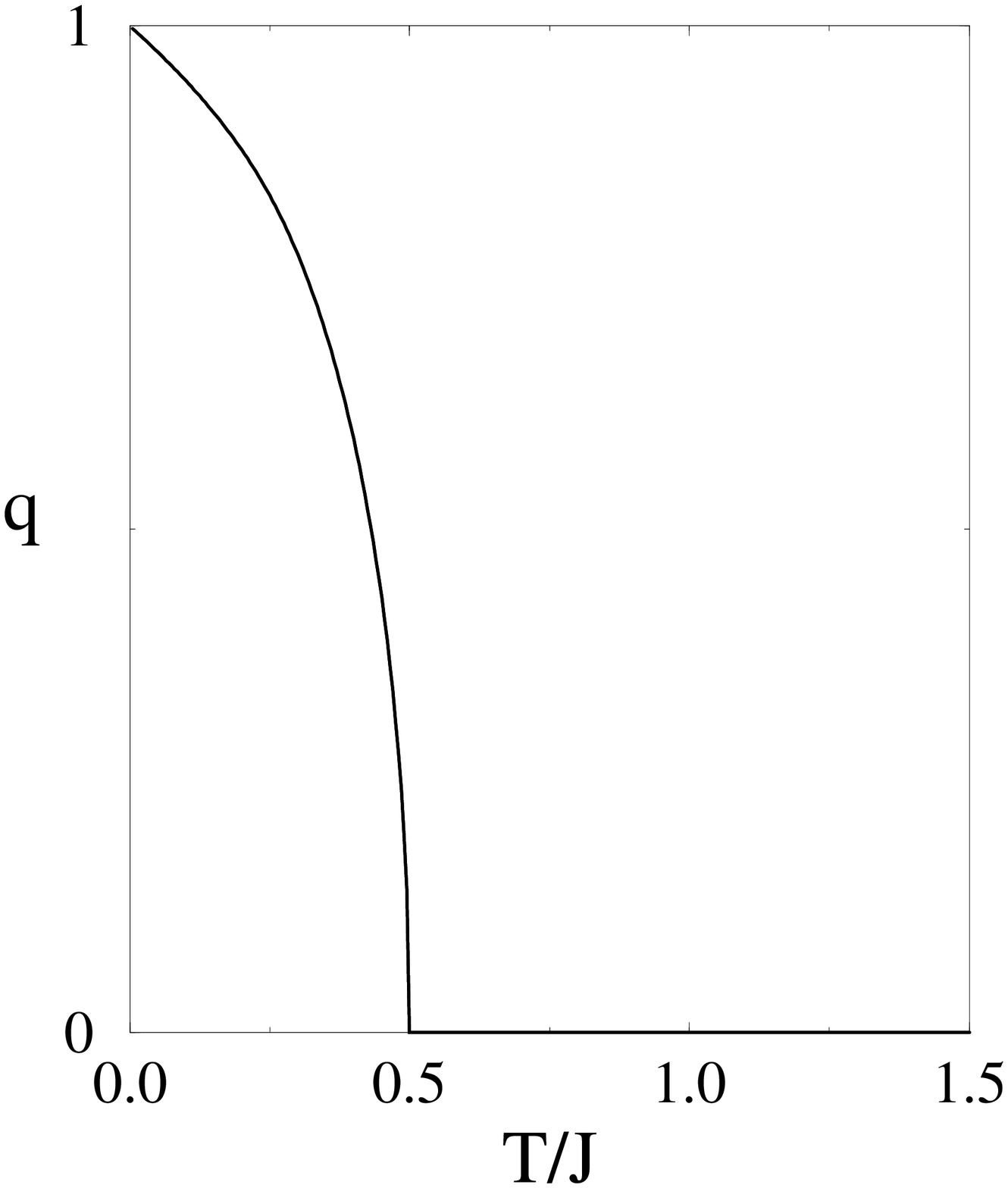}
\end{center}
\vspace*{-5mm}
\caption{The function $f(q)/T$ (left)
for networks of coupled oscillators with uniform synapses
$J_{ij}=J/N$, and for different choices of the re-scaled interaction strength
$J/T$ ($T=\beta^{-1}$):
$J/T=\minus\frac{5}{2},\minus 1,1,\frac{5}{2}$ (from top to bottom).
The right picture gives, for $J>0$, the location of
the non-negative minimum of $f(q)$ (which measures the overall
degree of global synchronisation in thermal equilibrium) as a function of
$T/J$. A transition to a synchronised state occurs at $T/J=\frac{1}{2}$. }
\label{fig:osc_energygraphs}
\end{figure}
\be
\lim_{N\rightarrow\infty}F/N=\min_{q\geq 0} f(q)~~~~~~~~~~~
\begin{array}{lll}
J>0: && \beta f(q)= \frac{1}{2}\beta|J|q^2-\log [2\pi I_0(\beta |J|q)]\\[2mm]
J<0: && \beta f(q)= \frac{1}{2}\beta|J|q^2-\log [2\pi I_0(i\beta |J|q)]
\end{array}
\label{eq:oscillatorsaddle}
\ee
in which the $I_n(z)$ are the Bessel functions (see e.g.
\cite{AbramStegun}).
The function $f(q)$
is shown in figure \ref{fig:osc_energygraphs}.
The equations from which to solve the minima are obtained by
differentiation, using $\frac{d}{dz}I_0(z)=I_1(z)$:
\be
J>0:~~~~ q= \frac{I_1(\beta|J|q)}{I_0(\beta|J|q)}
~~~~~~~~~~~~~~~~
J<0:~~~~ q=i ~\frac{I_1(i\beta |J|q)}{I_0(i\beta |J|q)}
\label{eq:kuramoto_saddle}
\ee
Again, in both cases the problem has been reduced to studying a single non-linear
equation.
The physical meaning of the solution follows from the identity
$\minus 2 \partial F/\partial J=\bra N^{-1}\sum_{i\neq j}\cos(\phi_i\minus
\phi_j)\ket$:
\bd
\lim_{N\to\infty}\bra[\frac{1}{N}\sum_{i}\cos(\phi_i)]^2\ket+
\lim_{N\to\infty}\bra[\frac{1}{N}\sum_{i}\sin(\phi_i)]^2\ket
=\sgn(J)~q^2
\ed
From this equation it also follows that $q\leq 1$.
Note: since $\partial f(q)/\partial q=0$ at the minimum, one only
needs to consider the explicit derivative of $f(q)$ with respect
to $J$.
If the synapses induce anti-synchronisation, $J<0$, the
only solution of (\ref{eq:kuramoto_saddle}) (and the minimum in
(\ref{eq:oscillatorsaddle})) is the trivial state $q=0$.
This also follows immediately from the equation which gave the physical meaning of $q$.
For synchronising forces, $J>0$, on the other hand, we again find the trivial solution
at high noise levels, but a globally  synchronised
 state with $q>0$ at low noise levels.
Here a phase transition occurs at $T=\frac{1}{2}J$
(a bifurcation of non-trivial solutions of (\ref{eq:kuramoto_saddle})),
and for $T<\frac{1}{2}J$ the minimum of (\ref{eq:oscillatorsaddle}) is found
at two non-zero values for $q$.
The critical noise level is again found upon expanding the
saddle-point equation, using $I_0(z)=1\plus\order(z^2)$ and $I_1(z)=\frac{1}{2}z\plus\order(z^3)$:
$q=\frac{1}{2}\beta Jq +\order(q^3)$. Precisely at $\beta J=2$ one finds a
de-stabilisation of the trivial solution $q=0$, together with the
creation of (two) stable non-trivial ones (see figure
\ref{fig:osc_energygraphs}). Note that, in view of (\ref{eq:oscillatorsaddle}), we are only
interested in non-negative values of $q$.  One can prove, using the properties
of the Bessel functions, that there are no other (discontinuous)
bifurcations of non-trivial solutions of the saddle-point
equation. Note, finally, that the absence of a state with global
anti-synchronisation for $J<0$ has the same origin as the absence
of an anti-ferromagnetic state for $J<0$ in the previous models
with binary neurons. Due to the long-range nature of the synapses $J_{ij}=J/N$
such states simply cannot exist: whereas any set of oscillators can be in
a fully synchronised state, if two oscillators are in
anti-synchrony it is already impossible for a third to be
simultaneously in anti-synchrony with the first two (since anti-synchrony
with one implies synchrony  with the other).

\subsection{Coupled Oscillator Attractor Networks}

\noindent{\em Intuition \& Definitions.}
Let us now turn to an alternative realisation of information storage
in a recurrent network based upon the creation of attractors. We will solve
models of coupled neural oscillators of the type
(\ref{eq:oscillators}), with zero natural frequencies (since we wish to
use equilibrium techniques), in which real-valued patterns are stored as
stable configurations of oscillator
phases, following \cite{Cook}. Let us, however, first find out how to store a
single pattern $\bxi\in[\minus \pi,\pi]^N$ in a noise-less
infinite-range oscillator network.
For simplicity  we will draw each component $\xi_i$ independently
at random from $[\minus\pi,\pi]$, with uniform probability
density. This allows us to use asymptotic properties such as
$|N^{-1}\sum_j e^{i\ell\xi_j}|=\order(N^{-\frac{1}{2}})$ for any integer $\ell$.
A sensible choice for the synapses
 would be $J_{ij}=\cos[\xi_i\minus \xi_j]$. To see this we
work out the corresponding Lyapunov function
(\ref{eq:lyapunov_osc}):
\bd
L[\bphi]=- \frac{1}{2N^2}\sum_{ij}\cos[\xi_i\minus\xi_j]\cos[\phi_i\minus\phi_j]
~~~~~~~~~~
L[\bxi]=- \frac{1}{2N^2}\sum_{ij}\cos^2[\xi_i\minus\xi_j]
=- \frac{1}{4}+\order(N^{-\frac{1}{2}})
\ed
(the factors of $N$ have been inserted to achieve
appropriate scaling in the $N\to\infty$ limit).
The function $L[\bphi]$, which is obviously bounded from below, must decrease monotonically
during the dynamics. To find out whether the state $\bxi$
is a stable fixed-point of the dynamics we have to calculate $L$ and derivatives of $L$ at $\bphi=\bxi$:
\bd
\left.\frac{\partial L}{\partial\phi_i}\right|_{\bxi}\!
= \frac{1}{2N^2}\!\sum_{j}\sin[2(\xi_i\minus\xi_j)]
~~~~~~~~~~
\left.\frac{\partial^2 L}{\partial\phi_i^2}\right|_{\bxi}\!=
 \frac{1}{N^2}\!\sum_{j}\cos^2[\xi_i\minus\xi_j]
~~~~~~~~~~
i\!\neq\! j:~\left.\frac{\partial^2 L}{\partial\phi_i\partial\phi_j}\right|_{\bxi}\!=
- \frac{1}{N^2}\cos^2[\xi_i\minus\xi_j]
\ed
Clearly $\lim_{N\to\infty}L[\bxi]=-\frac{1}{4}$.
Putting $\bphi=\bxi+\Delta\bphi$, with $\Delta\phi_i=\order(N^0)$, we find
\bd
L[\bxi\plus\Delta\bphi]-L[\bxi]
=\sum_i\Delta
\phi_i \frac{\partial L}{\partial\phi_i}|_{\bxi}+\frac{1}{2}\sum_{ij}\Delta\phi_i\Delta\phi_j\frac{\partial^2
L}{\partial\phi_i\partial\phi_j}|_{\bxi}+\order(\Delta\bphi^3)
~~~~~~~~~~~~~~~~~~~~~~~~~~
\ed
\bd
=\frac{1}{4N}\sum_{i}\Delta\phi_i^2
-\frac{1}{2N^2}\sum_{ij}\Delta\phi_i\Delta\phi_j\cos^2[\xi_i\minus\xi_j]
+\order(N^{-\frac{1}{2}}\!,\Delta\bphi^3)
\ed
\be
=\frac{1}{4}\left\{
\frac{1}{N}\!\sum_{i}\Delta\phi_i^2
-[\frac{1}{N}\!\sum_{i}\Delta\phi_i]^2
-[\frac{1}{N}\!\sum_{i}\Delta\phi_i\cos(2\xi_i)]^2
-[\frac{1}{N}\!\sum_{i}\Delta\phi_i\sin(2\xi_i)]^2
\right\}
+\order(N^{-\frac{1}{2}}\!,\Delta\bphi^3)
\label{eq:simple_oscill_minimum}
\ee
In leading order in $N$ the following three vectors in $\Re^N$ are
normalised and orthogonal:
\bd
{\bf e}_1\!=\!\frac{1}{\sqrt{N}}(1,1,\ldots,1),
~~~~~~
{\bf e}_2\!=\!\frac{\sqrt{2}}{\sqrt{N}}(\cos(2\xi_1),\ldots,\cos(2\xi_N)),
~~~~~~
{\bf e}_2\!=\!\frac{\sqrt{2}}{\sqrt{N}}(\sin(2\xi_1),\ldots,\sin(2\xi_N))
\ed
We may therefore use
$\Delta\bphi^2\geq (\Delta\bphi\inn{\bf e}_1)^2\plus (\Delta\bphi\inn{\bf e}_2)^2\plus (\Delta\bphi\inn{\bf
e}_3)^2$, insertion of which into (\ref{eq:simple_oscill_minimum})
leads to
\bd
L[\bxi\plus\Delta\bphi]-L[\bxi] ~\geq~
[\frac{1}{2N}\!\sum_{i}\Delta\phi_i\cos(2\xi_i)]^2
+[\frac{1}{2N}\!\sum_{i}\Delta\phi_i\sin(2\xi_i)]^2
+\order(N^{-\frac{1}{2}}\!,\Delta\bphi^3)
\ed
Thus for large $N$ the second derivative of $L$ is non-negative at $\bphi=\bxi$, and the phase pattern $\bxi$
has indeed become a fixed-point
attractor of the dynamics of the noise-free coupled oscillator network. The same is
found to be true for the states $\bphi=\pm\bxi\plus\alpha(1,\ldots,1)$ (for any $\alpha$).
\vsp

\noindent{\em Storing $p$ Phase Patterns: Equilibrium Order
Parameter Equations.} We next follow the strategy of the Hopfield model
and attempt to simply extend the above recipe for the synapses to
the case of having a finite number $p$ of phase
patterns $\bxi^\mu=(\xi_1^\mu,\ldots,\xi_N^\mu)\in[-\pi,\pi]^N$, giving
\be
J_{ij}=\frac{1}{N}\sum_{\mu=1}^{p}\cos[\xi_i^\mu-\xi_j^\mu]
\label{eq:synapses_ann_osc}
\ee
(the factor $N$, as before, ensures a proper limit $N\to\infty$ later).
In analogy with our solution of the Hopfield model
we define the following averages over pattern variables:
\bd
\bra g[\bxi]\ket_{\bxi}=\lim_{N\to\infty}\sum_i
g[\bxi_i],~~~~~~~~~~\bxi_i=(\xi_i^1,\ldots,\xi_i^p)\in[-\pi,\pi]^p
\ed
We can write the Hamiltonian $H(\bphi)$ of (\ref{eq:oscill_thermodynamics1}) in the form
\bd
H(\bphi)=- \frac{1}{2N}\sum_{\mu=1}^p\sum_{ij}\cos[\xi_i^\mu\minus\xi^\mu_j]\cos[\phi_i\minus\phi_j]
=- \frac{N}{2}\sum_{\mu=1}^p\left\{
m_{cc}^\mu(\bphi)^2\plus m_{cs}^\mu(\bphi)^2\plus m_{sc}^\mu(\bphi)^2\plus m_{ss}^\mu(\bphi)^2\right\}
\ed
in which
\nsp
\be
m_{cc}^\mu(\bphi)=\frac{1}{N}\!\sum_i\cos(\xi_i^\mu)\cos(\phi_i)~~~~~~~~
m_{cs}^\mu(\bphi)=\frac{1}{N}\!\sum_i\cos(\xi_i^\mu)\sin(\phi_i)
\nsp
\label{eq:osc_observables1}
\ee
\be
m_{sc}^\mu(\bphi)=\frac{1}{N}\!\sum_i\sin(\xi_i^\mu)\cos(\phi_i)~~~~~~~~
m_{ss}^\mu(\bphi)=\frac{1}{N}\!\sum_i\sin(\xi_i^\mu)\sin(\phi_i)
\label{eq:osc_observables2}
\ee
The free energy per oscillator can now be written as
\bd
F/N
=-\frac{1}{\beta N}\log \int\!\cdots\!\int\!d\bphi ~ e^{-\beta H(\bphi)}
=-\frac{1}{\beta N}\log \int\!\cdots\!\int\!d\bphi ~
e^{\frac{1}{2}\beta N\sum_{\mu}\sum_{\star\star}m_{\star\star}^\mu(\bphi)^2}
\ed
with $\star\star\in\{cc,ss,cs,sc\}$.
Upon introducing the notation
$\bm_{\star\star}=(m_{\star\star}^1,\ldots,m_{\star\star}^p)$ we can
again express the  free energy in terms of the density of states
${\cal D}(\{\bm_{\star\star}\})=
(2\pi)^{-N}\int\!\cdots\!\int\!d\bphi ~\prod_{\star\star}\delta[\bm_{\star\star}\minus \bm_{\star\star}(\bsigma)]$:
\be
F/N
=
-\frac{1}{\beta}\log (2\pi)
-\frac{1}{\beta N}\log \int\!\prod_{\star\star}d\bm_{\star\star}~{\cal D}(\{\bm_{\star\star}\})
e^{\frac{1}{2}\beta N\sum_{\star\star}\bm_{\star\star}^2}
\label{eq:oschopfreeenergy}
\ee
Since $p$ is finite, the leading contribution to the density of states (as
$N\to\infty$), which will give us the entropy,
can be calculated by writing the $\delta$-functions
in integral representation:
\bd
\lim_{N\rightarrow\infty}\frac{1}{N}\log {\cal D}(\{\bm_{\star\star}\}) =
\lim_{N\rightarrow\infty} \frac{1}{N}\log \int\!\prod_{\star\star}
\left[d\bx_{\star\star}~
e^{iN\bx_{\star\star}\cdot\bm_{\star\star}}\right]~
\times
~~~~~~~~~~~~~~~~~~~~~~~~~~~~~~~~~~~~~~~~~~~~~~~~~~~~~~~~~~~~~~~~~~~~~~~~~~~~~~~~~~~~~~~~~~~~~~~~~~~~~~~~~~~~~~
\nsp
\ed
\bd
\int\!\ldots\!\int\!\frac{d\bphi}{(2\pi)^N}~
e^{-i\sum_{i}\sum_{\mu}[x_{cc}^\mu\cos(\xi_i^\mu)\cos(\phi_i)
+x_{cs}^\mu\cos(\xi_i^\mu)\sin(\phi_i)
+x_{sc}^\mu\sin(\xi_i^\mu)\cos(\phi_i)
+x_{ss}^\mu\sin(\xi_i^\mu)\sin(\phi_i)]}
\ed
\bd
={\rm extr}_{\{\bx_{\star\star}\}}\left\{
i\sum_{\star\star}\bx_{\star\star}\inn\bm_{\star\star}
+\bra\log\int\!\frac{d\phi}{2\pi}
e^{-i\sum_{\mu}[x_{cc}^\mu\cos(\xi_\mu)\cos(\phi)
+x_{cs}^\mu\cos(\xi_\mu)\sin(\phi)
+x_{sc}^\mu\sin(\xi_\mu)\cos(\phi)
+x_{ss}^\mu\sin(\xi_\mu)\sin(\phi)]}\ket_{\bxi}\right\}
\ed
The relevant extremum is purely imaginary so we put $\bx_{\star\star}=i\beta\by_{\star\star}$
(see also our previous discussion for the Hopfield model)
and, upon inserting the density of states into our original expression for the
free energy per oscillator, arrive at
\bd
\lim_{N\to\infty}F/N
= {\rm extr}_{\{\bm_{\star\star},\by_{\star\star}\}}~ f(\{\bm_{\star\star},\by_{\star\star}\})
\ed
\bd
f(\{\bm_{\star\star},\by_{\star\star}\})=
-\frac{1}{\beta}\log (2\pi)-\frac{1}{2}\sum_{\star\star}\bm_{\star\star}^2
+\sum_{\star\star}\by_{\star\star}\inn\bm_{\star\star}
~~~~~~~~~~~~~~~~~~~~~~~~~~~~~~~~~~~~~~~~
\ed
\bd
-\frac{1}{\beta}\bra\log\int\!\frac{d\phi}{2\pi}
e^{\beta\sum_{\mu}[y_{cc}^\mu\cos(\xi_\mu)\cos(\phi)
+y_{cs}^\mu\cos(\xi_\mu)\sin(\phi)
+y_{sc}^\mu\sin(\xi_\mu)\cos(\phi)
+y_{ss}^\mu\sin(\xi_\mu)\sin(\phi)]}\ket_{\bxi}
\ed
Taking derivatives with respect to the order parameters $\bm_{\star\star}$
gives us $\by_{\star\star}=\bm_{\star\star}$, with which we can
eliminate the $\by_{\star\star}$. Derivation with respect
to the $\bm_{\star\star}$ subsequently gives the saddle-point
equations
\be
m_{cc}^\mu
=\bra ~\cos[\xi_\mu]\frac{\int\!d\phi~
\cos[\phi]e^{\beta \cos[\phi]\sum_{\nu}[m_{cc}^\nu \cos[\xi_\nu]+ m_{sc}^\nu \sin[\xi_\nu]]
+\beta\sin[\phi] \sum_{\nu}[m_{cs}^\nu \cos[\xi_\nu]+ m_{ss}^\nu \sin[\xi_\nu]]
}}
{\int\!d\phi~e^{\beta \cos[\phi]\sum_{\nu}[m_{cc}^\nu \cos[\xi_\nu]+ m_{sc}^\nu \sin[\xi_\nu]]
+\beta\sin[\phi] \sum_{\nu}[m_{cs}^\nu \cos[\xi_\nu]+ m_{ss}^\nu \sin[\xi_\nu]]
}}~
\ket_{\bxi}
\label{eq:osc_saddle1}
\ee
\be
m_{cs}^\mu
=\bra ~\cos[\xi_\mu]\frac{\int\!d\phi~
\sin[\phi]e^{\beta \cos[\phi]\sum_{\nu}[m_{cc}^\nu \cos[\xi_\nu]+ m_{sc}^\nu \sin[\xi_\nu]]
+\beta\sin[\phi] \sum_{\nu}[m_{cs}^\nu \cos[\xi_\nu]+ m_{ss}^\nu \sin[\xi_\nu]]
}}
{\int\!d\phi~e^{\beta \cos[\phi]\sum_{\nu}[m_{cc}^\nu \cos[\xi_\nu]+ m_{sc}^\nu \sin[\xi_\nu]]
+\beta\sin[\phi] \sum_{\nu}[m_{cs}^\nu \cos[\xi_\nu]+ m_{ss}^\nu \sin[\xi_\nu]]
}}~
\ket_{\bxi}
\label{eq:osc_saddle2}
\ee
\be
m_{sc}^\mu
=\bra ~\sin[\xi_\mu]\frac{\int\!d\phi~
\cos[\phi]e^{\beta \cos[\phi]\sum_{\nu}[m_{cc}^\nu \cos[\xi_\nu]+ m_{sc}^\nu \sin[\xi_\nu]]
+\beta\sin[\phi] \sum_{\nu}[m_{cs}^\nu \cos[\xi_\nu]+ m_{ss}^\nu \sin[\xi_\nu]]
}}
{\int\!d\phi~e^{\beta \cos[\phi]\sum_{\nu}[m_{cc}^\nu \cos[\xi_\nu]+ m_{sc}^\nu \sin[\xi_\nu]]
+\beta\sin[\phi] \sum_{\nu}[m_{cs}^\nu \cos[\xi_\nu]+ m_{ss}^\nu \sin[\xi_\nu]]
}}~
\ket_{\bxi}
\label{eq:osc_saddle3}
\ee
\be
m_{ss}^\mu
=\bra ~\sin[\xi_\mu]\frac{\int\!d\phi~
\sin[\phi]e^{\beta \cos[\phi]\sum_{\nu}[m_{cc}^\nu \cos[\xi_\nu]+ m_{sc}^\nu \sin[\xi_\nu]]
+\beta\sin[\phi] \sum_{\nu}[m_{cs}^\nu \cos[\xi_\nu]+ m_{ss}^\nu \sin[\xi_\nu]]
}}
{\int\!d\phi~e^{\beta \cos[\phi]\sum_{\nu}[m_{cc}^\nu \cos[\xi_\nu]+ m_{sc}^\nu \sin[\xi_\nu]]
+\beta\sin[\phi] \sum_{\nu}[m_{cs}^\nu \cos[\xi_\nu]+ m_{ss}^\nu \sin[\xi_\nu]]
}}~
\ket_{\bxi}
\label{eq:osc_saddle4}
\ee
The equilibrium values of the observables $\bm_{\star\star}$, as defined in
(\ref{eq:osc_observables1},\ref{eq:osc_observables2}),
are now given by the solution of the
coupled equations (\ref{eq:osc_saddle1}-\ref{eq:osc_saddle4})
which minimises
\be
f(\{\bm_{\star\star}\})=
\frac{1}{2}\sum_{\star\star}\bm_{\star\star}^2
-\frac{1}{\beta}\bra\log
\int\!d\phi~e^{\beta \cos[\phi]\sum_{\nu}[m_{cc}^\nu \cos[\xi_\nu]+ m_{sc}^\nu \sin[\xi_\nu]]
+\beta\sin[\phi] \sum_{\nu}[m_{cs}^\nu \cos[\xi_\nu]+ m_{ss}^\nu
\sin[\xi_\nu]]}\ket_{\bxi}
\label{eq:osc_freeenergysurface}
\ee
We can confirm that the relevant saddle-point must be a minimum by
inspecting the $\beta=0$ limit (infinite noise levels): $\lim_{\beta\to 0}
f(\{\bm_{\star\star}\})=
\frac{1}{2}\sum_{\star\star}\bm_{\star\star}^2-\frac{1}{\beta}\log
(2\pi)$.
\vsp

\noindent
{\em Analysis of Order Parameter Equations: Pure States.} From now on we will
restrict our analysis to phase pattern
components $\xi_i^\mu$ which have all been drawn independently at
random from $[\minus \pi,\pi]$, with uniform probability density,
so that $\bra g[\bxi]\ket_{\bxi}=(2\pi)^{-p}\int_{-\pi}^\pi
\!\ldots\!\int_{-\pi}^\pi\!d\bxi~g[\bxi]$. At $\beta=0$ ($T=\infty$) one finds only the
trivial state $m^\mu_{\star\star}=0$. It can be shown that
there will be no discontinuous transitions to a non-trivial state
as the noise level (temperature) is reduced. The continuous ones
follow upon expansion of the equations
(\ref{eq:osc_saddle1}-\ref{eq:osc_saddle4}) for small
$\{\bm_{\star\star}\}$,
which is found to give (for each $\mu$ and each combination $\star\star$):
\bd
m_{\star\star}^\mu
=\frac{1}{4}\beta m_{\star\star}^\mu
+\order(\{\bm_{\star\star}^2\})
\ed
Thus a continuous transition to recall states occurs at $T=\frac{1}{4}$.
Full classification of all solutions of (\ref{eq:osc_saddle1}-\ref{eq:osc_saddle4})
is ruled out. Here we will restrict ourselves to the most relevant
ones, such as the pure states, where $m^\mu_{\star\star}=m_{\star\star}\delta_{\mu\lambda}$
(for some pattern label $\lambda$). Here the oscillator phases are correlated with only
one of the stored phase patterns (if at all). Insertion into the above
expression for $f(\{\bm_{\star\star}\})$  shows that for such
solutions we have to minimise
\be
f(\{m_{\star\star}\})=
\frac{1}{2}\sum_{\star\star}m_{\star\star}^2
-\frac{1}{\beta}\int\!\frac{d\xi}{2\pi}~\log
\int\!d\phi~
e^{\beta \cos[\phi][m_{cc}\cos[\xi]+ m_{sc} \sin[\xi]]
+\beta\sin[\phi][m_{cs} \cos[\xi]+ m_{ss}\sin[\xi]]}
\label{eq:osc_purefreeenergy}
\ee
We anticipate solutions
corresponding to the (partial) recall of the stored phase pattern $\bxi^\lambda$
or its mirror image (modulo overall phase shifts $\xi_i\to\xi_i\plus\delta$, under which the synapses are
obviously invariant). Insertion into (\ref{eq:osc_saddle1}-\ref{eq:osc_saddle4})
of the state $\phi_i=\xi^\lambda_i\plus\delta$ gives
$(m_{cc},m_{sc},m_{cs},m_{ss})=\frac{1}{2}(\cos\delta,\minus\sin\delta,\sin\delta,\cos\delta)$.
Similarly, insertion into (\ref{eq:osc_saddle1}-\ref{eq:osc_saddle4})
of $\phi_i=\minus\xi^\lambda_i\plus\delta$ gives
$(m_{cc},m_{sc},m_{cs},m_{ss})=\frac{1}{2}(\cos\delta,\sin\delta,\sin\delta,\minus\cos\delta)$.
Thus we can identify retrieval states as those solutions which are of
the form
\bd
\begin{array}{lllll}
(i)&& {\rm retrieval~of}~\bxi^\lambda:       &&
(m_{cc},m_{sc},m_{cs},m_{ss})=m(\cos\delta,\minus\sin\delta,\sin\delta,\cos\delta)\\[1mm]
(ii)&&{\rm retrieval~of}~\minus\bxi^\lambda: &&
(m_{cc},m_{sc},m_{cs},m_{ss})=m(\cos\delta,\sin\delta,\sin\delta,\minus\cos\delta)
\end{array}
\ed
with full recall corresponding to $m=\frac{1}{2}$.
Insertion into the saddle-point equations and into
(\ref{eq:osc_purefreeenergy}), followed by an appropriate shift of the integration variable $\phi$,
 shows that the free energy is independent of $\delta$
(so the above two ans\"{a}tze solve the saddle-point equations for
any $\delta$) and that
\bd
m=\frac{1}{2}\frac{\int\!d\phi~
\cos[\phi]e^{\beta m\cos[\phi]}}
{\int\!d\phi~e^{\beta
m\cos[\phi]}},~~~~~~~~~~
f(m)=m^2
-\frac{1}{\beta}\log
\int\!d\phi~e^{\beta m\cos[\phi]}
\ed
Expansion in powers of $m$, using $\log(1\plus z)=z-\frac{1}{2}z^2+\order(z^3)$,
reveals that non-zero minima $m$
indeed bifurcate continuously at $T=\beta^{-1}=\frac{1}{4}$:
\be
f(m)+\frac{1}{\beta}\log[2\pi]
=(1\minus \frac{1}{4}\beta)m^2
+\frac{1}{64}\beta^3 m^4+\order(m^6)
\label{eq:osc_puref}
\ee
\begin{figure}[t]
\begin{center}\vspace*{-10mm}
\epsfxsize=55mm\epsfbox{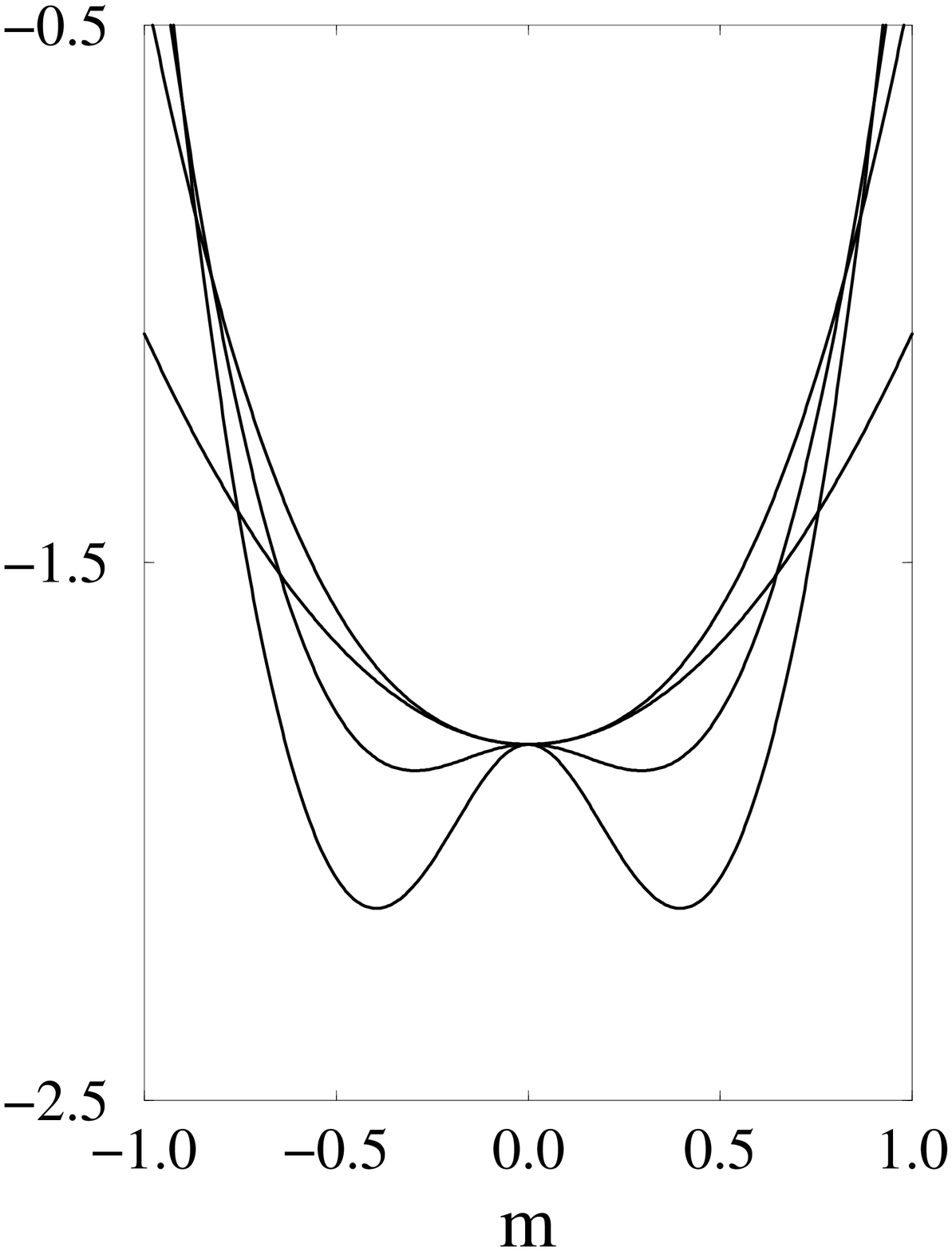}
\hspace*{3mm}
\epsfxsize=55mm\epsfbox{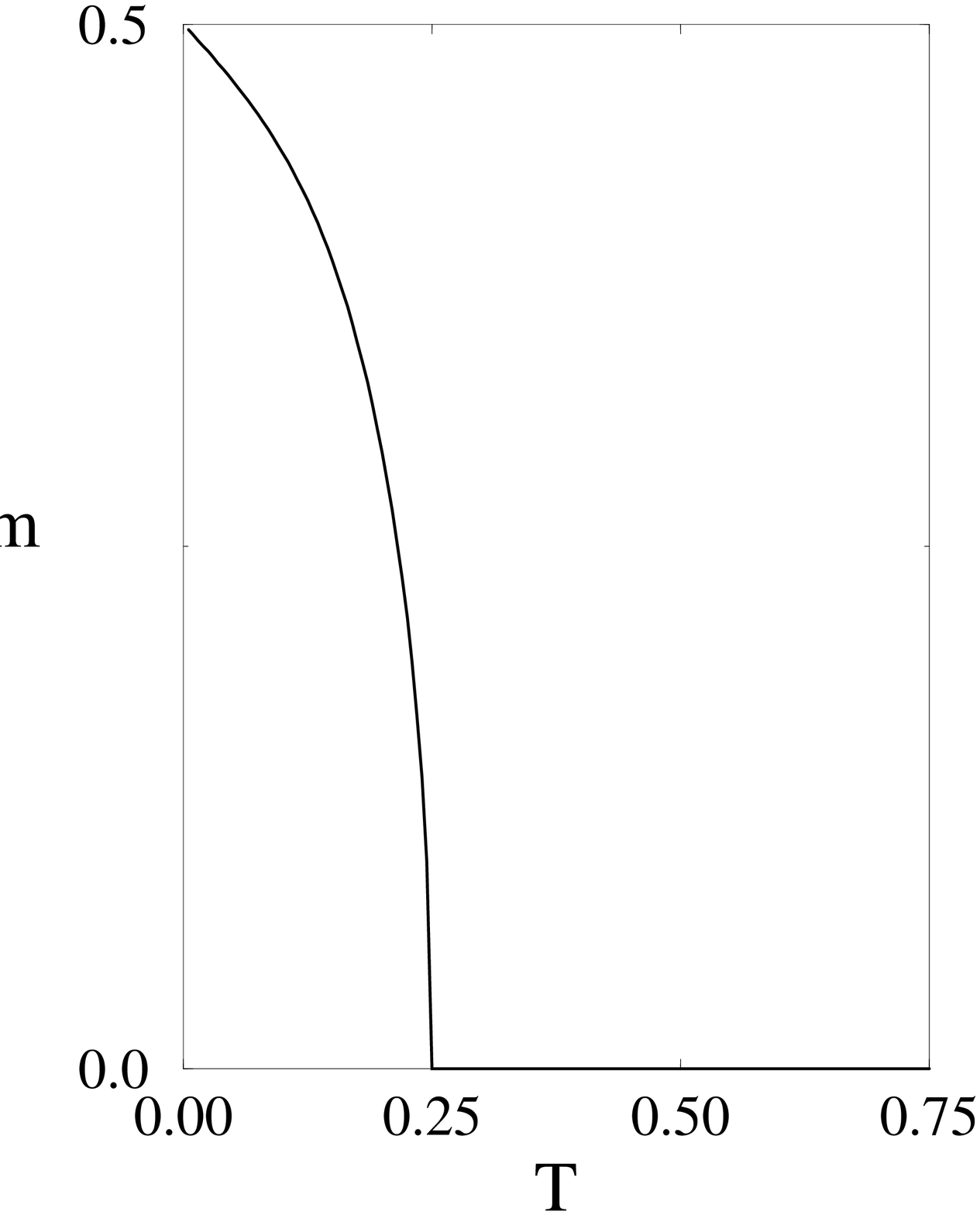}
\end{center}
\vspace*{-10mm}
\caption{The function $f(m)/T$ (left)
for networks of coupled oscillators with phase patterns stored via the synapses
$J_{ij}=N^{-1}\sum_{\mu}\cos[\xi_i^\mu\minus \xi_j^\mu]$, and for different choices of $\beta=T^{-1}$:
$\beta=1,3,5,7$ (from top to bottom).
The right picture gives the location of
the non-negative minimum of $f(m)$ (which measures the overall
degree of global synchronisation with one recalled phase pattern in thermal equilibrium) as a function of
$T$. A transition to a recall state occurs at $T=\frac{1}{4}$. }
\label{fig:osc_recallgraphs}
\end{figure}
Retrieval states are obviously not the only pure states that solve the saddle-point equations.
The function (\ref{eq:osc_purefreeenergy}) is invariant under the following
discrete (non-commuting) transformations:
\bd
\begin{array}{lllll}
{\rm I:}  && (m_{cc},m_{sc},m_{cs},m_{ss})& \to & (m_{cc},m_{sc},-m_{cs},-m_{ss})\\
{\rm II:} && (m_{cc},m_{sc},m_{cs},m_{ss})& \to & (m_{cs},m_{ss},m_{cc},m_{sc})
\end{array}
\ed
We expect these to induce solutions with specific symmetries. In
particular we anticipate the following symmetric and anti-symmetric states:
\bd
\begin{array}{lllll}
(iii)&& {\rm symmetric~under~I:}     && (m_{cc},m_{sc},m_{cs},m_{ss})=\sqrt{2}m(\cos\delta,\sin\delta,0,0)\\
(iv)&& {\rm antisymmetric~under~I:} && (m_{cc},m_{sc},m_{cs},m_{ss})=\sqrt{2}m(0,0,\cos\delta,\sin\delta)\\
(v)&&{\rm symmetric~under~II:}    && (m_{cc},m_{sc},m_{cs},m_{ss})
=m(\cos\delta,\sin\delta,\cos\delta,\sin\delta)\\
(vi)&&{\rm antisymmetric~under~II:}&& (m_{cc},m_{sc},m_{cs},m_{ss})
=m(\cos\delta,\sin\delta,-\cos\delta,-\sin\delta)
\end{array}
\ed
Insertion into the saddle-point equations and into
(\ref{eq:osc_purefreeenergy}) shows in all four cases the
parameter $\delta$ is arbitrary and that always
\bd
m
=\frac{1}{\sqrt{2}}\int\!\frac{d\xi}{2\pi} ~\cos[\xi]~\frac{\int\!d\phi~
\cos[\phi]e^{\beta m\sqrt{2}\cos[\phi]\cos[\xi]}}
{\int\!d\phi~e^{\beta m\sqrt{2}\cos[\phi]\cos[\xi]}},~~~~~~~~~~
f(m)=m^2-\frac{1}{\beta}\int\!\frac{d\xi}{2\pi}~\log
\int\!d\phi~e^{\beta m\sqrt{2}\cos[\phi]\cos[\xi]}
\ed
Expansion in powers of $m$ reveals that non-zero solutions $m$
here again bifurcate continuously at $T=\frac{1}{4}$:
\be
f(m)+\frac{1}{\beta}\log[2\pi]=(1\minus\frac{1}{4}\beta )m^2+\frac{3}{2}.\frac{1}{64}\beta^3m^4
+\order(m^6)
\label{eq:osc_otherf}
\ee
However, comparison with (\ref{eq:osc_puref}) shows that the free
energy of the pure recall states is lower. Thus the system will
prefer the recall states over the above solutions with specific
symmetries.

Note, finally, that the free energy and the order parameter equation for the
pure recall states can be written in terms of Bessel functions as
follows:
\bd
m=\frac{1}{2}\frac{I_1(\beta m)}{I_0(\beta m)},~~~~~~~~~~
f(m)=m^2
-\frac{1}{\beta}\log [2\pi I_0(\beta m)]
\ed
The behaviour of these equations and the observable $m$ for
different noise levels is shown in figure
\ref{fig:osc_recallgraphs}. One easily proves that $|m|\leq
\frac{1}{2}$, and that $\lim_{\beta\to\infty}m=\frac{1}{2}$.
Following the transition to a state with partial recall of a
stored phase pattern at $T=\frac{1}{4}$, further reduction of the
noise level $T$ gives a monotonic increase of retrieval quality
until retrieval is perfect at $T=0$.


\section{Networks with Gaussian Distributed Synapses}

The type of analysis presented so far to deal with attractor networks breaks
down if the number of patterns stored $p$ no longer remains finite for $N\rightarrow\infty$,
but scales as $p=\alpha N$
$(\alpha>0$).
Expressions such as (\ref{eq:hopfreeenergy},\ref{eq:parhopfreeenergy})
can no longer be evaluated by saddle-point methods, since the dimension
of the integral diverges at the same time as the exponent of the
integrand.
The number of local minima (ergodic components) of Hamiltonians such as
(\ref{eq:seq_thermodynamics1},\ref{eq:par_thermodynamics1})
 will diverge and we will encounter
phenomena reminiscent of complex disordered magnetic systems, i.e.  spin-glasses.
As a consequence we will need
corresponding methods of analysis, in the present case: replica
theory.

\subsection{Replica Analysis}

\noindent{\em Replica Calculation of the Disorder-Averaged Free
Energy}.
As an introduction to the replica technique we will
first discuss the equilibrium solution of a recurrent neural network model
with binary neurons $\sigma_i\in\{-1,1\}$ in which a single pattern $\bxi=(\xi_1,\ldots,\xi_N)\in\{-1,1\}^N$
has been stored
(via a Hebbian-type recipe) on a background of zero-average Gaussian synapses
(equivalent to the SK model, \cite{SK}):
\be
J_{ij}=\frac{J_0}{N}\xi_i\xi_j+\frac{J}{\sqrt{N}}z_{ij}
,~~~~~~~~\overline{z}_{ij}=0,~\overline{z^2}_{ij}=1
\label{eq:SKinteractions}
\ee
in which $J_0>0$ measures the embedding strength of the pattern, and the $z_{ij}$ ($i<j$) are
independent Gaussian random variables. We denote averaging over their distribution by
$\overline{\cdots}$ (the factors in (\ref{eq:SKinteractions}) involving $N$
ensure appropriate scaling and statistical
relevance of the two terms, and as always $J_{ii}=0$).
Here the Hamiltonian $H$ (\ref{eq:seq_thermodynamics1}),
corresponding to sequential dynamics
(\ref{eq:Ising_sequential}), becomes
\be
H(\bsigma)=
-\frac{1}{2}N J_0 m^2(\bsigma)+\frac{1}{2}J_0
-\frac{J}{\sqrt{N}}\sum_{i<j}\sigma_i\sigma_j z_{ij}
\label{eq:SKhamiltonian}
\ee
with the overlap $m(\bsigma)=\frac{1}{N}\sum_k\sigma_k\xi_k$ which measures pattern recall
quality.
We clearly cannot calculate the free energy for every given realization
of the synapses, furthermore it is to be expected
that for $N\to\infty$ macroscopic observables like the free energy per neuron and the overlap $m$
only depend on the statistics
of the synapses, not on their specific values.
We therefore average the free energy over the disorder distribution and
concentrate on
\be
\overline{F}=
-\frac{1}{\beta}\lim_{N\to\infty}\overline{\log Z},~~~~~~~~
Z=\sum_{\bsigma}e^{-\beta H(\bsigma)}
\label{eq:averagef}
\ee
The disorder average is transformed into an average of powers
of $Z$, with the identity
\be
\overline{\log Z}=\lim_{n\rightarrow0}\frac{1}{n}\left[\overline{Z^{n}}\minus 1\right]
~~~~~~~~~~
{\rm or,~ equivalently,}
~~~~~~~~~~
\overline{\log Z}=\lim_{n\rightarrow0}\frac{1}{n}\log
\overline{Z^{n}}
\label{eq:replicatrick}
\ee
The so-called `replica trick' consists in evaluating the averages
$\overline{Z^n}$ for integer values of $n$, and taking the limit
$n\rightarrow0$ afterwards, under the assumption that the resulting
expression is correct for non-integer values of $n$ as well. The
integer powers of $Z$ are written as a product of terms, each of which
can be interpreted as an equivalent copy, or `replica' of the
original system. The disorder-averaged free energy now becomes
\bd
\overline{F} = -\lim_{n\rightarrow0} \frac{1}{\beta n}\log\overline{Z^n}
= -\lim_{n\rightarrow0}\frac{1}{\beta n}\log\sum_{\bsigma^1\!\ldots\bsigma^n}\overline{e^{-\beta\sum_{\alpha=1}^n
H(\bsigma^{\alpha})}}
\ed
From now Roman indices will refer to {\rm sites}, i.e. $i=1\ldots N$, whereas
Greek indices will refer to replicas, i.e. $\alpha=1\ldots n$.
We introduce a  short-hand for the  Gaussian measure,
$Dz=(2\pi)^{-\frac{1}{2}}e^{-\frac{1}{2}z^2}dz$, and we will
repeatedly use the identity
$\int\! Dz~e^{xz}=e^{\frac{1}{2}x^2}$.
Upon insertion of the Hamiltonian (\ref{eq:SKhamiltonian}) we obtain
\bd
\begin{array}{ll}
\overline{F}\!\! & = -\frac{1}{\beta}N\log2  -\lim_{n\rightarrow0}
(\beta n)^{-1}
\log~\bra e^{\frac{\beta J_0}{N}\sum_{i<j}\xi_i\xi_j\sum_{\alpha}
\sigma_i^{\alpha}\sigma_j^{\alpha}}\prod_{i<j}\left[\int\!Dz~e^{\frac{\beta
J z}{\sqrt{N}}
\sum_{\alpha}
\sigma_i^{\alpha}\sigma_j^{\alpha}}\right]\ket_{\{\bsigma^{\alpha}\}}\room \\ \\
& = -\frac{1}{\beta}N\log2 - \lim_{n\rightarrow0}
(\beta n)^{-1}\log~
\bra e^{\frac{\beta J_0}{2N}\sum_{\alpha}\sum_{i\neq j}\xi_i\xi_j
\sigma_i^{\alpha}\sigma_j^{\alpha}
+\frac{\beta^2 J^2}{4N}\sum_{\alpha\gamma}\sum_{i\neq j}
\sigma_i^{\alpha}\sigma_j^{\alpha}\sigma_i^{\gamma}\sigma_j^{\gamma}}
\ket_{\{\bsigma^{\alpha}\}}\room
\end{array}
\ed
We now complete the sums over sites  in this expression,
\bd
\sum_{i\neq j}\sigma_i^{\alpha}
\sigma_j^{\alpha}=
[\sum_i\sigma_{i}^{\alpha}]^2\minus N,~~~~~~~~~~
\sum_{i\neq j}\sigma_i^{\alpha}\sigma_j^{\alpha}\sigma_i^{\gamma}\sigma_j^{\gamma}
=[\sum_i\sigma_{i}^{\alpha}\sigma_i^{\gamma}]^2\minus N
\ed
The averaging over the neuron states $\{\bsigma^{\alpha}\}$ in our expression for $\overline{F}$ will now factorize
nicely if we insert appropriate
$\delta$-functions (in their integral representations) to isolate the relevant terms, using
\bd
1=\int\!d\bq~\prod_{\alpha\beta}\delta\left[q_{\alpha\beta}\minus
\frac{1}{N}\sum_i\sigma_i^{\alpha}\sigma_i^{\beta}\right]
=\left[\frac{N}{2\pi}\right]^{n^2}\int\!d\bq
d\hat{\bq}~e^{iN\sum_{\alpha\beta}\hat{q}_{\alpha\beta}\left[q_{\alpha\beta}-\frac{1}{N}\sum_i\sigma_i^{\alpha}\sigma_i^{\beta}\right]}
\ed
\bd
1= \int\!d\bm~\prod_{\alpha}\delta\left[m_{\alpha}\minus
\frac{1}{N}\sum_i\xi_i\sigma_i^{\alpha}\right]
=\left[\frac{N}{2\pi}\right]^{n}\int\!d\bm d\hat{\bm}~e^{iN\sum_{\alpha}\hat{m}_{\alpha}
\left[m_{\alpha}-\frac{1}{N}\sum_i\xi_i\sigma_i^{\alpha}\right]}
\ed
The integrations are over the $n\times n$ matrices $\bq$ and
$\hat{\bq}$ and over the  $n$-vectors $\bm$ and $\hat{\bm}$. After
inserting these integrals we obtain
\bd
\lim_{N\rightarrow\infty}\overline{F}/N = -\frac{1}{\beta}\log2 - \lim_{N\rightarrow\infty}\lim_{n\rightarrow0}
\frac{1}{\beta N n}\log\left\{
\left[\frac{N}{2\pi}\right]^{n^2+n}
\!\int\! d\bq d\hat{\bq} d\bm
d\hat{\bm}
~e^{-\frac{1}{2}n\beta
J_0-\frac{1}{4}n^2\beta^2 J^2}
~~~~~~~~~~~~~~~~~~~~~~~~~~~~~~~~~~~~~~~~~~~~~~~~~~~~~~
\right.
\ed
\bd
\left.
~~~~~~~~\times~
e^{N\left[i\sum_{\alpha\gamma}\hat{q}_{\alpha\gamma}q_{\alpha\gamma}
+i\sum_{\alpha}\hat{m}_{\alpha}m_{\alpha}
+\frac{1}{2}\beta J_0 \sum_{\alpha}m_{\alpha}^2
+\frac{1}{4}\beta^2 J^2
\sum_{\alpha\gamma}q_{\alpha\gamma}^2\right]}
\bra
e^{-i\sum_i\left[\sum_{\alpha\gamma}\hat{q}_{\alpha\gamma}\sigma_i^{\alpha}\sigma_i^{\gamma}
+\sum_{\alpha}\hat{m}_{\alpha}\xi_i\sigma_i^{\alpha}\right]}
\ket_{\{\bsigma^{\alpha}\}}\right\}
\ed
The neuronal averages factorise and are therefore reduced to single-site ones. A simple transformation
$\sigma_i\to\xi_i\sigma_i$ for all $i$ eliminates the pattern components $\xi_i$
from our equations, and the remaining averages involve only
one $n$-replicated neuron $(\sigma_1,\ldots,\sigma_n)$.
Finally one assumes that the two limits $n\rightarrow0$ and
$N\rightarrow\infty$ commute. This allows us to evaluate the integral
with the steepest-descent method:
\be
\lim_{N\rightarrow\infty}\lim_{n\rightarrow
0}\frac{1}{Nn}\log \int\!d\bx~e^{N\Phi(\bx)} =
\lim_{n\rightarrow 0 }\lim_{N\rightarrow \infty}\frac{1}{Nn}\log e^{N\extr\Phi+\ldots}\room
= \lim_{n\rightarrow 0}\frac{1}{n}\extr\Phi \room
\label{eq:reverse}
\ee
The result of these manipulations is\room
\be
\lim_{N\rightarrow\infty}\overline{F}/N
 =\lim_{n\rightarrow0} \extr~f(\bq,\bm;\hat{\bq},\hat{\bm})\room
\label{eq:fullSKsaddle}
\ee
\bd
f(\bq,\bm;\hat{\bq},\hat{\bm})=
-\frac{1}{\beta}\log2
-\frac{1}{\beta n}\left[\log
\bra
e^{-i\sum_{\alpha\gamma}\hat{q}_{\alpha\gamma}\sigma_{\alpha}\sigma_{\gamma}-i\sum_{\alpha}\hat{m}_{\alpha}\sigma_{\alpha}}
\ket_{\bsigma}
~~~~~~~~~~~~~~~~~~~~~~~~~~~~~~~~
\right.
\ed
\be
\left.
~~~~~~~~~~~~~~~~~~~~~~~~~~~~~~~~
+i\sum_{\alpha\gamma}\hat{q}_{\alpha\gamma}q_{\alpha\gamma}
+i\sum_{\alpha}\hat{m}_{\alpha}m_{\alpha}
+\frac{1}{2}\beta J_0 \sum_{\alpha}m_{\alpha}^2
+\frac{1}{4}\beta^2 J^2
\sum_{\alpha\gamma}q_{\alpha\gamma}^2
\right]
\label{eq:fullSKsurface}
\ee
Variation of the parameters $\{q_{\alpha\beta}\}$ and $\{m_{\alpha}\}$ allows
us to eliminate immediately the conjugate parameters
$\{\hat{q}_{\alpha\beta}\}$ and $\{\hat{m}_{\alpha}\}$, since it leads
to the
saddle-point requirements
\be
\hat{q}_{\alpha\beta} =\frac{1}{2}i\beta^2 J^2 q_{\alpha\beta}
~~~~~~~~~~~~~~
\hat{m}_{\alpha} =i\beta J_0 m_{\alpha}
\label{eq:conjugates}
\ee
Upon elimination of $\{\hat{q}_{\alpha\beta},\hat{m}_{\alpha}\}$ according to
(\ref{eq:conjugates}) the result
(\ref{eq:fullSKsaddle},\ref{eq:fullSKsurface})
is simplified to
\be
\lim_{N\rightarrow\infty}\overline{F}/N =
\lim_{n\rightarrow0} \extr~f(\bq,\bm)
\label{eq:SKsaddle}
\ee
\be
f(\bq,\bm)=
-\frac{1}{\beta}\log2
+\frac{\beta J^2}{4n} \sum_{\alpha\gamma} q^2_{\alpha\gamma}
+\frac{J_0}{2n}\sum_{\alpha} m^2_{\alpha}
-\frac{1}{\beta n}\log
\bra  e^{\frac{1}{2}\beta^2 J^2
\sum_{\alpha\gamma}q_{\alpha\gamma}\sigma_{\alpha}\sigma_{\gamma}
+\beta J_0 \sum_{\alpha} m_{\alpha}\sigma_{\alpha}}
\ket_{\bsigma}
\label{eq:SKsurface}
\ee
Variation of the remaining parameters $\{q_{\alpha\beta}\}$ and
$\{m_{\alpha}\}$ gives the final saddle-point equations
\be
q_{\lambda\rho}=\frac{\bra \sigma_{\lambda}\sigma_{\rho} e^{\frac{1}{2}\beta^2 J^2
\sum_{\alpha\gamma}q_{\alpha\gamma}\sigma_{\alpha}\sigma_{\gamma}
+\beta J_0 \sum_{\alpha} m_{\alpha}\sigma_{\alpha}}
\ket_{\bsigma}}
{\bra  e^{\frac{1}{2}\beta^2 J^2
\sum_{\alpha\gamma}q_{\alpha\gamma}\sigma_{\alpha}\sigma_{\gamma}
+\beta J_0 \sum_{\alpha} m_{\alpha}\sigma_{\alpha}}
\ket_{\bsigma}}
\label{eq:eqn_qab}
\ee
\be
m_{\lambda}=\frac{\bra \sigma_{\lambda}e^{\frac{1}{2}\beta^2 J^2
\sum_{\alpha\gamma}q_{\alpha\gamma}\sigma_{\alpha}\sigma_{\gamma}
+\beta J_0 \sum_{\alpha} m_{\alpha}\sigma_{\alpha}}
\ket_{\bsigma}}
{\bra  e^{\frac{1}{2}\beta^2 J^2
\sum_{\alpha\gamma}q_{\alpha\gamma}\sigma_{\alpha}\sigma_{\gamma}
+\beta J_0 \sum_{\alpha} m_{\alpha}\sigma_{\alpha}}
\ket_{\bsigma}}
\label{eq:eqn_ma}
\ee
The diagonal elements are always $q_{\alpha\alpha}=1$.
For high noise levels, $\beta\rightarrow 0$, we obtain the
trivial result
\bd
q_{\alpha\gamma}=\delta_{\alpha\gamma},~~~~~~m_{\alpha}=0
\ed
Assuming a continuous transition to a non-trivial state as the
noise level is lowered, we can expand
the saddle-point equations (\ref{eq:eqn_qab},\ref{eq:eqn_ma}) in powers
of $\bq$ and $\bm$ and look for bifurcations, which gives ($\lambda\neq\rho$):
\bd
q_{\lambda\rho}=\beta^2 J^2
q_{\lambda\rho}+\order(\bq,\bm)^2
~~~~~~~~~~~~
m_{\lambda}=\beta J_0 m_{\lambda} +\order(\bq,\bm)^2\room
\ed
Therefore we expect transitions either at $T=J_0$ (if
$J_0>J$) or at $T=J$ (if $J>J_0$).
The remaining program is: find the saddle-point $(\bq,\bm)$ for $T<\max\{J_0,J\}$ which for integer $n$
minimises $f$, determine the corresponding minimum as a function of
$n$, and finally take the limit $n\rightarrow0$. This is in fact the
most complicated part of the procedure.

\subsection{Replica-Symmetric Solution and AT-Instability}

\noindent{\em Physical Interpretation of Saddle Points.}
To obtain a guide in how to select saddle-points we now turn to
a different (but equivalent) version of the replica trick
(\ref{eq:replicatrick}), which allows us to attach a physical meaning
to the saddle-points $(\bm,\bq)$. This version transforms
averages over a given measure $W$:
\bd
\frac{\sum_{\bsigma}\Phi(\bsigma)W(\bsigma)}{\sum_{\bsigma}W(\bsigma)}
=\lim_{n\rightarrow0}\sum_{\bsigma}\Phi(\bsigma)W(\bsigma)[\sum_{\bsigma}W(\bsigma)]^{n-1}
=\lim_{n\rightarrow0}\sum_{\bsigma^1\!\ldots\bsigma^n}\Phi(\bsigma^1)\prod_{\alpha=1}^n
W(\bsigma^{\alpha})
\vspace*{-2mm}
\ed
\be
=\lim_{n\rightarrow0}\frac{1}{n}\sum_{\gamma=1}^n\sum_{\bsigma^1\!\ldots\bsigma^n}\Phi(\bsigma^\gamma)\prod_{\alpha=1}^n
W(\bsigma^{\alpha})
\label{eq:secondreplicatrick}
\ee
The trick again consists in evaluating this quantity
for {\em integer} $n$, whereas the limit refers to non-integer $n$.
We use (\ref{eq:secondreplicatrick}) to write the distribution $P(m)$ of overlaps
 in equilibrium as
\bd
P(m)=
\frac{\sum_{\bsigma}\delta[m\minus\frac{1}{N}\sum_{i}\xi_i\sigma_i]e^{-\beta H(\bsigma)}}
{\sum_{\bsigma}e^{-\beta H(\bsigma)}}
=\lim_{n\rightarrow0}\frac{1}{n}\sum_{\gamma}\sum_{\bsigma^1\!\ldots\bsigma^n}
\delta[m\minus\frac{1}{N}\sum_{i}\xi_i\sigma_i^{\gamma}]
\prod_{\alpha}
e^{-\beta H(\bsigma^{\alpha})}
\ed
If we average this distribution over the disorder, we find identical
expressions to those encountered in evaluating the disorder averaged
free energy. By inserting the same delta-functions we arrive at the
steepest descend integration (\ref{eq:fullSKsaddle}) and find
\be
\overline{P(m)}=\lim_{n\rightarrow0}\frac{1}{n}\sum_{\gamma}
\delta\left[m\minus m_{\gamma}\right]
\label{eq:magdistribution}
\ee
where $\{m_{\gamma}\}$ refers to  the relevant solution
of (\ref{eq:eqn_qab},\ref{eq:eqn_ma}).
Similarly we can imagine {\em two} systems $\bsigma$ and
$\bsigma^{\prime}$ with
identical  synapses $\{J_{ij}\}$, both in
thermal equilibrium.
We now use (\ref{eq:secondreplicatrick}) to rewrite the distribution $P(q)$ for
the mutual overlap between the microstates
of the two systems
\bd
P(q)=
\frac{\sum_{\bsigma,\bsigma^{\prime}}\delta[q\minus\frac{1}{N}\sum_{i}\sigma_i\sigma_i^{\prime}]e^{-\beta
H(\bsigma)-\beta H(\bsigma^{\prime})}}
{\sum_{\bsigma,\bsigma^{\prime}}e^{-\beta H(\bsigma)-\beta H(\bsigma^{\prime})}}
~~~~~~~~~~~~~~~~~~~~~~~~~~~~~~~~~~~~~~~~~~~~~~
\vspace*{-2mm}
\ed
\bd
~~~~~~~~~~~~~~~~~~~~~~~~~~~~~~~~~~~~~~~~~~~~~~
=\lim_{n\rightarrow0}\frac{1}{n(n\minus 1)}\sum_{\lambda\neq
\gamma}\sum_{\bsigma^1\!\ldots\bsigma^n}
\delta[q\minus \frac{1}{N}\sum_{i}\sigma_i^{\lambda}\sigma_i^{\gamma}]\prod_{\alpha}
e^{-\beta H(\bsigma^{\alpha})}
\ed
Averaging over the disorder again leads to the steepest descend
integration (\ref{eq:fullSKsaddle}) and we find
\be
\overline{P(q)}=\lim_{n\rightarrow0}\frac{1}{n(n\minus1)}\sum_{\lambda\neq \gamma}
\delta\left[q-q_{\lambda\gamma}\right]
\label{eq:overlapdistribution}
\ee
where $\{q_{\lambda\gamma}\}$ refers to  the relevant solution
of (\ref{eq:eqn_qab},\ref{eq:eqn_ma}).
We can now partly interpret the saddle-points $(\bm,\bq)$, since
the shape of $\overline{P(q)}$
and $\overline{P(m)}$ gives direct information on the structure of
phase space with respect to ergodicity. The crucial observation is
that for an ergodic system one always has
\be
P(m)=\delta[m\minus\frac{1}{N}\sum_i\xi_i\bra\sigma_i\ket_{\rm eq}]~~~~~~~~
P(q)=\delta[q\minus\frac{1}{N}\sum_i\bra \sigma_i\ket^2_{\rm eq}]
\label{eq:ergodicsystems}
\ee
If, on the other hand, there are $L$ ergodic
components in our system, each of which corresponding to a pure Gibbs
state with microstate probabilities proportional to $\exp(-\beta
H)$ and thermal averages $\bra\ldots\ket_{\ell}$, and  if we denote the
probability of finding the system in component $\ell$ by
$W_{\ell}$, we find
\bd
P(m)=\sum_{\ell=1}^L W_{\ell}~\delta[m\minus
\frac{1}{N}\sum_i\xi_i\bra\sigma_i\ket_{\ell}]
~~~~~~~~~~~~~~~
P(q)=\sum_{\ell,\ell^{\prime}=1}^L W_{\ell}W_{\ell^{\prime}}~\delta[q\minus
\frac{1}{N}\sum_i\bra \sigma_i\ket_{\ell}\bra\sigma_i\ket_{\ell^{\prime}}]
\ed
For ergodic systems both $P(m)$ and $P(q)$ are
$\delta$-functions, for systems with a finite number of ergodic
components they are  finite sums of $\delta$-functions. A {\em
diverging}
number of ergodic components generally leads to
distributions with continuous pieces.
If we combine this interpretation with our results
(\ref{eq:magdistribution},\ref{eq:overlapdistribution}) we find that
ergodicity is equivalent to the relevant saddle-point being
of the form:
\be
q_{\alpha\beta}=\delta_{\alpha\beta}+q\left[1-\delta_{\alpha\beta}\right]
~~~~~~~~~~~~
m_{\alpha}=m
\label{eq:RSansatz}
\ee
which is called the `replica symmetry' (RS) ansatz. The
meaning of $m$ and $q$ is deduced from (\ref{eq:ergodicsystems})
(taking into account the transformation $\sigma_i\to\xi_i\sigma_i$
we performed along the way):
\bd
m=\frac{1}{N}\sum_i\overline{\xi_i \bra\sigma_i\ket_{\rm eq}}~~~~~~~~
q=\frac{1}{N}\sum_i\overline{\bra \sigma_i\ket^2_{\rm eq}}
\ed
\vsp

\noindent{\em Replica Symmetric Solution}. Having saddle-points of the simple form
(\ref{eq:RSansatz}) leads to an enormous simplification in our calculations.
Insertion of (\ref{eq:RSansatz}) as an ansatz into the equations
(\ref{eq:SKsurface},\ref{eq:eqn_qab},\ref{eq:eqn_ma}) gives
\bd
f(\bq,\bm)=
-\frac{1}{\beta}\log2
-\frac{1}{4}\beta J^2 (1\minus q)^2
+\frac{1}{2}J_0 m^2
-\frac{1}{\beta n}\log~
\bra
 e^{\frac{1}{2}q\beta^2 J^2 \left[\sum_{\alpha}\sigma_{\alpha}\right]^2
+\beta J_0 m \sum_{\alpha}\sigma_{\alpha}}
\ket_{\bsigma}+\order(n)
\ed
\bd
q=\frac{\bra \sigma_1\sigma_2 e^{\frac{1}{2}q\beta^2 J^2
\left[\sum_{\alpha}\sigma_{\alpha}\right]^2
+\beta J_0 m \sum_{\alpha}\sigma_{\alpha}}
\ket_{\bsigma}}
{\bra  e^{\frac{1}{2}q\beta^2 J^2
\left[\sum_{\alpha}\sigma_{\alpha}\right]^2
+\beta J_0 m\sum_{\alpha} \sigma_{\alpha}}
\ket_{\bsigma}}
~~~~~~~~~~~~~
m=\frac{\bra \sigma_1 e^{\frac{1}{2}q\beta^2 J^2
\left[\sum_{\alpha}\sigma_{\alpha}\right]^2
+\beta J_0 m \sum_{\alpha}\sigma_{\alpha}}
\ket_{\bsigma}}
{\bra  e^{\frac{1}{2}q\beta^2 J^2
\left[\sum_{\alpha}\sigma_{\alpha}\right]^2
+\beta J_0 m\sum_{\alpha} \sigma_{\alpha}}
\ket_{\bsigma}}
\ed
We linearise the terms
$\left[\sum_{\alpha}\sigma_{\alpha}\right]^2$ by introducing a Gaussian integral,
 and perform the average over the remaining
neurons. The solutions $m$ and $q$ turn out to be well defined for
$n\rightarrow0$ so we can take the limit:
\be
\lim_{n\rightarrow0} f(\bq,\bm)=
-\frac{1}{\beta}\log2
-\frac{1}{4}\beta J^2 (1\minus q)^2
+\frac{1}{2}J_0 m^2
-\frac{1}{\beta}\int\!Dz~\log \cosh\left[\beta J_0 m\plus\beta J z\sqrt{q}\right]
\label{eq:SKRSfreeenergy}
\ee
\be
q=\int\!Dz~\tanh^2\left[\beta J_0 m\plus\beta J z\sqrt{q}\right]
~~~~~~~~~~~~
m=\int\!Dz~\tanh\left[\beta J_0 m\plus\beta J z\sqrt{q}\right]
\label{eq:SKRSqm}
\ee
Writing the equation for $m$ in integral form gives
\bd
m=\beta J_0 m
\int_0^1\!d\lambda~\left[1\minus\int\!Dz~\tanh^2\left[\lambda\beta J_0
m\plus\beta J z\sqrt{q}\right]\right]
\ed
From this expression, in combination with (\ref{eq:SKRSqm}), we conclude:
\bd
T>J_0:~~ m=0~~~~~~~~~~~~
T>J_0~{\rm and}~ T>J: ~~ m=q=0
\ed
Linearisation of (\ref{eq:SKRSqm}) for small $q$ and $m$
shows the following continuous bifurcations:\vspace*{-2mm}
\bd
\begin{array}{llll}
                 & {\rm at}   & {\rm from} & {\rm to}     \room \\
J_0>J:           & T=J_0      & m=q=0      & m\neq0,~q>0   \\
J_0<J:           & T=J        & m=q=0      & m=0,~q>0      \\
T<\max\{J_0,J\}: & T=J_0[1-q] & m=0,~q>0   & m\neq 0,~q>0
\end{array}
\ed
\begin{figure}[t]
\begin{center}\vspace*{-5mm}
\epsfxsize=60mm\epsfbox{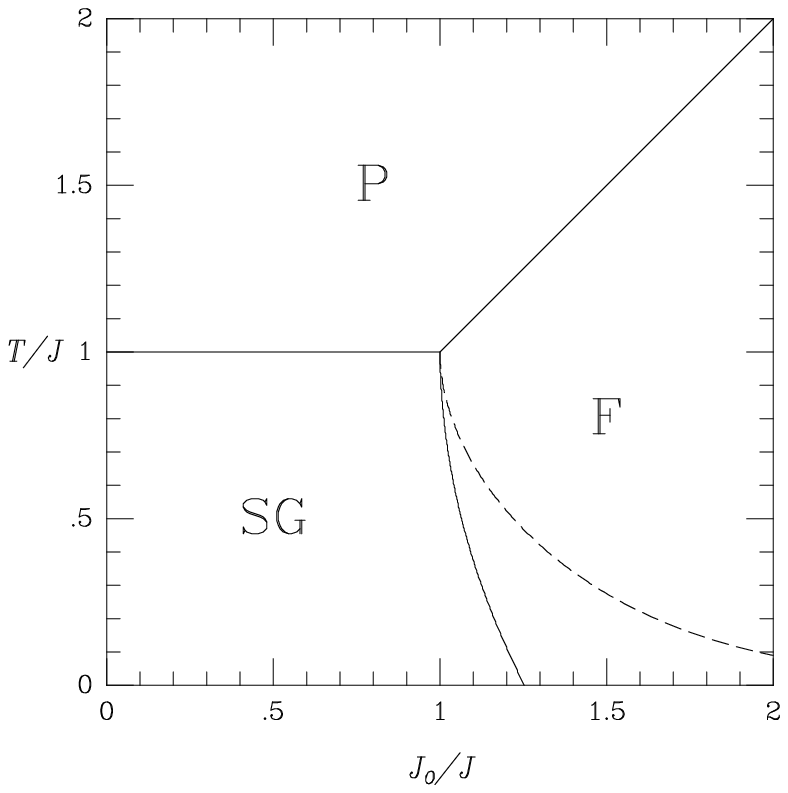}
\end{center}
\vspace*{-5mm}
\caption{Phase diagram of the model (\ref{eq:SKinteractions}) with Gaussian synapses, obtained from the
replica-symmetric solution. P: paramagnetic phase,
$m=q=0$ (more or less random evolution). SG: spin-glass phase, $m=0$, $q\neq 0$ (`frozen' equilibrium states without
pattern recall). F: recall (`ferro-magnetic')
phase, $m\neq 0$, $q\neq 0$. Solid lines: second-order transitions. Dashed: the AT instability.}
\label{fig:SKphasediagram}
\end{figure}
Solving numerically equations $T=J_0[1\minus q]$ and
(\ref{eq:SKRSqm}) leads to the phase diagram
shown in figure \ref{fig:SKphasediagram}.
\vsp

\noindent{\em Breaking of Replica Symmetry: the AT Instability.} If
for the replica symmetric solution we calculate the entropy
$S=\beta^2\partial F/\partial\beta$ numerically, we
find that for small temperatures it becomes negative. This is not
possible. Firstly, straightforward
differentiation shows
$\partial S/\partial\beta
=\beta[\bra H\ket_{\rm eq}^2\minus \bra H^2\ket_{\rm eq}]\leq 0$,
so $S$ increases with the noise level $T$.
Let us now
write $H(\bsigma)=H_0+\hat{H}(\bsigma)$, where $H_0$ is the ground
state energy and  $\hat{H}(\bsigma)\geq 0$ (zero only
for ground state configurations, the number of which we denote by
$N_0\geq 1$). We now find
\bd
\lim_{T\to 0}S
=\lim_{\beta\to\infty}\left\{\log\sum_{\bsigma}e^{-\beta
H(\bsigma)}+\beta\bra H\ket_{\rm eq}\right\}=
\lim_{\beta\rightarrow\infty}[\log \sum_{\bsigma}e^{-\beta
\hat{H}(\bsigma)}+\beta \bra \hat{H}\ket_{\rm eq}]\geq  \log N_0
\ed
We conclude that $S\geq 0$ for all $T$.
At small temperatures the RS ansatz (\ref{eq:RSansatz}) is apparently
incorrect in that it no longer corresponds to the minimum of
$f(\bq,\bm)$ (\ref{eq:SKsurface}).
If saddle-points without replica symmetry bifurcate continuously from
the RS one, we can locate the occurrence of this `replica symmetry
breaking' (RSB) by studying the effect on $f(\bq,\bm)$ of small fluctuations
around the RS solution. It was shown \cite{AT}
that the `dangerous' fluctuations are of the form
\be
q_{\alpha\beta}\rightarrow\delta_{\alpha\beta}+q\left[1-\delta_{\alpha\beta}\right]+\eta_{\alpha\beta},~~~~~~~~~~~~~~
\sum_{\beta}\eta_{\alpha\beta}=0~~~\forall\alpha
\label{eq:repliconmode}
\ee
in which $q$ is the solution of
(\ref{eq:SKRSqm}) and $\eta_{\alpha\beta}=\eta_{\beta\alpha}$. We now calculate the
resulting change in $f(\bq,\bm)$, away from the RS value $f(\bq_{\rm
RS},\bm_{\rm RS})$,
the leading order of which is quadratic in the fluctuations
$\{\eta_{\alpha\beta}\}$ since the RS
solution of (\ref{eq:SKRSqm}) is a saddle-point:
\bd
f(\bq,\bm)-f(\bq_{\rm RS},\bm_{\rm RS})
=
\frac{\beta J^2}{4n} \sum_{\alpha\neq \gamma} \eta^2_{\alpha\gamma}
-\frac{\beta^3 J^4}{8n}
\sum_{\alpha\neq\gamma}\sum_{\rho\neq\lambda}
\eta_{\alpha\gamma}\eta_{\rho\lambda}G_{\alpha\gamma\rho\lambda}
\ed
with
\bd
G_{\alpha\gamma\rho\lambda}=
\frac{
\bra
\sigma_{\alpha}\sigma_{\gamma}\sigma_{\rho}\sigma_{\lambda}
e^{\frac{1}{2}q\beta^2 J^2\left[\sum_{\alpha}\sigma_{\alpha}\right]^2
+\beta  m J_0 \sum_{\alpha}\sigma_{\alpha}}
\ket_{\bsigma}}
{\bra
e^{\frac{1}{2}q\beta^2 J^2\left[\sum_{\alpha}\sigma_{\alpha}\right]^2
+\beta  m J_0\sum_{\alpha} \sigma_{\alpha}}
\ket_{\bsigma}}
\ed
Because of the index permutation symmetry in the above average we can
write for $\alpha\neq\gamma$ and $\rho\neq\lambda$:
\bd
G_{\alpha\gamma\rho\lambda}=
\delta_{\alpha\rho}\delta_{\gamma\lambda}\plus
\delta_{\alpha\lambda}\delta_{\gamma\rho}
+ G_4\left[1\minus\delta_{\alpha\rho}\right]\left[1\minus\delta_{\gamma\lambda}\right]\left[1\minus\delta_{\alpha\lambda}\right]\left[1\minus
\delta_{\gamma\rho}\right]\room
\ed
\bd
~~~~+ ~G_2\left\{\delta_{\alpha\rho}\left[1\minus\delta_{\gamma\lambda}\right]\plus
\delta_{\gamma\lambda}\left[1\minus\delta_{\alpha\rho}\right]\plus
\delta_{\alpha\lambda}\left[1\minus\delta_{\gamma\rho}\right]\plus
\delta_{\gamma\rho}\left[1\minus\delta_{\alpha\lambda}\right]\right\}
\ed
with
\bd
G_{\ell}=
\frac{\int\!Dz~\tanh^\ell\left[\beta J_0 m\plus\beta J
z\sqrt{q}\right]\cosh^n \left[\beta J_0 m\plus\beta J z\sqrt{q}\right]}
{\int\!Dz~\cosh^n\left[\beta J_0 m\plus\beta J z\sqrt{q}\right]}
\ed
Only terms which involve precisely two $\delta$-functions can contribute,
because of the requirements $\alpha\neq\gamma$, $\rho\neq\lambda$ and $\sum_{\beta}\eta_{\alpha\beta}=0$.
As a result:
\bd
f(\bq,\bm)-f(\bq_{\rm RS},\bm_{\rm RS})
=
\frac{\beta J^2}{4n}\left[
1\minus\beta^2 J^2\left(1\minus 2G_2\plus G_4\right)\right]
\sum_{\alpha\neq\gamma}
\eta^2_{\alpha\gamma}
\ed
The condition for the RS solution to minimise $f(\bq,\bm)$, if compared
to the so called `replicon' fluctuations (\ref{eq:repliconmode}), is
therefore
\bd
1>\beta^2 J^2\lim_{n\rightarrow0}\left(1-2G_2+G_4\right)\room
\ed
After taking the limit in the expressions $G_{\ell}$
this condition can be written as
\be
1> \beta^2 J^2\int\!Dz~\cosh^{-4}\left[\beta J_0 m\plus\beta J z\sqrt{q}\right]
\label{eq:ATline}
\ee
The so-called AT line in the phase diagram where this condition ceases to be met,
indicates a continuous transition to a complex `spin-glass' state where ergodicity is
broken (i.e. the distribution $\overline{P(q)}$
(\ref{eq:overlapdistribution}) is no longer a $\delta$-function). It is
shown in figure \ref{fig:SKphasediagram} as a dashed line for
$J_0/J>1$, and coincides with the line $T/J=1$ for $J_0<1$.

\section{The Hopfield Model Near Saturation}

\subsection{Replica Analysis}

We now turn to the Hopfield model with an extensive number of stored patterns, i.e. $p=\alpha N$ in
(\ref{eq:hebbinteractions}). We can still write the free
energy in the form (\ref{eq:hopfreeenergy}), but this will not be of help
since here it involves integrals over an extensive number of
variables, so that steepest descent integration does not apply.
Instead, following the approach of the previous model (\ref{eq:SKinteractions}),
we assume \cite{AGS2} that we can average the free energy over
the distribution of the patterns, with help of the replica-trick (\ref{eq:replicatrick}):
\bd
\overline{F} = -\lim_{n\rightarrow0}\frac{1}{\beta n}\log \sum_{\bsigma^1\!\ldots\bsigma^n}\overline{e^{-\beta\sum_{\alpha=1}^n
H(\bsigma^{\alpha})}}
\ed
Greek indices will denote either replica labels or pattern labels (it
will be clear from the context), i.e.
$\alpha,\beta=1,\ldots,n$ and $\mu,\nu=1,\ldots,p$.
The $p\times N$ pattern components $\{\xi_i^{\mu}\}$ are
assumed to be drawn independently at random from $\{-1,1\}$.
\vsp

\noindent{\em Replica Calculation of the Disorder-Averaged Free Energy}.
We first add to the Hamiltonian of (\ref{eq:seq_thermodynamics1}) a finite number $\ell$ of generating terms, that
will allow us to obtain expectation values of the overlap order
parameters $m_{\mu}$ (\ref{eq:overlaps}) by differentiation of the
free energy (since all patterns are equivalent in the calculation we
may choose these $\ell$ nominated patterns arbitrarily):
\be
H\rightarrow H + \sum_{\mu=1}^{\ell}\lambda_{\mu}\sum_i \sigma_i\xi_i^{\mu}
~~~~~~~~
\bra m_{\mu}(\bsigma)\ket_{\rm
eq}=\lim_{\blambda\rightarrow\bnul}\frac{\partial}{\partial\lambda_{\mu}}F/N
\label{eq:hopgenerators}
\ee
We know how to deal with a finite number of overlaps and corresponding
patterns, therefore we average only over the disorder that is
responsible for the complications: the patterns
$\{\bxi^{\ell+1},\ldots,\bxi^{p}\}$ (as in the previous section
we denote this disorder-averaging by $\overline{\cdots}$).
Upon inserting the extended Hamiltonian into the replica-expression
for the free energy, and assuming that the
order of the limits $N\rightarrow\infty$ and $n\rightarrow 0$ can be
interchanged, we obtain for large $N$:
\bd
\overline{F}/N = \frac{1}{2}\alpha - \frac{1}{\beta}\log2 -\lim_{n\rightarrow0} \frac{1}{\beta N
n}\log
\bra
e^{-\beta\sum_{\mu\leq \ell}\sum_{\alpha}\left[\lambda_{\mu}\sum_i\sigma^{\alpha}_i\xi_i^{\mu}
-\frac{1}{2N}\left[\sum_i\sigma_i^{\alpha}\xi_i^{\mu}\right]^2\right]}
\overline{e^{\frac{\beta}{2N}\sum_{\alpha}\sum_{\mu>\ell}\left[\sum_i\sigma_i^{\alpha}\xi_i^{\mu}\right]^2}}\ket_{\{\bsigma^{\alpha}\}}
\ed
We linearise the $\mu\leq \ell$ quadratic term using the
identity (\ref{eq:gaussians}), leading to $n\times\ell$ Gaussian
integrals with $D\bm=(Dm^1_1,\ldots,Dm^{\ell}_n)$:
\bd
\overline{F}/N = \frac{1}{2}\alpha - \frac{1}{\beta}\log2
-\lim_{n\rightarrow0} \frac{1}{\beta N n}\log\int\!D\bm
\bra
e^{\sum_{\mu\leq \ell}\sum_{\alpha}
\sum_i\sigma_i^{\alpha}\xi_i^{\mu}\left[\sqrt{\frac{\beta}{N}}
m_{\alpha}^{\mu}-\beta\lambda_{\mu}\right]}
\overline{e^{\frac{\beta}{2N}\sum_{\alpha}\sum_{\mu>\ell}\left[\sum_i\sigma_i^{\alpha}\xi_i^{\mu}\right]^2}}
\ket_{\{\bsigma^{\alpha}\}}
\ed
Anticipating that only terms exponential in the system size $N$ will
retain statistical relevance in the limit $N\rightarrow\infty$, we
rescale the $n\times\ell$ integration variables $\bm$ according to
$\bm\rightarrow\bm\sqrt{\beta N}$:
\bd
\overline{F}/N = \frac{1}{2}\alpha - \frac{1}{\beta}\log2
-\lim_{n\rightarrow0} \frac{1}{\beta N n}\log\left\{\left[\frac{\beta
N}{2\pi}\right]^{\frac{n\ell}{2}}\int\!d\bm~e^{-\frac{1}{2}\beta N\bm^2}\times
~~~~~~~~~~~~~~~~~~~~~~~~~~~~~~~~~~~
\right.
\ed
\be
\left.
~~~~~~~~~~~~~~~~~~~~~~~~~~
\bra
e^{\beta\sum_{\mu\leq \ell}\sum_{\alpha}
\sum_i\sigma_i^{\alpha}\xi_i^{\mu}\left[m_{\alpha}^{\mu}-\lambda_{\mu}\right]}
~\overline{e^{\frac{\beta}{2N}\sum_{\alpha}\sum_{\mu>\ell}\left[\sum_i\sigma_i^{\alpha}\xi_i^{\mu}\right]^2}}
\ket_{\{\bsigma^{\alpha}\}}
\right\}
\label{eq:intermediate}
\ee
Next we turn to the disorder average, where we again linearise the exponent containing the pattern
components using the identity (\ref{eq:gaussians}),
with
$D\bz=(Dz_1,\ldots,Dz_n)$:
\bd
\overline{e^{\frac{\beta}{2N}\sum_{\alpha}\sum_{\mu>\ell}\left[\sum_i\sigma_i^{\alpha}\xi_i^{\mu}\right]^2}}
=\left\{
\overline{e^{\frac{1}{2}\sum_{\alpha}\left[
\left(\frac{\beta}{N}\right)^{\frac{1}{2}}
\sum_i\sigma_i^{\alpha}\xi_i\right]^2}}\right\}^{p-\ell}
=\left\{ \int\!D\bz~
\overline{e^{\left(\frac{\beta}{N}\right)^{\frac{1}{2}}\sum_{\alpha}z_{\alpha}\sum_i\sigma_i^{\alpha}\xi_i}}\right\}^{p-\ell}
\vspace*{-3mm}
\ed
\be
=
\left\{
\int\!D\bz~\prod_i
\cosh\left[\left(\frac{\beta}{N}\right)^{\frac{1}{2}}\sum_{\alpha}z_{\alpha}\sigma_i^{\alpha}\right]\right\}^{p-\ell}
=
\left\{ \int\!D\bz~
e^{\frac{\beta}{2N}\sum_{\alpha\beta}z_{\alpha}z_{\beta}
\sum_i\sigma_i^{\alpha}\sigma_i^{\beta}+\order(\frac{1}{N})}
\right\}^{p}
\label{eq:disav}
\ee
We are now as in the previous case
led to introducing the replica order parameters $q_{\alpha\beta}$:
\bd
1 =\int\!d\bq~\prod_{\alpha\beta}\delta\left[q_{\alpha\beta}\minus
\frac{1}{N}\sum_i\sigma_i^{\alpha}\sigma_i^{\beta}\right]\room
 =\left[\frac{N}{2\pi}\right]^{n^2}
\int\!d\bq d\hat{\bq}~
e^{iN\sum_{\alpha\beta}\hat{q}_{\alpha\beta}\left[q_{\alpha\beta}-\frac{1}{N}\sum_i\sigma_i^{\alpha}\sigma_i^{\beta}\right]}\room
\ed
Inserting (\ref{eq:disav}) and the above identities into
(\ref{eq:intermediate}) and assuming that the limits
$N\rightarrow\infty$ and $n\rightarrow0$ commute gives:
\vspace*{-2mm}
\bd
\lim_{N\rightarrow\infty}\overline{F}/N = \frac{1}{2}\alpha - \frac{1}{\beta}\log2
-\lim_{N\to\infty}\lim_{n\rightarrow0} \frac{1}{\beta N n}\log\int\!d\bm d\bq
d\hat{\bq}~
e^{N\left[i\sum_{\alpha\beta}\hat{q}_{\alpha\beta}q_{\alpha\beta}
-\frac{1}{2}\beta\bm^2 +\alpha \log
\int\!D\bz~
e^{\frac{\beta}{2}\sum_{\alpha\beta}z_{\alpha}z_{\beta}q_{\alpha\beta}}
\right]}
\ed
\bd
~~~~~~~~~~~~~~~~~~~~~~~~~~~~~~~~~~~~~~~~~~~~
\times~
\bra
e^{\beta\sum_{\mu\leq \ell}\sum_{\alpha}
\sum_i\sigma_i^{\alpha}\xi_i^{\mu}\left[m_{\alpha}^{\mu}-\lambda_{\mu}\right]
-i\sum_{\alpha\beta}\hat{q}_{\alpha\beta}\sum_i\sigma_i^{\alpha}\sigma_i^{\beta}}\ket_{\{\bsigma^{\alpha}\}}
\ed
The $n$-dimensional Gaussian integral over $\bz$ factorises in the standard way after appropriate rotation
of the integration variables $\bz$, with the result:
\bd
\log \int\!D\bz~
e^{\frac{\beta}{2}\sum_{\alpha\beta}z_{\alpha}z_{\beta}q_{\alpha\beta}}
= -\frac{1}{2} \log \det\left[\one-\beta\bq\right]
\ed
in which $\one$ denotes the $n\times n$ identity matrix.
The neuron averages factorise and are reduced to single-site ones over
the $n$-replicated neuron $\bsigma=(\sigma_1,\ldots,\sigma_n)$:
\bd
\lim_{N\rightarrow\infty}\overline{F}/N =
 \frac{1}{2}\alpha - \frac{1}{\beta}\log2
-\lim_{N\to\infty}\lim_{n\rightarrow0} \frac{1}{\beta N n}\log \int\!d\bm d\bq
d\hat{\bq}~
e^{N\left[i\sum_{\alpha\beta}\hat{q}_{\alpha\beta}q_{\alpha\beta}
-\frac{1}{2}\beta\bm^2-\frac{1}{2}\alpha\log\det\left[\one-\beta\bq\right]\right]}
\ed
\bd
~~~~~~~~~~~~~~~~~~~~~~~~~~
\times~
\prod_i
\bra e^{\beta\sum_{\mu\leq \ell}\sum_{\alpha}\sigma_{\alpha}\xi_i^{\mu}\left[
m_{\alpha}^{\mu}-\lambda_{\mu}\right]-i\sum_{\alpha\beta}\hat{q}_{\alpha\beta}\sigma_{\alpha}\sigma_{\beta}}\ket_{\bsigma}
\ed
and we arrive at integrals that can
be evaluated by steepest descent, following the manipulations
(\ref{eq:reverse}).
If we denote averages over the remaining
$\ell$ patterns in the familiar way
\bd
\bxi=(\xi_1,\ldots,\xi_{\ell})\room~~~~~~~~\bra\Phi(\bxi)\ket_{\bxi}=2^{-\ell}\!\!\sum_{\bxi\in\{-1,1\}^{\ell}}\Phi(\bxi)
\ed
we can write the final result in the form
\be
\lim_{N\rightarrow\infty}\overline{F}/N =
\lim_{n\rightarrow0} \extr~f(\bm,\bq,\hat{\bq})
\room
\label{eq:fullhopfieldsaddle}
\ee
\bd
f(\bm,\bq,\hat{\bq})=\frac{1}{2}\alpha
-\frac{1}{\beta}\log 2
-\frac{1}{\beta n}\left[\bra \log \bra
e^{\beta\sum_{\mu\leq \ell}\sum_{\alpha}\sigma_{\alpha}\xi_{\mu}\left[
m_{\alpha}^{\mu}-\lambda_{\mu}\right]-i\sum_{\alpha\beta}\hat{q}_{\alpha\beta}\sigma_{\alpha}\sigma_{\beta}}\ket_{\bsigma}\ket_{\bxi}
\right.
\ed
\bd
\left.
+i\sum_{\alpha\beta}\hat{q}_{\alpha\beta}q_{\alpha\beta}
-\frac{1}{2}\beta\bm^2
-\frac{1}{2}\alpha \log \det\left[\one\minus \beta\bq\right]
\right]
\ed
Having arrived at a saddle-point problem we now
first identify the expectation values of the overlaps with
(\ref{eq:hopgenerators}) (note: extremisation with respect to the
saddle-point variables and differentiation with respect to $\blambda$ commute):
\bd
\overline{\bra m_{\mu}(\bsigma)\ket_{\rm
eq}}
=\lim_{n\rightarrow0}
\lim_{\blambda\rightarrow\bnul}\frac{\partial}{\partial\lambda_{\mu}}
\extr~f(\bm,\bq,\hat{\bq})
\ed
\be
=\lim_{n\rightarrow0}
\bra \xi_{\mu}
\frac{
\bra \frac{1}{n}\sum_{\alpha} \sigma_{\alpha}
e^{\beta\sum_{\mu\leq \ell}\sum_{\alpha}\sigma_{\alpha}\xi_{\mu}
m_{\alpha}^{\mu}-i\sum_{\alpha\beta}\hat{q}_{\alpha\beta}\sigma_{\alpha}\sigma_{\beta}}
\ket_{\bsigma}}
{
\bra e^{\beta\sum_{\mu\leq \ell}\sum_{\alpha}\sigma_{\alpha}\xi_{\mu}
m_{\alpha}^{\mu}-i\sum_{\alpha\beta}\hat{q}_{\alpha\beta}\sigma_{\alpha}\sigma_{\beta}}\ket_{\bsigma}}
\ket_{\bxi}
\label{eq:overlapidentification}
\ee
which is to be evaluated in the $\blambda=\bnul$ saddle-point.
Having served their purpose, the generating fields $\lambda_{\mu}$ can be
set to zero and we can restrict ourselves to the
$\blambda=\bnul$ saddle-point problem:
\bd
f(\bm,\bq,\hat{\bq})=\frac{1}{2}\alpha
-\frac{1}{\beta}\log2 -\frac{1}{\beta n}\left[
i\sum_{\alpha\beta}\hat{q}_{\alpha\beta}q_{\alpha\beta}
-\frac{1}{2}\beta\bm^2
-\frac{1}{2}\alpha\log \det\left[\one\minus\beta\bq\right]
~~~~~~~~~~~~~~~~~~~~~~~~~~~~~~
\right.
\ed
\be
\left.
~~~~~~~~~~~~~~~~~~~~~~~~~~~~~~~~~~~~~~~~
+ \bra \log \bra e^{\beta\sum_{\mu\leq
\ell}\sum_{\alpha}\sigma_{\alpha}\xi_{\mu}m_{\alpha}^{\mu}-i\sum_{\alpha\beta}\hat{q}_{\alpha\beta}\sigma_{\alpha}\sigma_{\beta}}\ket_{\bsigma}
\ket_{\bxi}
\right]
\label{eq:fullhopfieldsurface}
\ee
Variation of the parameters
$\{m_{\alpha}^{\mu},~\hat{q}_{\alpha\beta},~q_{\alpha\beta}\}$ gives
the saddle-point equations:
\be
m_{\alpha}^{\mu}=
\bra \xi_{\mu}
\frac{
\bra \sigma_{\alpha}
e^{\beta\sum_{\mu\leq \ell}\sum_{\alpha}\sigma_{\alpha}\xi_{\mu}
m_{\alpha}^{\mu}-i\sum_{\alpha\beta}\hat{q}_{\alpha\beta}\sigma_{\alpha}\sigma_{\beta}}
\ket_{\bsigma}}
{
\bra e^{\beta\sum_{\mu\leq \ell}\sum_{\alpha}\sigma_{\alpha}\xi_{\mu}
m_{\alpha}^{\mu}-i\sum_{\alpha\beta}\hat{q}_{\alpha\beta}\sigma_{\alpha}\sigma_{\beta}}\ket_{\bsigma}}
\ket_{\bxi}
\label{eq:hopsaddlemamu}
\ee
\be
q_{\lambda\rho}=
\bra
\frac{
\bra \sigma_{\lambda}\sigma_{\rho}
e^{\beta\sum_{\mu\leq \ell}\sum_{\alpha}\sigma_{\alpha}\xi_{\mu}
m_{\alpha}^{\mu}-i\sum_{\alpha\beta}\hat{q}_{\alpha\beta}\sigma_{\alpha}\sigma_{\beta}}
\ket_{\bsigma}}
{
\bra e^{\beta\sum_{\mu\leq \ell}\sum_{\alpha}\sigma_{\alpha}\xi_{\mu}
m_{\alpha}^{\mu}-i\sum_{\alpha\beta}\hat{q}_{\alpha\beta}\sigma_{\alpha}\sigma_{\beta}}\ket_{\bsigma}}
\ket_{\bxi}
\label{eq:hopsaddleqab}
\ee
\be
\hat{q}_{\lambda\rho}=\frac{1}{2}i\alpha\beta
\frac{\int\!d\bz~z_{\lambda}z_{\rho} e^{-\frac{1}{2}\bz\cdot\left[\one-\beta
\bq\right]\bz}}
{\int\!d\bz~e^{-\frac{1}{2}\bz\cdot\left[\one-\beta \bq\right]\bz}}
\label{eq:hopsaddleqhatab}
\ee
furthermore,
\be
\overline{\bra m_{\mu}(\bsigma)\ket_{\rm
eq}}=\lim_{n\rightarrow 0}\frac{1}{n}\sum_{\alpha}m_{\alpha}^{\mu}
\label{eq:finaloverlapidentification}
\ee
replaces the identification (\ref{eq:overlapidentification}).
As expected, one always has $q_{\alpha\alpha}=1$. The
diagonal elements $\hat{q}_{\alpha\alpha}$ drop out of
(\ref{eq:hopsaddlemamu},\ref{eq:hopsaddleqab}), their values are simply
given as functions of the remaining parameters by
(\ref{eq:hopsaddleqhatab}). \vsp

\noindent{\em Physical Interpretation of Saddle Points.} We proceed
along the lines of the Gaussian model (\ref{eq:SKinteractions}). If we apply
the alternative version (\ref{eq:secondreplicatrick}) of the replica
trick to the Hopfield
model, we can write the distribution of the $\ell$ overlaps
$\bm=(m_1,\ldots,m_{\ell})$ in equilibrium as
\bd
P(\bm)=\lim_{n\rightarrow0}\frac{1}{n}\sum_{\gamma}\sum_{\bsigma^1\!\ldots\bsigma^n}
\delta[\bm\minus\frac{1}{N}\sum_{i}\sigma_i^{\gamma}\bxi_i]
\prod_{\alpha}
e^{-\beta H(\bsigma^{\alpha})}
\ed
with $\bxi_i=(\xi_i^1,\ldots,\xi_i^{\ell})$.
Averaging this distribution over the disorder leads to expressions
identical to those encountered in evaluating the disorder averaged
free energy. By inserting the same delta-functions we arrive at the
saddle-point integration (\ref{eq:fullhopfieldsaddle},\ref{eq:fullhopfieldsurface}) and find
\be
\overline{P(\bm)}=\lim_{n\rightarrow0}\frac{1}{n}\sum_{\gamma}
\delta\left[\bm\minus \bm_{\gamma}\right]
\label{eq:hopmagdistribution}
\ee
where $\bm_{\gamma}=(m_{\gamma}^1,\ldots,m_{\gamma}^{\ell})$ refers to  the relevant solution
of (\ref{eq:hopsaddlemamu},\ref{eq:hopsaddleqab},\ref{eq:hopsaddleqhatab}).

Similarly we imagine two systems $\bsigma$ and
$\bsigma^{\prime}$ with
identical realisation of the interactions $\{J_{ij}\}$, both in
thermal equilibrium,
and use (\ref{eq:secondreplicatrick}) to rewrite the distribution $P(q)$ for
the mutual overlap between the microstates
of the two systems
\bd
P(q)=
\lim_{n\rightarrow0}\frac{1}{n(n\minus 1)}\sum_{\lambda\neq
\gamma}\sum_{\bsigma^1\!\ldots\bsigma^n}
\delta[q\minus \frac{1}{N}\sum_{i}\sigma_i^{\lambda}\sigma_i^{\gamma}]\prod_{\alpha}
e^{-\beta H(\bsigma^{\alpha})}
\ed
Averaging over the disorder again leads to the steepest descend
integration (\ref{eq:fullhopfieldsaddle},\ref{eq:fullhopfieldsurface}) and we find
\be
\overline{P(q)}=\lim_{n\rightarrow0}\frac{1}{n(n\minus1)}\sum_{\lambda\neq \gamma}
\delta\left[q-q_{\lambda\gamma}\right]
\label{eq:hopoverlapdistribution}
\ee
where $\{q_{\lambda\gamma}\}$ refers to  the relevant solution
of (\ref{eq:hopsaddlemamu},\ref{eq:hopsaddleqab},\ref{eq:hopsaddleqhatab}).

Finally we analyse the physical meaning of the conjugate
parameters $\{\hat{q}_{\alpha\beta}\}$
for $\alpha\neq\beta$.
We will do this in more detail, the analysis being rather specific
for the Hopfield model and slightly
different from the derivations above.
Again we imagine two systems $\bsigma$ and
$\bsigma^{\prime}$ with
identical interactions $\{J_{ij}\}$, both in
thermal equilibrium. We now use (\ref{eq:secondreplicatrick}) to
evaluate the covariance of the overlaps corresponding to
non-nominated patterns:
\be
r=\frac{1}{\alpha}\sum_{\mu=\ell+1}^p\overline{
\bra\frac{1}{N}\sum_i\sigma_i\xi_i^{\mu}\ket_{\rm eq}
\bra\frac{1}{N}\sum_i\sigma^{\prime}_i\xi_i^{\mu}\ket_{\rm eq}
}~~~~~~~~~~~~~~~~~~~~~~~~~~~~~~~~~~~
\label{eq:orderparameterr}
\ee
\bd
=\lim_{n\rightarrow0}\frac{N\minus \ell/\alpha}{n(n\minus 1)}\sum_{\lambda\neq
\gamma}
\sum_{\bsigma^1\!\ldots\bsigma^n}\overline{
\left[\frac{1}{N}\sum_i\sigma^{\lambda}_i\xi_i^{p}\right]
\left[\frac{1}{N}\sum_i\sigma^{\gamma}_i\xi_i^{p}\right]
\prod_{\alpha} e^{-\beta H(\bsigma^{\alpha})}}
\ed
(using the equivalence of all such patterns). We next perform
the same manipulations as in calculating the free energy. Here the
disorder average involves
\bd
\overline{
\left[\frac{1}{\sqrt{N}}\sum_i\sigma^{\lambda}_i\xi_i^{p}\right]
\left[\frac{1}{\sqrt{N}}\sum_i\sigma^{\gamma}_i\xi_i^{p}\right]
e^{\frac{\beta}{2N}\sum_{\alpha}\sum_{\mu>\ell}\left[\sum_i\sigma_i^{\alpha}\xi_i^{\mu}\right]^2}}
\ed
\bd
=\left\{
\int\!D\bz~
\overline{e^{\left(\frac{\beta}{N}\right)^{\frac{1}{2}}\sum_{\alpha}z_{\alpha}\sum_i\sigma_i^{\alpha}\xi_i}}
\right\}^{p-\ell-1}
\!\int\!\frac{D\bz}{\beta}\frac{\partial^2}{\partial z_{\lambda}\partial z_{\gamma}}
\overline{e^{\left(\frac{\beta}{N}\right)^{\frac{1}{2}}\sum_{\alpha}z_{\alpha}\sum_i\sigma_i^{\alpha}\xi_i}}
\ed
\bd
=\left\{
\int\!D\bz~
\overline{e^{\left(\frac{\beta}{N}\right)^{\frac{1}{2}}\sum_{\alpha}z_{\alpha}\sum_i\sigma_i^{\alpha}\xi_i}}
\right\}^{p-\ell-1}
\!\int\!D\bz \frac{z_{\lambda}z_{\gamma}}{\beta}
\overline{e^{\left(\frac{\beta}{N}\right)^{\frac{1}{2}}\sum_{\alpha}z_{\alpha}\sum_i\sigma_i^{\alpha}\xi_i}}
\ed
(after partial integration). We finally obtain an expression which
involves the surface (\ref{eq:fullhopfieldsurface}):
\bd
r=\frac{1}{\beta}\lim_{n\rightarrow0}\frac{1}{n(n\minus
1)}\sum_{\lambda\neq
\rho}
\lim_{N\rightarrow\infty}\frac
{\int\!d\bm d\bq d\hbq
\left[\frac{\int\!d\bz~z_{\lambda}z_{\rho}~e^{-\frac{1}{2}\bz\cdot\left[\one-\beta
\bq\right]\bz}}
{\int\!d\bz~e^{-\frac{1}{2}\bz\cdot\left[\one-\beta \bq\right]\bz}}\right]
~e^{-\beta nNf(\bm,\bq,\hat{\bq})}}
{\int\!d\bm d\bq d\hbq~
~e^{-\beta nNf(\bm,\bq,\hat{\bq})}}
\ed
The normalisation of the above integral over $\{\bm,\bq,\hbq\}$
follows from using the replica procedure to
rewrite unity.
The integration being dominated by the minima of $f$, we
can use the saddle-point equations (\ref{eq:hopsaddleqhatab}) to arrive
at
\be
\lim_{n\rightarrow 0}\frac{1}{n(n\minus
1)}\sum_{\lambda\neq \rho}\hat{q}_{\lambda\rho}=\frac{1}{2}i\alpha \beta^2 r
\label{eq:hatinterpretation}
\ee
The result (\ref{eq:orderparameterr},\ref{eq:hatinterpretation})
provides a physical interpretation of the order parameters
$\{\hat{q}_{\alpha\beta}\}$.

Ergodicity implies that the distributions $\overline{P(q)}$ and
$\overline{P(\bm)}$ are $\delta$-functions, this is equivalent to the relevant saddle-point being
of the form:
\be
m^{\mu}_{\gamma}=m_{\mu}~~~~~~q_{\gamma\rho}=\delta_{\gamma\rho}+
q\left[1\minus\delta_{\gamma\rho}\right]~~~~~~
\hat{q}_{\gamma\rho}=\frac{1}{2}i\alpha\beta^2\left[R\delta_{\gamma\rho}+
r\left[1\minus\delta_{\gamma\rho}\right]\right]
\label{eq:RShopfield}
\ee
which is the `replica symmetry' (RS) ansatz for the Hopfield model.
The RS form for $\{q_{\alpha\beta}\}$ and $\{m_{\alpha}^{\mu}\}$ is a
direct consequence of the corresponding distributions being
$\delta$-functions, whereas the RS form for
$\{\hat{q}_{\alpha\beta}\}$ subsequently follows from
(\ref{eq:hopsaddleqhatab}). The physical
meaning of $m_{\mu}$ and $q$ is
\bd
m_{\mu}=\overline{\bra m_{\mu}(\bsigma)\ket_{\rm
eq}}~~~~~~~~
q=\frac{1}{N}\sum_i\overline{\bra \sigma_i\ket^2_{\rm eq}}
\ed
Before proceeding with a full analysis of the RS saddle-point
equations, we finally make a few tentative statements on the phase
diagram.
For $\beta=0$ we obtain the trivial result
$q_{\lambda\rho}=\delta_{\lambda\rho}$, $\hat{q}_{\lambda\rho}=0$,
$m_{\alpha}^{\mu}=0$. We can
identify continuous bifurcations to a non-trivial state by expanding
the saddle-point equations in first order in the relevant parameters:
\bd
m_{\alpha}^{\mu}=\beta m_{\alpha}^{\mu}+\ldots,~~~~~~~
q_{\lambda\rho}=-2i\hat{q}_{\lambda\rho}+\ldots~~(\lambda\neq\rho),~~~~~~~
\hat{q}_{\lambda\rho}=
\frac{1}{2}\frac{i\alpha\beta}{1\minus \beta}
\left[\delta_{\lambda\rho}\plus \frac{\beta}{1\minus \beta}q_{\lambda\rho}
\left[1\minus\delta_{\lambda\rho}\right]\right]
+\ldots
\ed
Combining the equations for $\bq$ and $\hbq$ gives
$q_{\lambda\rho}=
\alpha\left[\frac{\beta}{1-\beta}\right]^2
q_{\lambda\rho}\plus\ldots$.
Thus we expect a continuous transition at $T=1\plus \sqrt{\alpha}$
from the trivial state to an ordered state where $q_{\lambda\rho}\neq 0$, but still
$\bra m_{\mu}\ket_{\rm eq}=0$
(a spin-glass state).

\subsection{Replica Symmetric Solution and AT-Instability}

The symmetry of the ansatz
(\ref{eq:RShopfield}) for the saddle-point allows us to diagonalise
the matrix $\bLambda=\one\minus \beta\bq$ which we encountered in the saddle-point
problem,
$\Lambda_{\alpha\beta}=[1\minus \beta(1\minus q)]\delta_{\alpha\beta}\minus \beta
q$:
\bd
\begin{array}{llc}
{\rm eigenspace:} & {\rm eigenvalue:} & {\rm multiplicity:}
\\[2mm]
\bx=(1,\ldots,1) & 1\minus\beta(1\minus q)\minus\beta q n & 1  \\
\sum_{\alpha}x_{\alpha}\!=\!0 & 1\minus\beta(1\minus q) & n\minus1
\end{array}
\ed
so that
\bd
\log\det\bLambda=\log\left[1\minus\beta(1\minus
q)\minus \beta q
n\right]
+(n\minus 1)\log\left[1\minus\beta(1\minus q)\right]
~~~
\ed
\bd
~~~
=n\left[\log\left[1\minus\beta(1\minus q)\right] - \frac{\beta
q}{1\minus \beta(1\minus q)}\right] +\order(n^2)
\ed
Inserting the RS ansatz
(\ref{eq:RShopfield}) for the saddle-point into
(\ref{eq:fullhopfieldsurface}), utilising the above expression for the
determinant and the short-hand $\bm=(m_1,\ldots,m_{\ell})$, gives
\bd
f(\bm_{\rm RS},\bq_{\rm RS},\hat{\bq}_{\rm RS})=
-\frac{1}{\beta}\log2
+\frac{1}{2}\alpha\left[1\plus \beta r(1\minus q)\right]
+\frac{1}{2}\bm^2
+\frac{\alpha}{2\beta}\left[
\log\left[1\minus \beta(1\minus q)\right] - \frac{\beta q}{1\minus
\beta(1\minus q)}\right]
\ed
\bd
-\frac{1}{\beta n}
\bra \log
\bra
e^{\beta\bm\cdot\bxi\sum_{\alpha}\sigma_{\alpha}+\frac{1}{2}\alpha
r\beta^2\left[\sum_{\alpha}\sigma_{\alpha}\right]^2}\ket_{\bsigma}
\ket_{\bxi}
+\order(n)
\ed
We now linearise the squares in the neuron averages with
(\ref{eq:gaussians}), subsequently average over the replicated
neuron
$\bsigma$, use $\cosh^n[x]=1+n\log\cosh[x]+\order(n^2)$, and take the
limit $n\rightarrow0$:
\bd
\lim_{N\rightarrow\infty}\overline{F}_{\rm RS}/N=
\lim_{n\rightarrow0}
f(\bm_{\rm RS},\bq_{\rm RS},\hat{\bq}_{\rm
RS})~~~~~~~~~~~~~~~~~~~~~~~~~~~~~~~~~~~~~~~~~~~~~~~~~~~~~~~~~~~~~~~~~~~
\ed
\be
=\frac{1}{2}\bm^2
+\frac{1}{2}\alpha\left[
1\plus \beta r(1\minus q)\plus \frac{1}{\beta}\log\left[1\minus \beta(1\minus q)\right]\minus \frac{q}{1\minus
\beta(1\minus q)}\right]
-\frac{1}{\beta}
\bra \int\!Dz~\log 2\cosh\beta\left[\bm\cdot\bxi\plus z\sqrt{\alpha
r}\right]\ket_{\bxi}
\label{eq:RShopfreeenergy}
\ee
The saddle-point equations for $\bm$, $q$ and $r$ can be obtained either by
insertion of the RS ansatz (\ref{eq:RShopfield}) into
(\ref{eq:hopsaddlemamu},\ref{eq:hopsaddleqab},\ref{eq:hopsaddleqhatab})
and subsequently taking the $n\rightarrow0$ limit,
or by variation of the RS expression (\ref{eq:RShopfreeenergy}). The
latter route is the fastest one. After performing partial
integrations where appropriate we obtain the final result:
\be
\bm=
\bra \bxi \int\!Dz~\tanh\beta\left[\bm\cdot\bxi\plus z\sqrt{\alpha
r}\right]\ket_{\bxi}
\label{eq:RShopm}
\ee
\be
q=\bra \int\!Dz~\tanh^2\beta\left[\bm\cdot\bxi\plus z\sqrt{\alpha
r}\right]\ket_{\bxi}
~~~~~~~~~~~~~
r=q\left[1\minus\beta(1\minus q)\right]^{-2}
\label{eq:RShopqr}
\ee
By substitution of the equation for $r$ into the remaining equations this
set can easily be further reduced, should the need arise.
In case of multiple solutions of
(\ref{eq:RShopm},\ref{eq:RShopqr}) the relevant
saddle-point is the one that minimises (\ref{eq:RShopfreeenergy}).
Clearly for $\alpha=0$ we recover our previous results
(\ref{eq:overlapeqns},\ref{eq:freeenergysurface}).
\vsp

\noindent{\em Analysis of RS Order Parameter Equations and Phase
Diagram}. We first establish an upper bound for the temperature $T=1/\beta$ for
non-trivial solutions of the set
(\ref{eq:RShopm},\ref{eq:RShopqr}) to exist, by writing
(\ref{eq:RShopm}) in integral form:
\bd
m_{\mu}=
\beta\bra\xi_{\mu}\left(\bxi\cdot\bm\right)\int_0^1\!d\lambda\!\int\!Dz
\left[1\minus\tanh^2\beta\left(\lambda\bxi\cdot\bm\plus
z\sqrt{\alpha r}\right)\right]\ket_{\bxi}
\ed
from which we deduce
\bd
0 =\bm^2\minus
\beta\bra\left(\bxi\cdot\bm\right)^2\int_0^1\!d\lambda\!\int\!Dz\left[1\minus
\tanh^2\beta\left(\lambda\bxi\cdot\bm\plus
z\sqrt{\alpha r}\right)\right]\ket_{\bxi} \geq
\bm^2\minus
\beta\bra\left(\bxi\cdot\bm\right)^2\ket_{\bxi}=\bm^2\left[1\minus \beta\right]
\ed
Therefore $\bm=0$ for $T>1$. If $T>1$ we obtain in turn from
(\ref{eq:RShopqr}), using $\tanh^2(x)\leq x^2$ and
$0\leq q\leq 1$:
$q=0$ or $q\leq 1\plus \sqrt{\alpha}\minus T$.
We conclude that $q=0$
for $T>1+\sqrt{\alpha}$. Secondly, for the free energy
(\ref{eq:RShopfreeenergy}) to be well defined we must require
$q>1-T$.
Linearisation of (\ref{eq:RShopm},\ref{eq:RShopqr}) for small $q$ and
$\bm$ shows the continuous bifurcations:
\bd
\begin{array}{llclclc}
                 && {\rm at}           && {\rm from} && {\rm to} \\[2mm]
\alpha>0:  ~     && T=1\plus \sqrt{\alpha} ~&& \bm=\bnul,~q=0  ~  && \bm=\bnul,~q>0   \\
\alpha=0:  ~     && T=1               ~&& \bm=\bnul,~q=0  ~  && \bm\neq\bnul,~q>0
\end{array}
\ed
The upper bound $T=1\plus\sqrt{\alpha}$ turns out to be the critical
noise level indicating (for $\alpha>0$) a continuous transition
to a spin-glass state, where there is no significant alignment of the
neurons
in the direction of one particular pattern, but still a certain degree
of local freezing. Since $\bm=\bnul$ for $T>1$ this spin-glass state
persists at least down to $T=1$. The quantitative details of the
spin-glass state are obtained by inserting $\bm=\bnul$ into
(\ref{eq:RShopqr}) (since (\ref{eq:RShopm}) is
fulfilled automatically).

The impact on the saddle-point equations
(\ref{eq:RShopm},\ref{eq:RShopqr}) of having $\alpha>0$, a smoothening
of the hyperbolic tangent by convolution with a Gaussian
kernel, can be viewed as noise caused by interference between the
attractors. The natural strategy for solving
(\ref{eq:RShopm},\ref{eq:RShopqr}) is therefore to make an ansatz for
the nominated overlaps $\bm$ of the type (\ref{eq:mixturestates}) (the
mixture states). Insertion of this ansatz into the saddle-point
equations indeed leads to self-consistent solutions.
One can solve numerically the remaining equations for the amplitudes of
the mixture states and evaluate their stability by calculating the
eigenvalues of the second derivative of $f(\bm,\bq,\hbq)$, in the same
way as for
$\alpha=0$. The calculations are just more involved. It then turns out
that even mixtures are again unstable for any $T$ and $\alpha$, whereas odd
mixtures can become locally stable for sufficiently small $T$ and
$\alpha$. Among the mixture states, the pure
states, where the vector $\bm$ has only one nonzero component,
are the first to stabilise as the temperature is lowered. These pure
 states, together with the spin-glass state ($\bm=0,~q>0)$, we will
study in more detail.

Let us first calculate the second derivatives of
(\ref{eq:RShopfreeenergy}) and evaluate them in the spin-glass saddle-point.
One finds, after elimination of $r$ with (\ref{eq:RShopqr}):
\bd
\partial^2 \!f/\partial m_{\mu}\partial
m_{\nu}= \delta_{\mu\nu}\left[1\minus \beta(1\minus q)\right]
~~~~~~~~~~~
\partial^2 \!f/\partial m_{\mu}\partial q=0
\ed
The $(\ell\plus 1)\times(\ell\plus 1)$ matrix of second derivatives
with respect to variation of $(\bm,q)$, evaluated in the spin-glass
saddle-point, thereby acquires a diagonal form
\bd
\partial^2 f =
\left(\!
\begin{array}{cccc}
1\minus\beta(1\minus q)  &        &                         & \\
                         & \ddots &                         & \\
                         &        & 1\minus\beta(1\minus q) & \\
 &  &  & \partial^2 f/\partial q^2
\end{array}\!\right)
\ed
and the eigenvalues can simply be read off. The $\ell$-fold degenerate
eigenvalue $1\minus\beta(1\minus q)$ is always positive (otherwise
(\ref{eq:RShopfreeenergy}) would not even exist), implying stability of the
spin-glass state in the
direction of the nominated patterns. The remaining eigenvalue
measures
the stability of the spin-glass state with respect to variation in the
amplitude $q$. Below the critical noise level $T=1\plus\sqrt{\alpha}$ it
turns out to be positive for the spin-glass solution of (\ref{eq:RShopqr}) with
nonzero $q$.
One important difference between the previously studied case $\alpha=0$ and the present
case $\alpha>0$ is that there is now a $\bm=\bnul$ spin-glass solution
which is {\em stable}
for all $T<1\plus\sqrt{\alpha}$. In terms of
information processing this implies that for $\alpha>0$ an initial
state must have a certain non-zero overlap with a pattern to evoke a
final state with $\bm\neq \bnul$, in order to avoid ending up in the
$\bm=\bnul$ spin-glass state. This is clearly consistent with the observations in figure \ref{fig:flows}.
In contrast, for $\alpha=0$, the state with $\bm=\bnul$ is
unstable, so {\em any} initial state will eventually lead to a final
state with $\bm\neq\bnul$.

Inserting the pure state ansatz
$\bm=m(1,0,\ldots,0)$ into our RS equations gives
\be
m= \int\!Dz~\tanh
\left[\beta m+\frac{z\beta\sqrt{\alpha q}}{1\minus\beta(1\minus q)}\right]
~~~~~~~~~~~~
q=\int\!Dz~\tanh^2
\left[\beta m+\frac{z\beta\sqrt{\alpha q}}{1\minus\beta(1\minus q)}\right]
\label{eq:satmq}
\ee
\be
f=\frac{1}{2}m^2
+\frac{1}{2}\alpha\left[
(1\minus q)\frac{1\plus\beta(1\minus q)(\beta\minus 2)}{\left[1\minus\beta(1\minus q)\right]^2}
+\frac{1}{\beta}\log\left[1\minus \beta(1\minus q)\right]
\right]
-\frac{1}{\beta}
\int\!Dz~\log 2\cosh
\left[\beta m \plus
\frac{z\beta \sqrt{\alpha q}}{1\minus\beta(1\minus q)}\right]
\label{eq:satf}
\ee
\begin{figure}[t]
\begin{center}\vspace*{-4mm}
\epsfxsize=80mm\epsfbox{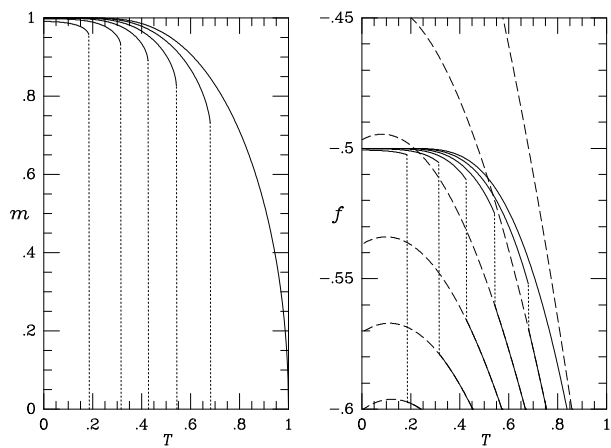}
\end{center}
\vspace*{-5mm}
\caption{Left: RS amplitudes $m$ of the pure
states of the Hopfield model versus temperature. From top to bottom:
$\alpha=0.000~-~0.125$ ($\Delta\alpha=0.025$).
Right, solid lines: `free
energies' $f$ of the pure states. From bottom to top:
$\alpha=0.000~-~0.125$ ($\Delta\alpha=0.025$). Right, dashed lines:
`free energies' of the spin-glass state $\bm=0$ (for
comparison). From top to bottom: $\alpha=0.000~-~0.125$
($\Delta\alpha=0.025$).}
\label{fig:satamplitudes}
\end{figure}
\begin{figure}[t]
\begin{center}\vspace*{-3mm}
\epsfxsize=80mm\epsfbox{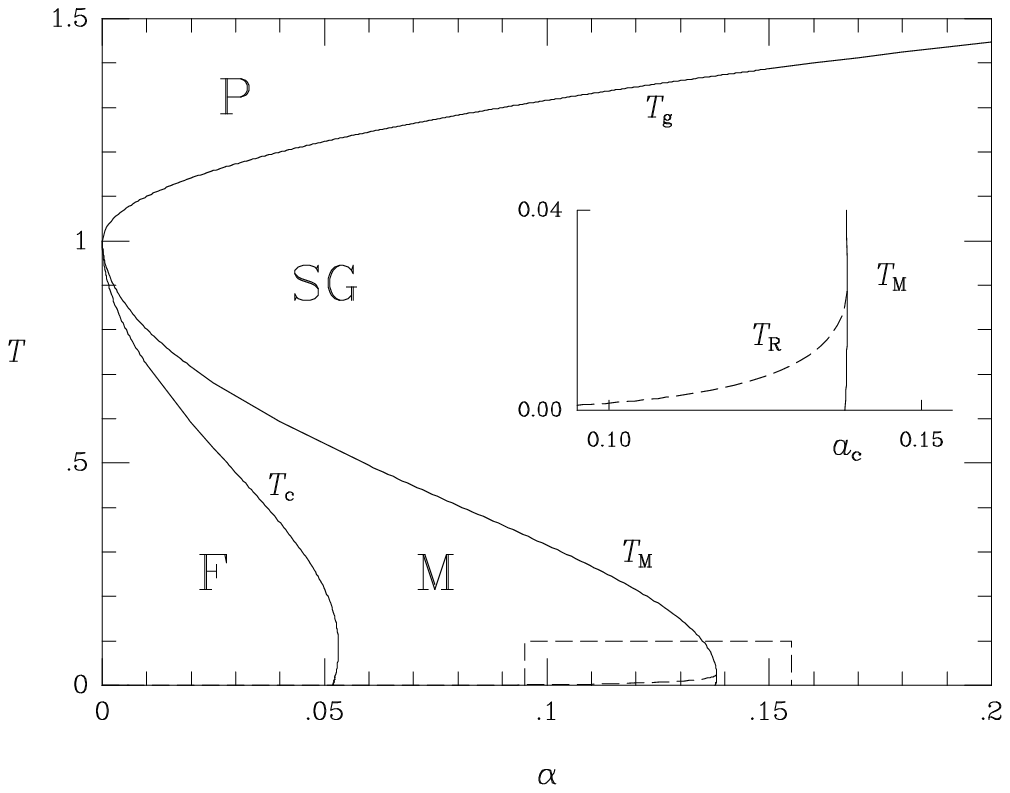}
\end{center}
\vspace*{-5mm}
\caption{Phase diagram of the Hopfield model. P: paramagnetic phase,
$m=q=0$ (no recall). SG: spin-glass phase, $m=0$, $q\neq 0$ (no recall). F: pattern recall
phase (recall states minimise $f$), $m\neq 0$, $q\neq 0$. M: mixed
phase (recall states are local but not global minima of $f$).
Solid lines: separations of the above phases ($T_g$: second order,
$T_M$ and $T_c$: first order).
Dashed: the AT instability for the recall solutions ($T_R$).
Inset: close-up of the low temperature region.}
\label{fig:hopfieldphasediagram}
\end{figure}
If we solve the equations
(\ref{eq:satmq}) numerically for different values of
$\alpha$, and calculate the corresponding `free energies' $f$
(\ref{eq:satf}) for the pure states and the
spin-glass state $\bm=0$,
we obtain figure \ref{fig:satamplitudes}. For
$\alpha>0$ the nontrivial solution $m$ for the amplitude of the pure
state appears {\em discontinously} as the temperature is lowered, defining a
critical temperature $T_M(\alpha)$. Once the pure state appears, it
turns out to be locally stable (within the RS ansatz). Its `free
energy' $f$, however,  remains larger than the one corresponding to the
spin-glass
state, until
the temperature is further reduced to below a second critical temperature
$T_c(\alpha)$. For $T<T_c(\alpha)$ the pure states are therefore the
equilibrium states in the thermodynamics sense.

By drawing these critical lines in the $(\alpha,T)$ plane, together
with the line $T_g(\alpha)=1+\sqrt{\alpha}$ which signals the second
order transition from the paramagnetic to the spin-glass state, we
obtain the RS phase diagram of the Hopfield model, depicted in figure
\ref{fig:hopfieldphasediagram}.
Strictly speaking the line $T_M$ would appear meaningless in the thermodynamic
picture, only the saddle-point that minimises $f$ being relevant.
However, we have to keep in mind the physics behind the formalism.
The occurrence of multiple locally stable saddle-points
is the manifestation of ergodicity breaking in the limit
$N\rightarrow\infty$.
The thermodynamic analysis, based on ergodicity,
therefore applies only within a single ergodic component. Each locally
stable saddle-point is indeed relevant for appropriate initial conditions and
time-scales.
\vsp

\noindent{\em Zero Temperature, Storage Capacity}.
The storage capacity $\alpha_c$ of the
Hopfield model is defined as the largest $\alpha$ for which locally stable pure
states exist. If for the moment we neglect the low temperature
re-entrance peculiarities in the phase diagram
(\ref{fig:hopfieldphasediagram}) to which we will come back later, the critical temperature $T_M(\alpha)$,
where the pure states
appear decreases monotonically with $\alpha$, and the storage capacity is
reached for $T=0$. Before we can put $T\rightarrow0$ in
$(\ref{eq:satmq})$, however,  we will have to rewrite these
equations in terms
of quantities with well defined $T\rightarrow0$ limits, since
$q\rightarrow1$. A suitable quantity is $C=\beta(1\minus q)$, which
obeys $0\leq C\leq 1$ for the free energy (\ref{eq:RShopfreeenergy})
to exist. The saddle-point equations can now be written in the form
\bd
m=
\int\!Dz~\tanh
\left[\beta m\plus\frac{z\beta\sqrt{\alpha q}}{1\minus C}\right]
~~~~~~~~~~
C=\frac{\partial}{\partial m}
\int\!Dz~\tanh
\left[\beta m+\frac{z\beta\sqrt{\alpha q}}{1\minus C}\right]
\ed
in which the limit $T\rightarrow0$ simply corresponds to $\tanh(\beta
x)\rightarrow\sgn(x)$ and $q\rightarrow1$. After having taken the
limit we perform the Gaussian integral:
\bd
m=\erf\left[\frac{m(1-C)}{\sqrt{2\alpha}}\right]~~~~~~~~~~
C=(1\minus C)\sqrt{\frac{2}{\alpha\pi}}e^{-m^2(1-C)^2/2\alpha}
\ed
\begin{figure}[t]
\centering
\vspace*{53mm}
\hbox to \hsize{\hspace*{-10mm}\includegraphics{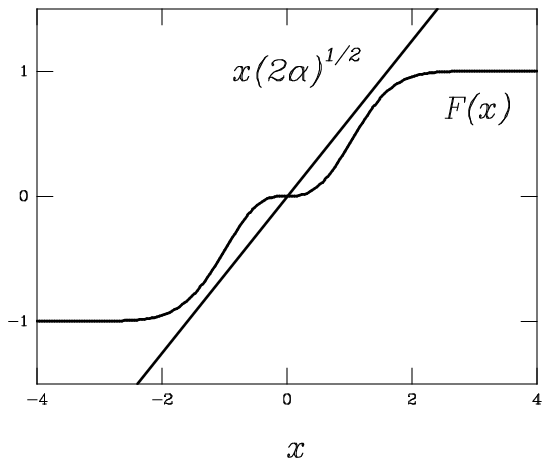}\hspace*{0mm}}
\vspace*{-52.5mm}\hspace*{60mm}
\epsfxsize=83mm\epsfbox{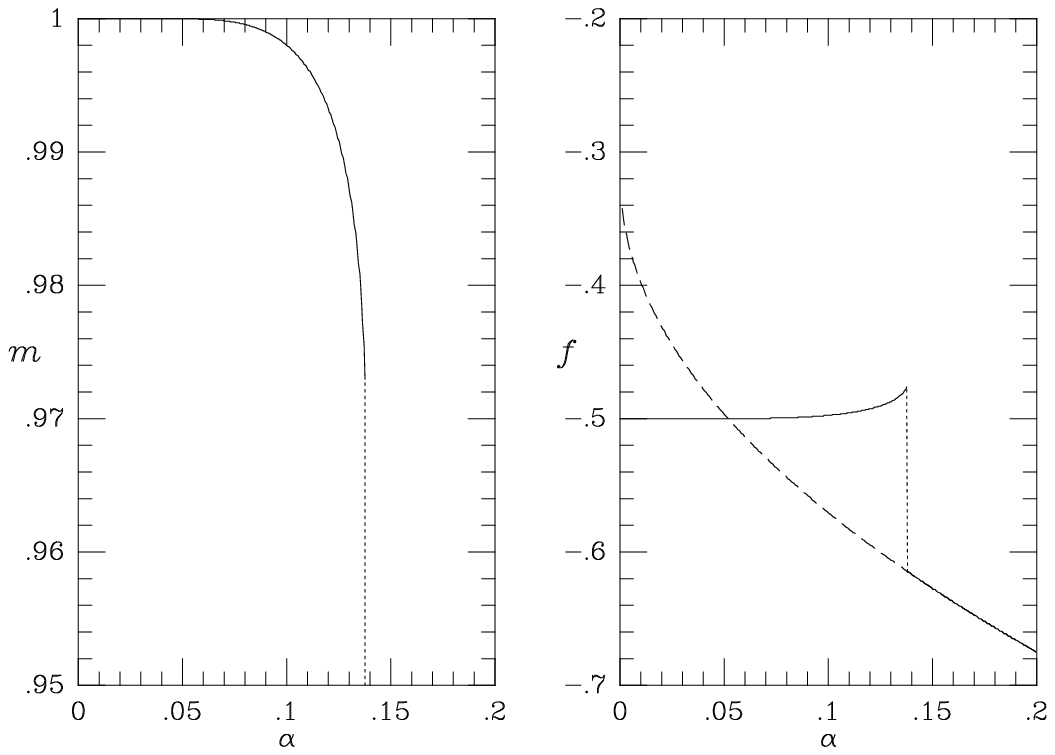}
\vspace*{-0mm}
\caption{Left: solution of the transcendental equation
$F(x)=x\protect\sqrt{2\alpha}$, where $x=\protect\erf^{\rm inv}(m)$.
The storage
capacity $\alpha_c\sim 0.138$ of the Hopfield model is the largest $\alpha$ for which
solutions $x\neq 0$ exist.
Middle picture: RS amplitudes $m$ of the pure
states of the Hopfield model for $T=0$ as
a function of $\alpha=p/N$. The location of the discontinuity, where
$m$ vanishes, defines the storage capacity $\alpha_c\sim0.138$.
Right picture, solid line: $T=0$ `free
energy' $f$ of the pure states.
Dashed lines:
$T=0$ `free energy' of the spin-glass state $\bm=\bnul$ (for
comparison).}
\label{fig:groundstatenergies}
\end{figure}
This set can be reduced to a single transcendental equation by introducing
$x=m(1\minus C)/\sqrt{2\alpha}$:
\be
x\sqrt{2\alpha}=F(x)
~~~~~~~~~~~~~~
F(x)=\erf(x)-\frac{2x}{\sqrt{\pi}}e^{-x^2}
\label{eq:storagecap}
\ee
Equation (\ref{eq:storagecap}) is solved numerically (see figure
\ref{fig:groundstatenergies}). Since $F(x)$ is anti-symmetric, solutions
 come in pairs $(x,-x)$ (reflecting the symmetry of the Hamiltonian of
the system with respect to an overall state-flip
$\bsigma\rightarrow-\bsigma$).  For $\alpha<\alpha_c\sim 0.138$ there
indeed exist pure state solutions $x\neq 0$. For $\alpha>\alpha_c$
there is only the spin-glass solution $x=0$.
Given a solution $x$ of (\ref{eq:storagecap}), the zero temperature
values for the order parameters follow from
\bd
\lim_{T\rightarrow 0}m=\erf[x]~~~~~~~~~~~~
\lim_{T\rightarrow 0}C=\left[1\plus \sqrt{\frac{\alpha\pi}{2}}e^{x^2}\right]^{-1}
\ed
with which in turn we can take the zero temperature limit in our
expression (\ref{eq:satf}) for the free energy:
\bd
\lim_{T\rightarrow 0} f=\frac{1}{2}\erf^2[x]
+\frac{1}{\pi}e^{-x^2}\!
-\frac{2}{\pi}\left[e^{-x^2}\!\plus\sqrt{\frac{\alpha\pi}{2}}\right]
\left[x\sqrt{\pi}~\erf[x]\plus e^{-x^2}\room\right]
\ed
Comparison of the values for $\lim_{T\rightarrow0}f$ thus obtained, for the pure state
$m>0$ and the spin-glass state $m=0$ leads to figure
\ref{fig:groundstatenergies}, which clearly shows that for
sufficiently small $\alpha$ the pure states
are the true ground states of the system.
\vsp

\noindent{\em The AT-Instability.}
As in the case of the Gaussian model (\ref{eq:SKinteractions}), the above RS solution again generates negative
entropies at sufficiently low temperatures, indicating that
replica-symmetry must be broken. We can locate continuous replica symmetry
breaking by studying the effect on $f(\bm,\bq,\hat{\bq})$
(\ref{eq:fullhopfieldsurface}) of small
replicon \cite{AT} fluctuations
around the RS solution:
\be
q_{\alpha\beta}\rightarrow\delta_{\alpha\beta}+q\left[1\minus
\delta_{\alpha\beta}\right]+\eta_{\alpha\beta},
~~~~~~~~~~
\eta_{\alpha\beta}=\eta_{\beta\alpha}~~~~~\eta_{\alpha\alpha}=0~~~~~\sum_{\alpha}\eta_{\alpha\beta}=0
\label{eq:repliconproperties}
\ee
The variation of $\bq$
induces a similar variation in the conjugate parameters $\hat{\bq}$ through equation
(\ref{eq:hopsaddleqhatab}):
\bd
\hat{q}_{\alpha\beta}\rightarrow\frac{1}{2}i\alpha\beta^2\left[ R
\delta_{\alpha\beta}+ r\left[1\minus \delta_{\alpha\beta}\right]+\hat{\eta}_{\alpha\beta}\right]
~~~~~~~~~~~~
\hat{\eta}_{\alpha\beta}=\frac{1}{2}\sum_{\gamma\delta}\eta_{\gamma\delta}\left[g_{\alpha\beta\gamma\delta}-g_{\alpha\beta}g_{\gamma\delta}\right]
\vspace*{-2mm}
\ed
with
\bd
g_{\alpha\beta\gamma\delta}=
\frac{\int\!d\bz~z_{\alpha}z_{\beta}z_{\gamma}z_{\delta} e^{-\frac{1}{2}\bz\cdot\left[\one-\beta
\bq_{\rm RS}\right]\bz}}
{\int\!d\bz~e^{-\frac{1}{2}\bz\cdot\left[\one-\beta \bq_{\rm RS}\right]\bz}}
~~~~~~~~~~
g_{\alpha\beta}=
\frac{\int\!d\bz~z_{\alpha}z_{\beta} e^{-\frac{1}{2}\bz\cdot\left[\one-\beta
\bq_{\rm RS}\right]\bz}}
{\int\!d\bz~e^{-\frac{1}{2}\bz\cdot\left[\one-\beta \bq_{\rm RS}\right]\bz}}
\ed
Wick's theorem (see e.g. \cite{Zinn-Justin}) can now be used to write everything in terms of second
moments of the Gaussian integrals only:
\bd
g_{\alpha\beta\gamma\delta}=
g_{\alpha\beta}g_{\gamma\delta}+
g_{\alpha\gamma}g_{\beta\delta}+
g_{\alpha\delta}g_{\beta\gamma}
\ed
with which we can express the replicon variation in $\hat{\bq}$, using
the symmetry of $\{\eta_{\alpha\beta}\}$ and the saddle-point equation (\ref{eq:hopsaddleqhatab}), as
\be
\hat{\eta}_{\alpha\beta}=\sum_{\gamma\delta}
g_{\alpha\gamma}\eta_{\gamma\delta}g_{\delta\beta}
 =\beta^2\sum_{\gamma\neq\delta}\left[R\delta_{\alpha\gamma}\plus
r\left[1\minus
\delta_{\alpha\gamma}\right]\right]\eta_{\gamma\delta}\left[R\delta_{\delta\beta}\plus
r\left[1\minus \delta_{\delta\beta}\right]\right]
 =\beta^2(R\minus r)^2\eta_{\alpha\beta}
\label{eq:etarelation}
\ee
since only those terms can contribute which involve precisely two
$\delta$-symbols, due to $\sum_{\alpha}\eta_{\alpha\beta}=0$.
We can now calculate the
 change in $f(\bm,\bq,\hat{\bq})$, away from the RS value $f(\bm_{\rm
RS},\bq_{\rm RS},\hat{\bq}_{\rm RS})$,
the leading order of which must be quadratic in the fluctuations
$\{\eta_{\alpha\beta}\}$ since the RS
solution
is a saddle-point:
\bd
f(\bm_{\rm RS},\bq,\hat{\bq})-f(\bm_{\rm RS},\bq_{\rm
RS},\hat{\bq}_{\rm RS})
=\frac{1}{\beta
n}\left[\frac{1}{2}\alpha
\log \frac{\det\left[\one\minus\beta(\bq_{\rm RS}\plus\bEta)\right]}
{\det\left[\one\minus\beta\bq_{\rm RS}\right]}
-i{\rm Tr}\left[\hat{\bq}_{\rm RS}.\bEta\right]
~~~~~~~~~~~~~~~~~~~~
\right.
\ed
\be
\left.
~~~~~~~~~~~~~~~~~~~~
+\frac{1}{2}\alpha\beta^2{\rm Tr}\left[
\hat{\bEta}.\bEta\plus
\hat{\bEta}.\bq_{\rm RS}\right]
-\bra
\log \frac{\bra
e^{\beta\bxi\cdot\bm_{\rm
RS}\sum_{\alpha}\sigma_{\alpha}-i\bsigma\cdot[\hat{\bq}_{\rm
RS}+\frac{1}{2}i\alpha\beta^2\hat{\bEta}]\bsigma}
\ket_{\bsigma}}
{\bra
e^{\beta\bxi\cdot\bm_{\rm
RS}\sum_{\alpha}\sigma_{\alpha}-i\bsigma\cdot\hat{\bq}_{\rm RS}\bsigma}\ket_{\bsigma}}
\ket_{\bxi}
\right]
\label{eq:energychange}
\ee
Evaluating (\ref{eq:energychange})
is simplified by the fact that the matrices $\bq_{\rm RS}$ and
$\bEta$ commute, which is a direct consequence of the properties
(\ref{eq:repliconproperties}) of the replicon fluctuations and the
form of the replica-symmetric saddle-point. If we define the $n\!\times
\!n$ matrix $\bP$ as the
projection onto the vector $(1,\ldots,1)$, we have
\be
P_{\alpha\beta}=n^{-1}~~~~~~~~\bP.\bEta=\bEta.\bP=0
~~~~~~~~
\bq_{\rm
RS}=(1\minus q)\one+nq \bP
~~~~~~~~
\bq_{\rm RS}.\bEta=\bEta.\bq_{\rm RS}=(1\minus q)\bEta
\label{eq:commrelations}
\ee
\bd
\left[\one\minus\beta\bq_{\rm RS}\right]^{-1}=
\frac{1}{1\minus\beta(1\minus q)}
\one
\plus\frac{\beta nq}{\left[1\minus\beta(1\minus q)\minus \beta
nq\right]\left[1\minus\beta(1\minus q)\right]}\bP
\ed
We can now simply expand the relevant terms, using the identity
$\log\det M = {\rm Tr}~\log M$:
\bd
\log \frac{\det\left[\one-\beta(\bq_{\rm RS}+\bEta)\right]}
{\det\left[\one-\beta\bq_{\rm RS}\right]}
={\rm Tr}
 \log\left[\one-\beta\bEta\left[\one-\beta\bq_{\rm
RS}\right]^{-1}\right]
~~~~~~~~~~~~~~~~~~~~~~~~~~~~~~~~~~~~~~~
\ed
\bd
={\rm Tr}\left\{
-\beta\bEta\left[\one-\beta\bq_{\rm RS}\right]^{-1}
-\frac{1}{2}\beta^2\left[\bEta\left[\one-\beta\bq_{\rm
RS}\right]^{-1}\right]^2
\right\}+\order(\bEta^3)
\ed
\be
=
-\frac{1}{2}\frac{\beta^2}{\left[1\minus\beta(1\minus q)\right]^2}
{\rm Tr}~\bEta^2+\order(\bEta^3)
\label{eq:logdetterm}
\ee
Finally we address the remaining term in (\ref{eq:energychange}),
again using the RS saddle-point equations (\ref{eq:RShopm},\ref{eq:RShopqr}) where appropriate:
\bd
\bra\log \frac{\bra
e^{\beta\bxi\cdot\bm_{\rm
RS}\sum_{\alpha}\sigma_{\alpha}-i\bsigma\cdot\hat{\bq}_{\rm
RS}\bsigma}
\left[1\plus \frac{1}{2}\alpha\beta^2\bsigma\cdot\hat{\bEta}\bsigma
\plus\frac{1}{8}\alpha^2\beta^4(\bsigma\cdot\hat{\bEta}\bsigma)^2\plus\ldots\right]
\ket_{\bsigma}}
{\bra
e^{\beta\bxi\cdot\bm_{\rm
RS}\sum_{\alpha}\sigma_{\alpha}-i\bsigma\cdot\hat{\bq}_{\rm RS}\bsigma}\ket_{\bsigma}}\ket_{\bxi}
\ed
\be
=\frac{1}{2}\alpha\beta^2{\rm Tr}[\hat{\bEta}.\bq_{\rm
RS}]+\frac{1}{8}\alpha^2\beta^4\sum_{\alpha\beta\gamma\delta}\hat{\eta}_{\alpha\beta}\hat{\eta}_{\gamma\delta}[G_{\alpha\beta\gamma\delta}\minus
H_{\alpha\beta\gamma\delta}]
+\ldots
\label{eq:spinterm}
\ee
with
\bd
G_{\alpha\beta\gamma\delta}=
\bra\frac{\bra\sigma_{\alpha}\sigma_{\beta}\sigma_{\gamma}\sigma_{\delta}
e^{\beta\bxi\cdot\bm_{\rm
RS}\sum_{\alpha}\sigma_{\alpha}-i\bsigma\cdot\hat{\bq}_{\rm
RS}\bsigma}
\ket_{\bsigma}}
{\bra
e^{\beta\bxi\cdot\bm_{\rm
RS}\sum_{\alpha}\sigma_{\alpha}-i\bsigma\cdot\hat{\bq}_{\rm RS}\bsigma}\ket_{\bsigma}}\ket_{\bxi}
\ed
\bd
H_{\alpha\beta\gamma\delta}=
\bra
\frac{\bra\sigma_{\alpha}\sigma_{\beta}
e^{\beta\bxi\cdot\bm_{\rm
RS}\sum_{\alpha}\sigma_{\alpha}\minus i\bsigma\cdot\hat{\bq}_{\rm
RS}\bsigma}
\ket_{\bsigma}}
{\bra
e^{\beta\bxi\cdot\bm_{\rm
RS}\sum_{\alpha}\sigma_{\alpha}\minus i\bsigma\cdot\hat{\bq}_{\rm
RS}\bsigma}\ket_{\bsigma}}
\frac{\bra\sigma_{\gamma}\sigma_{\delta}
e^{\beta\bxi\cdot\bm_{\rm
RS}\sum_{\alpha}\sigma_{\alpha}\minus i\bsigma\cdot\hat{\bq}_{\rm
RS}\bsigma}
\ket_{\bsigma}}
{\bra
e^{\beta\bxi\cdot\bm_{\rm
RS}\sum_{\alpha}\sigma_{\alpha}\minus i\bsigma\cdot\hat{\bq}_{\rm
RS}\bsigma}\ket_{\bsigma}}
\ket_{\bxi}
\ed
Inserting the ingredients
(\ref{eq:etarelation},\ref{eq:commrelations},\ref{eq:logdetterm},\ref{eq:spinterm})
into expression (\ref{eq:energychange}) and rearranging terms shows
that the linear terms indeed cancel, and that the term involving
$H_{\alpha\beta\gamma\delta}$ does not contribute (since the elements
$H_{\alpha\beta\gamma\delta}$ don't depend on the indices for
$\alpha\neq\beta$ and $\gamma\neq\delta$), and we are left with:
\bd
f(\bm_{\rm RS},\bq,\hat{\bq})-f(\bm_{\rm RS},\bq_{\rm
RS},\hat{\bq}_{\rm RS})
=
\frac{1}{\beta
n}\left[-\frac{1}{4}
\frac{\alpha\beta^2}{\left[1\minus\beta(1\minus q)\right]^2}
{\rm Tr}~\bEta^2
+\frac{1}{2}\alpha\beta^4(R\minus r)^2{\rm Tr}~\bEta^2
~~~~~~~~~~
\right.
\ed
\bd
\left.
~~~~~~~~~~~~~~~~~~~~~~~~~~~~~~~~~~~~~~~~
-\frac{1}{8}\alpha^2\beta^8(R\minus r)^4
\sum_{\alpha\beta\gamma\delta}\eta_{\alpha\beta}\eta_{\gamma\delta}G_{\alpha\beta\gamma\delta}
\right]
+\ldots
\ed
Because of the index permutation symmetry in the neuron average we can
write for $\alpha\neq\gamma$ and $\rho\neq\lambda$:
\bd
G_{\alpha\gamma\rho\lambda}=
\delta_{\alpha\rho}\delta_{\gamma\lambda}\plus
\delta_{\alpha\lambda}\delta_{\gamma\rho}
+ G_4\left[1\minus\delta_{\alpha\rho}\right]\left[1\minus\delta_{\gamma\lambda}\right]\left[1\minus\delta_{\alpha\lambda}\right]\left[1\minus
\delta_{\gamma\rho}\right]
~~~~~~~~~~~~~~~~~~~~
\ed
\bd
~~~~~~~~~~~~~~~~~~~~
+ ~G_2\left\{\delta_{\alpha\rho}\left[1\minus\delta_{\gamma\lambda}\right]\plus
\delta_{\gamma\lambda}\left[1\minus\delta_{\alpha\rho}\right]\plus
\delta_{\alpha\lambda}\left[1\minus\delta_{\gamma\rho}\right]\plus
\delta_{\gamma\rho}\left[1\minus\delta_{\alpha\lambda}\right]\right\}
\ed
with
\bd
G_{\ell}=
\bra\frac{\int\!Dz~\tanh^\ell\beta\left[\bm\cdot\bxi\plus
z\sqrt{\alpha r}\right]\cosh^n \beta\left[\bm\cdot\bxi\plus
z\sqrt{\alpha r}\right]}
{\int\!Dz~\cosh^n \beta\left[\bm\cdot\bxi\plus
z\sqrt{\alpha r}\right]}\ket_{\bxi}
\ed
Only terms which involve precisely two $\delta$-functions can contribute,
because of the replicon properties (\ref{eq:repliconproperties}).
As a result:
\bd
f(\bm_{\rm RS},\bq,\hat{\bq})-f(\bm_{\rm RS},\bq_{\rm
RS},\hat{\bq}_{\rm RS})
=
\frac{1}{\beta
n}{\rm Tr}~\bEta^2\left[
-\frac{1}{4}
\frac{\alpha\beta^2}{\left[1\minus\beta(1\minus q)\right]^2}
+\frac{1}{2}\alpha\beta^4(R\minus r)^2
\right.
~~~~~~~~~~~~~~~~~~~~
\ed
\bd
\left.
~~~~~~~~~~~~~~~~~~~~~~~~~~~~~~~~~~~~~~~~~~~~~~~~~~~~~~~~~~~~
-\frac{1}{4}\alpha^2\beta^8(R\minus r)^4\left[1\minus 2G_2\plus G_4\right]
\right]
+\ldots
\ed
Since ${\rm Tr}~\bEta^2=\sum_{\alpha\beta}\eta_{\alpha\beta}^2$, the condition for the RS solution to minimise $f(\bm,\bq,\hat{\bq})$, if compared
to the `replicon' fluctuations, is
therefore
\be
-\frac{1}{\left[1\minus\beta(1\minus q)\right]^2}
+2\beta^2(R\minus r)^2
-\alpha\beta^6(R\minus r)^4\left[1\minus 2G_2\plus G_4\right]
>0
\label{eq:ATlinebeforelimit}
\ee
After taking the limit in the expressions $G_{\ell}$ and after
evaluating
\bd
\lim_{n\rightarrow0}R= \frac{1}{\beta}\lim_{n\rightarrow0}g_{\alpha\alpha}=\lim_{n\rightarrow0}\frac{1}{n\beta}
\frac{\int\!d\bz~\bz^2 e^{-\frac{1}{2}\bz\cdot\left[\one-\beta
\bq_{\rm RS}\right]\bz}}
{\int\!d\bz~e^{-\frac{1}{2}\bz\cdot\left[\one-\beta \bq_{\rm
RS}\right]\bz}}~~~~~~~~~~~~~~~~~~~~~~~~~~~~~~~~~~~~~~~~~~~
\ed
\bd
=\lim_{n\rightarrow0}\frac{1}{n\beta}\left[
\frac{n-1}{1\minus\beta(1\minus
q)}+\frac{1}{1\minus\beta(1\minus q\plus nq)}
\right]=\frac{1}{\beta}\frac{1\minus\beta\plus 2\beta
q}{\left[1\minus\beta(1\minus q)\right]^2}
\ed
and using (\ref{eq:RShopqr}), the condition (\ref{eq:ATlinebeforelimit}) can be written as
\be
[1\minus\beta(1\minus q)]^2 ~>~ \alpha\beta^2
\bra\int\!Dz~\cosh^{-4}\beta\left[\bm\cdot\bxi\plus
z\sqrt{\alpha r}\right]\ket_{\bxi}
\label{eq:hopATline}
\ee
The AT line in the phase diagram, where this condition ceases to be met,
indicates a second-order transition to a spin-glass state where ergodicity is
broken in the sense that the distribution
$\overline{P(q)}$ (\ref{eq:hopoverlapdistribution}) is no longer a
$\delta$-function.
In the paramagnetic regime of the phase diagram, $\bm=\bnul$ and $q=0$, the
AT condition reduces precisely to $T>T_{g}=1\plus\sqrt{\alpha}$. Therefore the
paramagnetic solution is stable. The AT line coincides with the
boundary between the paramagnetic and spin-glass phase. Numerical
evaluation of (\ref{eq:hopATline}) shows that the RS spin-glass
solution remains unstable for all $T<T_g$, but that the retrieval
solution $\bm\neq\bnul$ is unstable only for very low temperatures
$T<T_{R}$ (see figure \ref{fig:hopfieldphasediagram}).

\section{Epilogue}

In this paper I have tried to give a self-contained expos\'{e} of
the main issues, models and mathematical techniques relating to
the equilibrium statistical mechanical analysis of  recurrent neural networks.
I have included networks of binary neurons and
networks of coupled (neural) oscillators, with various degrees of synaptic
complexity (albeit always fully connected), ranging from uniform synapses, via synapses storing a
small number of patterns, to  Gaussian synapses and synapses
encoding an extensive number of stored patterns.
The latter (complex) cases I only worked out for binary neurons; similar
calculations can be done for coupled oscillators (see \cite{Cook}).
Networks of graded
response neurons could not be included, because these are found
never to go to (detailed balance) equilibrium, ruling out
equilibrium statistical mechanical analysis.
All analytical results and predictions have later also been confirmed
comprehensively by numerical simulations.
Over the years we have learned an
impressive amount about the operation of recurrent networks by
thinking in terms of free energies and phase transitions, and by
having been able to derive explicit analytical solutions (since a good
theory always supersedes an infinite number of simulation experiments
...).
I have given a number
of key references along the way; many could have been added but
were left out for practical reasons. Instead I will just mention a
number of textbooks in which more science as well as more references to research papers
can be found. Any such selection is obviously highly subjective,
and I wish to apologize beforehand to the authors which I regret
to have omitted.
Several relevant review papers dealing with the statistical mechanics
of neural networks can be found scattered over the three volumes
\cite{DHS1,DHS2,DHS3}. Textbooks which attempt to take the
interested but non-expert reader towards the expert level are \cite{Peretto,CoolenSherr}.
Finally, a good introduction to the methods and backgrounds of
replica theory, together with a good collection of reprints of original
papers, can be found in \cite{Mezardetal}.

What should we expect for the
next decades, in the equilibrium statistical mechanics of
recurrent neural networks ? Within the confined area of large symmetric and fully connected
 recurrent
networks with simple neuron types we can now deal with fairly complicated
choices for the synapses, inducing complicated energy landscapes
with many stable states, but this involves non-trivial and
cutting-edge mathematical techniques.
If our basic driving force next is  the
aim to bring our models closer to biological reality, balancing
the need to retain mathematical solvability with the desire to
bring in more details of the various electro-chemical processes known
to occur in neurons and synapses and spatio-temporal characteristics of dendrites,
the boundaries of what can be done with equilibrium statistical
mechanics
are, roughly speaking, set by the three key issues of (presence or absence of)
detailed balance, system size, and
synaptic interaction range.
The first issue is vital: no detailed balance immediately implies no
equilibrium statistical mechanics. This generally rules out
networks with non-symmetric synapses and all
networks of graded response neurons (even when the latter are equipped with
symmetric synapses). The issue of system size is slightly less
severe; models of networks with  $N<\infty$ neurons can often be solved in leading order
in $N^{-\frac{1}{2}}$, but  a price will have to be paid in
the form of a reduction of our ambition elsewhere (e.g. we might have to restrict ourselves
to simpler choices of
synaptic interactions). Finally,  we know how to deal with
fully connected models (such as those discussed in this paper), and
also with models having dendritic structures which cover a long (but not infinite) range, provided
they  vary smoothly with
distance. We can also deal with short-range dendrites in one-dimensional (and to a lesser  extent two
dimensional)
networks; however, since even the relatively  simple Ising model
(mathematically equivalent to a network of binary neurons
with uniform synapses connecting only nearest-neighbour neurons)
has so far not yet  been solved in
three dimensions,  it is not realistic to assume that analytical
solution will be possible soon of general recurrent neural network models with short range
interactions.
On balance, although there are still many interesting puzzles to keep theorists
happy for years to come, and although many of the model types discussed in this text will continue
to be useful building blocks in explaining at a basic and qualitative level the operation of specific
recurrent brain regions (such as the CA3 region of the hippocampus), one is therefore led to the conclusion that
equilibrium statistical mechanics has by now brought us as far as can be
expected with regard to increasing our understanding of biological neural networks.
Dale's law already rules out synaptic symmetry, and thereby equilibrium statistical mechanics altogether,
so we are forced to turn to dynamical techniques if we wish to
improve biological realism.

\subsection*{Acknowledgements}

I is my great pleasure to thank David Sherrington and Nikos Skantzos for
their contributions to content and presentation of this review.


\end{document}

%% file: biophyscommands.tex
\newcommand{\room}{\rule[-0.3cm]{0cm}{0.8cm}}

\newcommand{\vsp}{\vspace*{3mm}}
\newcommand{\hsp}{\hspace*{3mm}}
\newcommand{\nsp}{\vspace*{-3mm}}
\newcommand{\be}{\begin{equation}}
\newcommand{\ee}{\end{equation}}
\newcommand{\bd}{\begin{displaymath}}
\newcommand{\ed}{\end{displaymath}}
\newcommand{\bea}{\begin{eqnarray}}
\newcommand{\eea}{\end{eqnarray}}
\newcommand{\bean}{\begin{eqnarray*}}
\newcommand{\eean}{\end{eqnarray*}}

\newcommand{\one}{1\!\!{\rm I}}
\newcommand{\sgn}{~{\rm sgn}}
\newcommand{\extr}{~{\rm extr}}

\newcommand{\bra}{\langle}
\newcommand{\ket}{\rangle}

\newcommand{\order}{{\cal O}}
\newcommand{\minus}{\!-\!}
\newcommand{\plus}{\!+\!}
\newcommand{\erf}{{\rm erf}}

\newcommand{\inn}{{\vspace*{-0.1mm}\cdot\vspace*{-0.1mm}}}
\newcommand{\bnul}{\mbox{\boldmath $0$}}

\newcommand{\bbf}{\mbox{\protect\boldmath $f$}}

\newcommand{\bm}{\mbox{\protect\boldmath $m$}}
\newcommand{\bq}{\mbox{\protect\boldmath $q$}}

\newcommand{\bu}{\mbox{\protect\boldmath $u$}}

\newcommand{\bx}{\mbox{\protect\boldmath $x$}}
\newcommand{\by}{\mbox{\protect\boldmath $y$}}
\newcommand{\bz}{\mbox{\protect\boldmath $z$}}

\newcommand{\bP}{\mbox{\protect\boldmath $P$}}

\newcommand{\bsigma}{\mbox{\protect\boldmath $\sigma$}}

\newcommand{\bphi}{\mbox{\protect\boldmath $\phi$}}

\newcommand{\bxi}{\mbox{\protect\boldmath $\xi$}}

\newcommand{\blambda}{\mbox{\protect\boldmath $\lambda$}}

\newcommand{\bEta}{\mbox{\protect\boldmath $\eta$}}

\newcommand{\bLambda}{\mbox{\protect\boldmath $\Lambda$}}

\newcommand{\hbq}{\hat{\mbox{\protect\boldmath $q$}}}